\documentclass[emulateapj]{aastex62}

\usepackage{amsmath, amssymb}
\usepackage{color}
\received{\today}
\revised{}
\accepted{}

\shortauthors{Chuss et al.}

\begin{document}

\title{HAWC+/SOFIA Multiwavelength Polarimetric Observations of OMC-1}

\correspondingauthor{David T. Chuss}
\email{david.chuss@villanova.edu}

\author[0000-0003-0016-0533]{David T. Chuss}
\affil{Department of Physics, Villanova University, 800 E. Lancaster Ave., Villanova, PA 19085, USA}

\author{B-G Andersson}
\affil{SOFIA Science Center/Universities Space Research Association}
\affil{NASA Ames Research Center, M.S. N232-12, Moffett Field, CA, 94035, USA}

\author{John Bally}
\affil{Astrophysical and Planetary Sciences Department, University of Colorado, UCB 389 Boulder, Colorado 80309, USA}

\author[0000-0003-4206-5649]{Jessie L. Dotson}
\affil{NASA Ames Research Center, M.S. N232-12, Moffett Field, CA, 94035, USA}

\author{C. Darren Dowell}
\affil{NASA Jet Propulsion Laboratory, California Institute of Technology, 4800 Oak Grove Drive, Pasadena, CA 91109, USA}

\author[0000-0001-8819-9648]{Jordan A. Guerra}
\affil{Department of Physics, Villanova University, 800 E. Lancaster Ave., Villanova, PA 19085, USA}

\author{Doyal A. Harper}
\affil{Department of Astronomy and Astrophysics, University of Chicago, Chicago, IL 60637, USA}
\affil{Yerkes Observatory, Williams Bay, WI, USA}

\author[0000-0003-4420-8674]{Martin Houde}
\affil{Department of Physics and Astronomy, University of Western Ontario, 1151 Richmond Street, London, ON N6A 3K7, Canada}

\author{Terry Jay Jones}
\affil{Minnesota Institute for Astrophysics, University of Minnesota, Minneapolis, MN 55455, USA}

\author{A. Lazarian}
\affil{Department of Astronomy, University of Wisconsin, Madison, WI 53706, USA}

\author{Enrique Lopez Rodriguez}
\affil{SOFIA Science Center/Universities Space Research Association}
\affil{NASA Ames Research Center, M.S. N232-12, Moffett Field, CA, 94035, USA}

\author[0000-0003-3503-3446]{Joseph M. Michail}
\affil{Department of Astrophysics and Planetary Science, Villanova University, 800 E. Lancaster Ave., Villanova, PA 19085, USA}
\affil{Department of Physics, Villanova University, 800 E. Lancaster Ave., Villanova, PA 19085, USA}

\author[0000-0002-6753-2066]{Mark R. Morris}
\affil{Department of Physics and Astronomy, University of California, Los Angeles, Box 951547, Los Angeles, CA 90095-1547 USA}

\author{Giles Novak}
\affil{Center for Interdisciplinary Exploration and Research in Astrophysics (CIERA), and Department of Physics \& Astronomy, Northwestern University, 2145 Sheridan Rd, Evanston, IL, 60208, USA}

\author{Javad Siah}
\affil{Department of Physics, Villanova University, 800 E. Lancaster Ave., Villanova, PA 19085, USA}

\author{Johannes Staguhn}
\affil{Dept. of Physics \& Astronomy, Johns Hopkins University, Baltimore, MD 21218, USA}
\affil{NASA Goddard Space Flight Center, Greenbelt, MD 20771, USA}

\author{John E. Vaillancourt}
\affil{Lincoln Laboratory, Massachusetts Institute of Technology, 244 Wood Street, Lexington, Massachusetts 02420-9108, USA}

\author{C. G. Volpert}
\affil{University of Chicago, Chicago, IL 60637, USA}

\author{Michael Werner}
\affil{NASA Jet Propulsion Laboratory, California Institute of Technology, 4800 Oak Grove Drive, Pasadena, CA 91109, USA}

\author[0000-0002-7567-4451]{Edward J. Wollack}
\affil{NASA Goddard Space Flight Center, Greenbelt, MD 20771, USA}

\author[0000-0002-9884-4206]{Dominic J. Benford}
\affil{NASA Headquarters, 300 E Street SW, Washington DC  20546, USA}

\author{Marc Berthoud}
\affil{Yerkes Observatory, Williams Bay, WI, USA}

\author{Erin G. Cox}
\affil{Center for Interdisciplinary Exploration and Research in Astrophysics (CIERA), and Department of Physics \& Astronomy, Northwestern University, 2145 Sheridan Rd, Evanston, IL, 60208, USA}

\author{Richard Crutcher}
\affil{Department of Astronomy, University of Illinois, 1002 West Green Street, Urbana, IL 61801, USA}

\author{Daniel A. Dale}
\affil{Department of Physics \& Astronomy, University of Wyoming, Laramie, WY, USA}

\author{L. M. Fissel}
\affil{National Radio Astronomy Observatory, 520 Edgemont Rd, Charlottesville, VA, USA}

\author{Paul F. Goldsmith}
\affil{NASA Jet Propulsion Laboratory, California Institute of Technology, 4800 Oak Grove Drive, Pasadena, CA 91109, USA}

\author[0000-0001-6350-2209]{Ryan T. Hamilton}
\affil{Lowell Observatory, 1400 W Mars Hill Rd, Flagstaff, AZ 86001, USA}

\author[0000-0002-8702-6291]{Shaul Hanany}
\affil{School of Physics and Astronomy, University of Minnesota / Twin Cities,Minneapolis, MN, 55455, USA}

\author{Thomas K. Henning}
\affil{Max Planck Institute for Astronomy, K\"{o}nigstuhl 17, D-69117 Heidelberg, Germany}

\author[0000-0002-4540-6587]{Leslie W. Looney}
\affil{Department of Astronomy, University of Illinois, 1002 West Green Street, Urbana, IL 61801, USA}

\author{S. Harvey Moseley}
\affil{NASA Goddard Space Flight Center, Greenbelt, MD 20771, USA}

\author[0000-0002-9650-3619]{Fabio P. Santos}
\affil{Max Planck Institute for Astronomy, K\"{o}nigstuhl 17, D-69117 Heidelberg, Germany}

\author{Ian Stephens}
\affil{Harvard-Smithsonian Center for Astrophysics, 60 Garden Street, Cambridge, MA, USA}

\author[0000-0002-8831-2038]{Konstantinos Tassis}
\affil{Department of Physics and ITCP, University of Crete, Voutes, GR-71003 Heraklion, Greece}
\affil{IESL and Institute of Astrophysics,
Foundation for Research and Technology-Hellas, PO Box 1527, GR-71110 Heraklion, Greece}

\author{Christopher Q. Trinh}
\affil{USRA/SOFIA, NASA Armstrong Flight Research Center, Building 703, Palmdale, CA 93550, USA}

\author{Eric Van Camp}
\affil{Center for Interdisciplinary Exploration and Research in Astrophysics (CIERA), and Department of Physics \& Astronomy, Northwestern University, 2145 Sheridan Rd, Evanston, IL, 60208, USA}

\author{Derek Ward-Thompson}
\affil{Jeremiah Horrocks Institute, University of Central Lancashire, Preston PR1 2HE, United Kingdom}

\collaboration{(HAWC+ Science Team)}



\begin{abstract}
 We report new polarimetric and photometric maps of the massive star-forming region OMC-1 using the HAWC+ instrument on the Stratospheric Observatory for Infrared Astronomy (SOFIA).  We present continuum polarimetric and photometric measurements of this region at 53, 89, 154, and 214 $\mu$m at angular resolutions of 5.1\arcsec, 7.9\arcsec, 14.0\arcsec, and 18.7\arcsec\ for the four bands, respectively. The photometric maps enable the computation of improved SEDs for the region. We find that at the longer wavelengths, the inferred magnetic field configuration matches the ``hourglass'' configuration seen in previous studies, indicating magnetically-regulated star formation. The field morphology differs at the shorter wavelengths. The magnetic field inferred at these wavelengths traces the bipolar structure of the explosive  Becklin-Neugebauer (BN)/Kleinman-Low (KL) outflow emerging from OMC-1 behind the Orion Nebula. Using statistical methods to estimate the field strength in the region, we find that the explosion dominates the magnetic field near the center of the feature. Farther out, the magnetic field is close to energetic equilibrium with the ejecta and may be providing confinement to the explosion. The correlation between polarization fraction and the local polarization angle dispersion indicates that the depolarization as a function of unpolarized intensity is a result of intrinsic field geometry as opposed to decreases in grain alignment efficiency in denser regions.

\end{abstract}


\section{Introduction} \label{sec:intro}
Located at a distance of 390 pc \citep{Kounkel2017}, the Orion Nebula has been well studied as the nearest example of a region of massive star formation. The OMC-1 part of this complex is located behind an \ion{H}{2} region that is ionized by the Trapezium cluster of O-B stars.  The main feature on the west side of OMC-1 is the Molecular Ridge, which is oriented roughly North-South and contains the Becklin-Neugebauer (BN) object \citep{Becklin1967}, a massive young stellar object, and the Kleinman-Low Nebula (KL) \citep{Kleinmann1967} that consists of molecular gas and dust surrounding additional massive stars. 

The BN/KL region contains a bi-directional outflow \citep{Allen1993} oriented approximately perpendicular to the molecular ridge having a total kinetic energy of 2-6 $\times 10^{47}$ ergs \citep{Bally2011}.  This outflow is traced by CO and H$_2$ emission \citep{Bally2011,Bally2017} and is thought to have been produced by the dynamical decay of stellar orbits near the center of the explosion roughly 500 years ago. This explosion has been identified with the same dynamical event that ejected several massive stars, including BN, from the core.

To the Southeast of the Molecular Ridge and \ion{H}{2} region created by the Trapezium stars is the Orion Bar, which bounds the \ion{H}{2} region and contains a Photon Dominated Region (PDR) at the boundary between the \ion{H}{2} region and the molecular material. 
The dynamical importance of the magnetic field in OMC-1 is of interest, in part because of the relatively high ($\sim$mG) fields estimated in the region by previous studies \citep{Johnston1989, Heiles1993, Pattle2017}.

A key technique for studying magnetic fields in star forming regions is far-infrared and submillimeter polarimetry \citep{Hildebrand2000}. 
Interstellar dust grains can become aligned with their long axis perpendicular to the magnetic field direction via a process known as radiative alignment torque (RAT) \citep{dolginov1976,draine1997,Lazarian2007}. In this scenario, an anisotropic radiation field at wavelengths less than the grain diameter imparts an angular momentum to the grains.  For grains with paramagnetic bulk properties, solid body rotation is traded for quantum spin-flips in the nuclei of the constituent atoms - lowering the total energy of the system, while conserving angular momentum - a process known as the Barnett effect.  The resulting magnetization of the grain causes the angular momentum of the grain to undergo Larmor precession around the external magnetic field direction and, under the continued radiative torques, to align the grain angular momentum with the field. Because grains preferentially rotate about their axis of greatest moment of inertia, the observed polarization direction is perpendicular to the magnetic field direction projected on the plane of the sky.  In regions of extremely strong radiation fields (or for non-paramagnetic grains), the reference direction of the alignment can shift from the magnetic field ($B$-RAT), to that of the radiation field $k$-vector ($k$-RAT) as discussed by \citet{Lazarian2007}. 

\citet{Schleuning1998} mapped OMC-1 using far-infrared polarimetry at 100 \micron\ and submillimeter polarimetry at 350 \micron\ with angular resolutions of 35\arcsec\ and 18\arcsec, respectively. These authors suggested that the magnetic field in this region is highly ordered with a general direction oriented northwest-southeast. The field also exhibits a ``pinch'' in the orthogonal direction. This ``hourglass'' shape has been interpreted to indicate that the star formation in  OMC-1 is magnetically regulated. That is, the field supports the cloud against gravitational collapse in the direction perpendicular to the magnetic field direction. \citet{Vallee1999} measured the polarization at eight positions in OMC-1 at 760 \micron, finding a similar inferred magnetic field direction. \citet{Houde2004} presented a larger map of the OMC-1 region at 350 \micron\ and found general verification of the hourglass pattern. These authors also note that the polarization of the Bar does not follow the hourglass shape and note the low polarization, suggesting poor grain alignment as an explanation.

More recently, SCUBA-2/JCMT has measured the polarization at 850 \micron\ with an angular resolution of 14\arcsec\  \citep{Ward-Thompson2017}. These authors suggest that the low polarization in the Bar could be due to variation in the magnetic field structure (e.g., a helical structure in the PDR). They also measure a field parallel to the Northwest Filament and connect this result to the work of \citet{Soler2013}, who identify a statistical trend of magnetic field direction perpendicular to dense filamentary structures and parallel to low-density filamentary structures as an indicator of sub-Alfvenic dynamics. \citet{Pattle2017} estimate the field strength from the 850 \micron\ data through the use of the Davis-Chandrasekhar-Fermi \citep{Davis1951,Chandrasekhar1953} technique in combination with a technique related to unsharp masking to separate the turbulent contribution to the angular dispersion from the large-scale field. They find the resulting field strength to be $6.6\pm4.7$ mG. These authors also conclude that the BN/KL outflow is regulated by the field and that the outflow is not responsible for creating the hourglass geometry.

\citet{Tang2010} have measured polarization over a small area in the core of the BN/KL region using the SMA with the highest resolution to date (1\arcsec\ at 870 \micron).  These results indicate that the grains are most likely magnetically-aligned and that magnetic field structure has features below the typical resolution of single dish polarimeters. \cite{poid11} mapped the region using visual and near-infrared polarimetry of stars that mostly samples the magnetic field geometry in the foreground of OMC-1 at visual wavelengths and the lines of sight to bright, embedded sources such as BN at near-infrared wavelengths.

In this work, we present far-infrared polarimetry and photometry in four bands from 53 to 214 $\mu$m from the High-resolution Airborne Wideband Camera-Plus (HAWC+) \citep{Harper2018} on the Stratospheric Observatory for Infrared Astronomy (SOFIA).  
Section~\ref{sec:data} describes the data and  signal-to-noise cuts. Section~\ref{sec:analysis} describes new maps of OMC-1 temperature, column density, and dust emissivity index based on spectral energy distributions created from the HAWC+ photometry and complementary data sets. We also describe data cuts based on estimates of the effects of reference beam intensity for all four bands. 
We explore the polarization fraction as a function of intensity for all four HAWC+ bands.   Finally, we use the Davis-Chandrasekhar-Fermi technique (DCF; \citet{Davis1951,Chandrasekhar1953}) to estimate field strengths in the Becklin-–Neugebauer (BN)/Kleinman–Low (KL) region \citep{Becklin1967,Kleinmann1967}, the Orion Bar, and the intercloud medium surrounding the Trapezium Cluster. We examine the field geometry around the BN/KL explosion \citep{Bally2011, Bally2017} as illuminated by the 53 \micron\ HAWC+ data. We summarize our findings in Section~\ref{sec:summary}.

\section{Observations} \label{sec:data} 

Photometry (total intensity) and polarimetry data on the OMC-1 region were obtained using the HAWC+ camera on SOFIA \citep{Harper2018}.  For the purpose of mapping total intensity, raster scans of the source in all four bands were done in December 2016 on SOFIA flight 354. The observing time per band ranged from 9 minutes at 53 \micron\ to 2 minutes at 214 \micron. For each scan, the band-specific half-wave plate was in place for optical similarity of the photometric measurements to the polarization observations discussed later in this section.  
The scan photometry data were reduced using CRUSH \textsc{v2.4.2-alpha1} \citep{Kovacs2006, Kovacs2008} with non-default reduction options. In particular, the ``bright" keyword was used to stop possible clipping of data near brighter regions, notably close to BN/KL. Different combinations of the ``sourcesize" and ``rounds" keywords were used to recover spatial scales beyond the default reduction. For the 53, 89, and 214 \micron\ bands, a ``sourcesize" of 100\arcsec\ was used. This keyword was not used for the 154 \micron\ band. The map-making process was iterated 20 times for 53 \micron, 70 times for 89 \micron, 15 times for 154 \micron, and 100 times for 214 \micron\ by setting the ``rounds" parameter.  In addition, the ``stability" parameter was only changed for the 214 \micron\ band to 3 seconds from the CRUSH default of 5 seconds to remove any large-scale emission remaining in the map. The final scan reductions have effective resolutions of 5.1\arcsec, 7.9\arcsec, 14.0\arcsec, and 18.7\arcsec\ for the 53, 89, 154, and 214 \micron\ bands, respectively. Due to the relatively small fractional bandwidth of the filters, $\Delta\lambda / \lambda \approx 0.2$, no color corrections are made to the data. 
Based on the variance of HAWC+ planet measurements (from scan mode, analyzed with CRUSH), 
we adopt a 15\% calibration uncertainty for the 53, 89, and 154 \micron\ bands and 20\% for the 214 \micron\ band. 


Polarimetry data in the 53, 154, and 214 \micron\ bands were obtained in October-November 2017, and polarimetry at 89 \micron \ was performed in both December 2016 and October-November 2017. Additional polarimetry data at 53~\micron\ were obtained in September 2018. Polarimetry observations were done using the standard chop-nod-dither observing method \citep{Harper2018}. The chop throw ranged between 7.6\arcmin~and 8.0\arcmin, and the chop/nod angle was 125$^\circ$, measured west of north.  The observing times were approximately 3.5, 2.4, 0.5, and 0.5 hours at 53, 89, 154, and 214 \micron, respectively.  The data were reduced using the \textsc{v1.3.0-beta3} (April 2018) version of the HAWC+ data reduction pipeline, with some particular settings and enhancements as noted below.  Our 89 $\mu$m maps are qualitatively very similar to the ones in the SOFIA Data Cycle System (DCS) June 2018 data release, but have some differences in detail in both the signal and noise maps.  As is standard, HAWC+ obtained total intensity data in chop-nod mode simultaneously with the polarimetry data. We utilize these data for our study of polarization as a function of intensity in Section~\ref{sec:pvi} to take advantage of the accurate registration of the polarization and intensity. For photometry data elsewhere in the paper, we use the maps from the scan mode, reduced with CRUSH as described above.


Due to an intermittent vacuum leak in the HAWC+ instrument in 2016, the 89 \micron\ data on SOFIA flight 355 in December 2016 suffered from condensation of a helium film on the detectors; two fields to the east of BN/KL and one to the west ($\sim30\%$ of all 89 \micron\ data) were affected.  The presence of this helium increased the thermal time constants of the detectors, thereby changing the amplitude and phase response of the system to the 10 Hz chop.  To calibrate these data, we measured the time constants of each detector from the 3 Hz internal calibrator flashes interspersed with the chop-nod-dither data and generated new phase and gain correction tables by scaling to 10 Hz, assuming the detector time constant acts as a single-pole filter.  As a result of this correction, we noticed a significant improvement in the internal consistency of the 89 \micron\ measurements, especially in Stokes I, for which the flight 355 data no longer produced noticeable artifacts in maps of $\chi^2$.

Our $\chi ^2$-based analysis of observations of other, fainter targets indicated that the dither map products from the \textsc{v1.3.0-beta3} pipeline have calculated noise uncertainty which is typically $\sim$25\% below the true uncertainty; therefore, we have increased the uncertainties in the I, Q, and U maps by this amount to compensate. This increase in the uncertainties has been found to be a satisfactory way of treating residual systematic effects (including correlated noise) in similar polarimetric systems \citep{Novak2011}.  Instrumental polarization (IP) based on ``polarization skydips,'' with median over the focal plane ranging from 1.8--2.0\% across the bands \citep{Harper2018}, has been removed from the measurements for each pixel by subtracting the reduced Stokes parameters of the IP from the measured parameters \citep{Hildebrand2000}.  In the merging of the measurements into combined maps, we use relative background subtraction (three offsets applied to each input $I$, $Q$, and $U$ map to minimize the standard deviation of the output map) and smoothing with a Gaussian kernel having full width half maximum equal to half that of the diffraction-limited beam for each HAWC+ band \citep{Harper2018}; both of these are standard parts of the pipeline.

To minimize isolated ``spikes" present in the $I$, $Q$, and $U$ maps, we used a deglitching algorithm that operates in the map domain.  Each measurement is compared with 20 neighboring measurements, for which the mean, spatial slope, and standard deviation are calculated.  Measurements that differ by more than 3$\sigma$ (statistical) from the neighbor model are eliminated. Approximately 1--3\% of measurements were removed by the deglitcher.

We examined the telescope tracking data for each integration.  Three integrations on SOFIA flights 450 and 454 with unstable, oscillatory tracking were discarded.  Due to an error in the telescope control software, the two nod positions on flights 442, 444, and 447 were displaced by approximately $\pm$3\arcsec\ (northwest and southeast) from the desired position; this affects primarily the northern half of the 53 \micron\ map.  The effective point spread function is larger and asymmetric in that part of the map.

\begin{figure}
    \centering
    \includegraphics[width=3.5in]{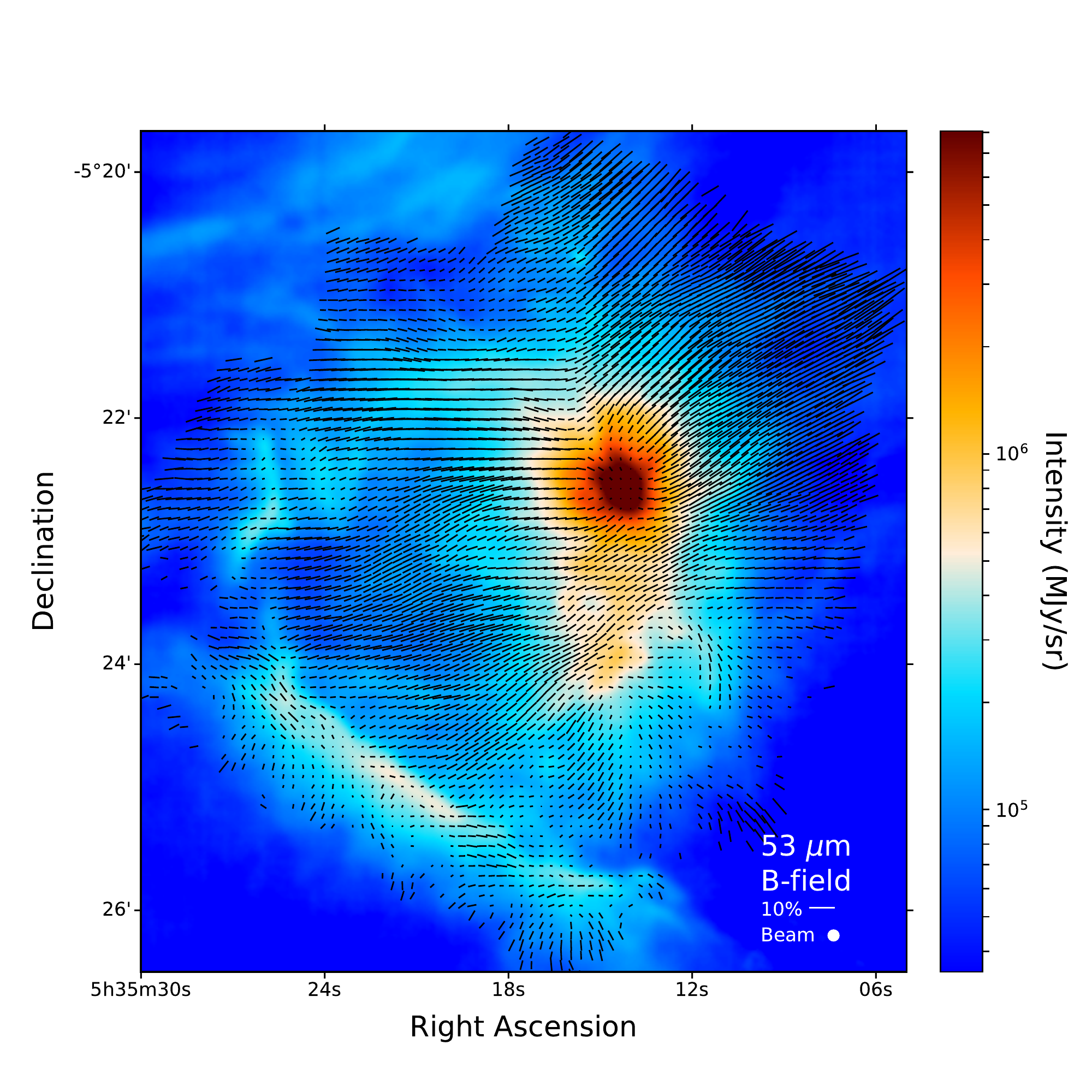}
    \includegraphics[width=3.5in]{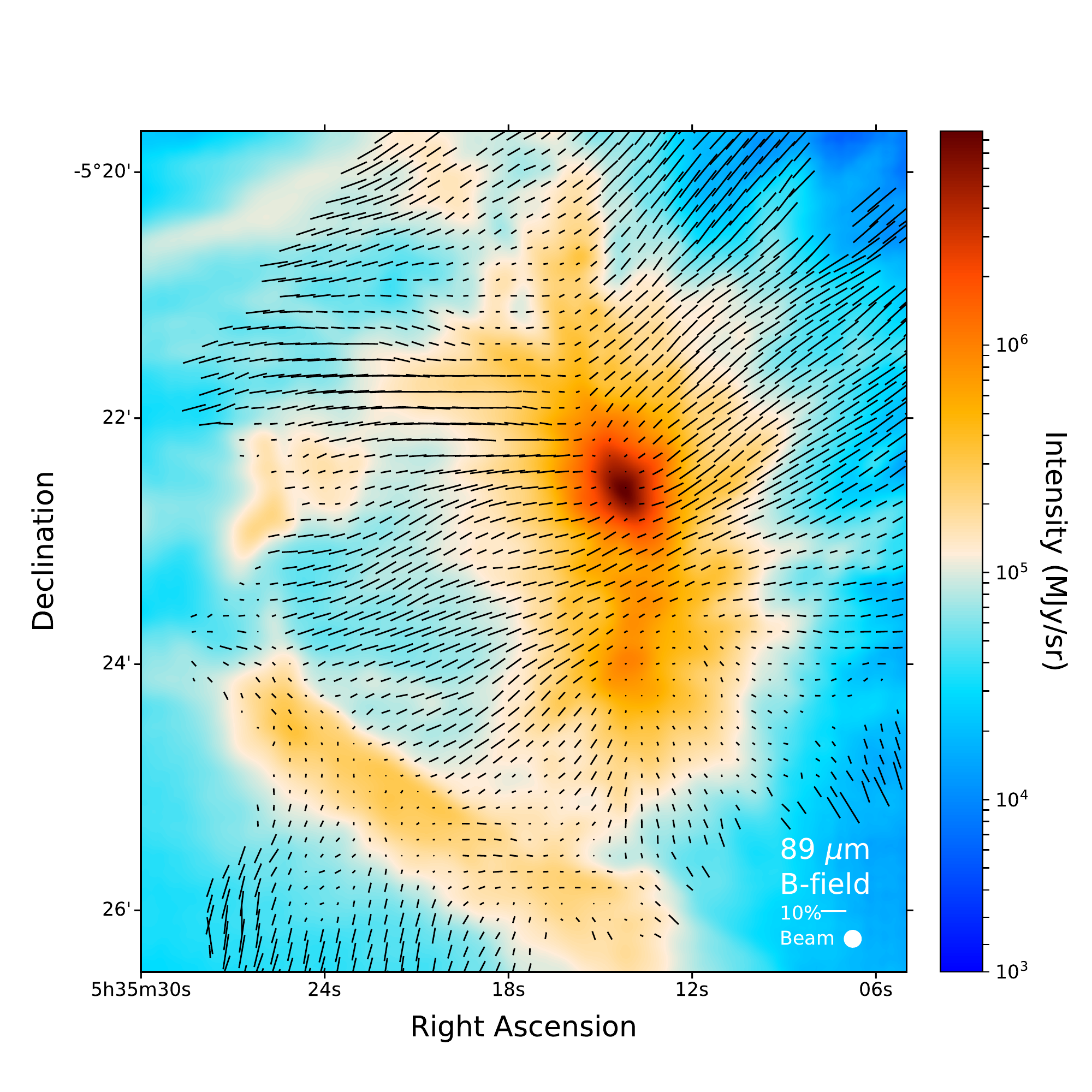}\\
    \includegraphics[width=3.5in]{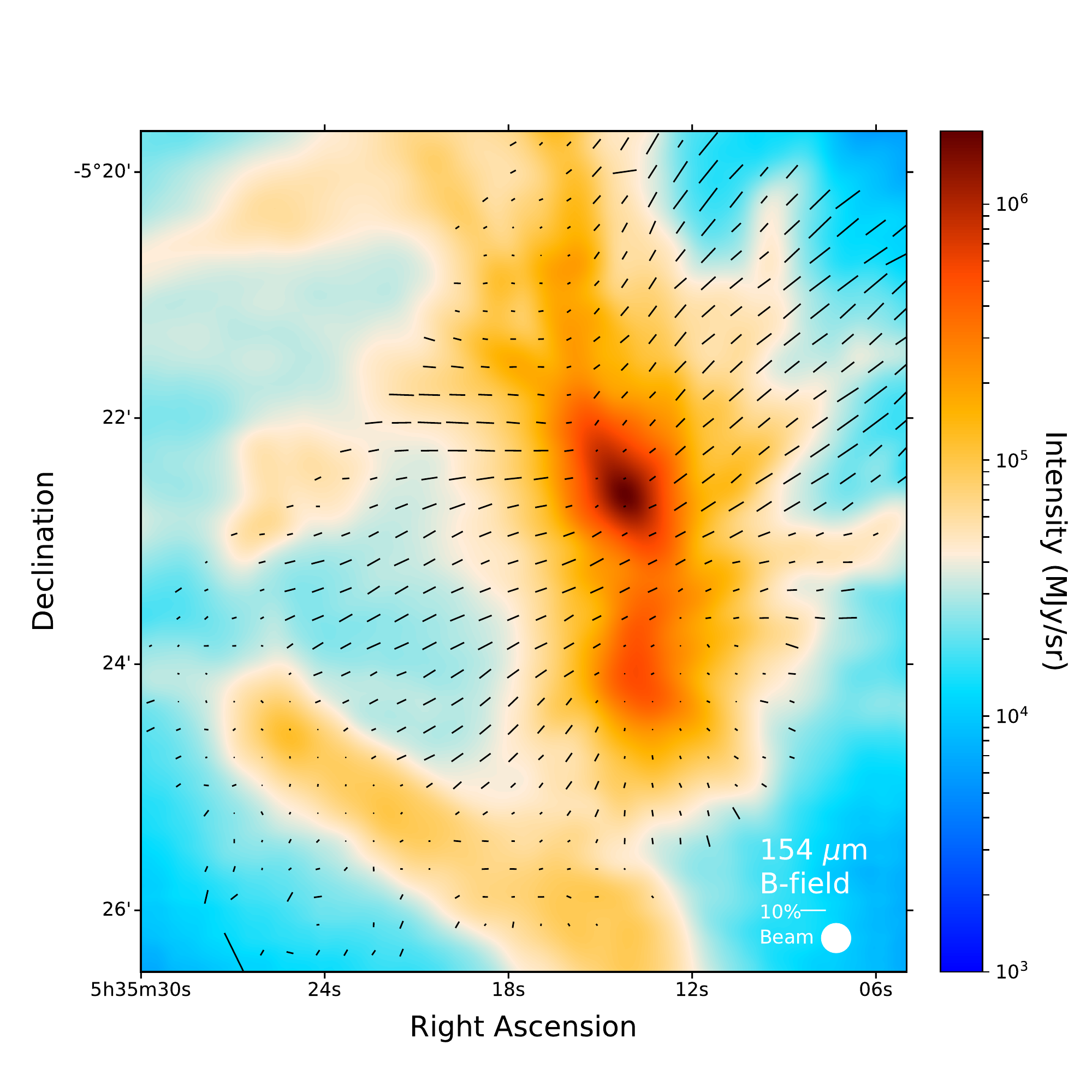}
    \includegraphics[width=3.5in]{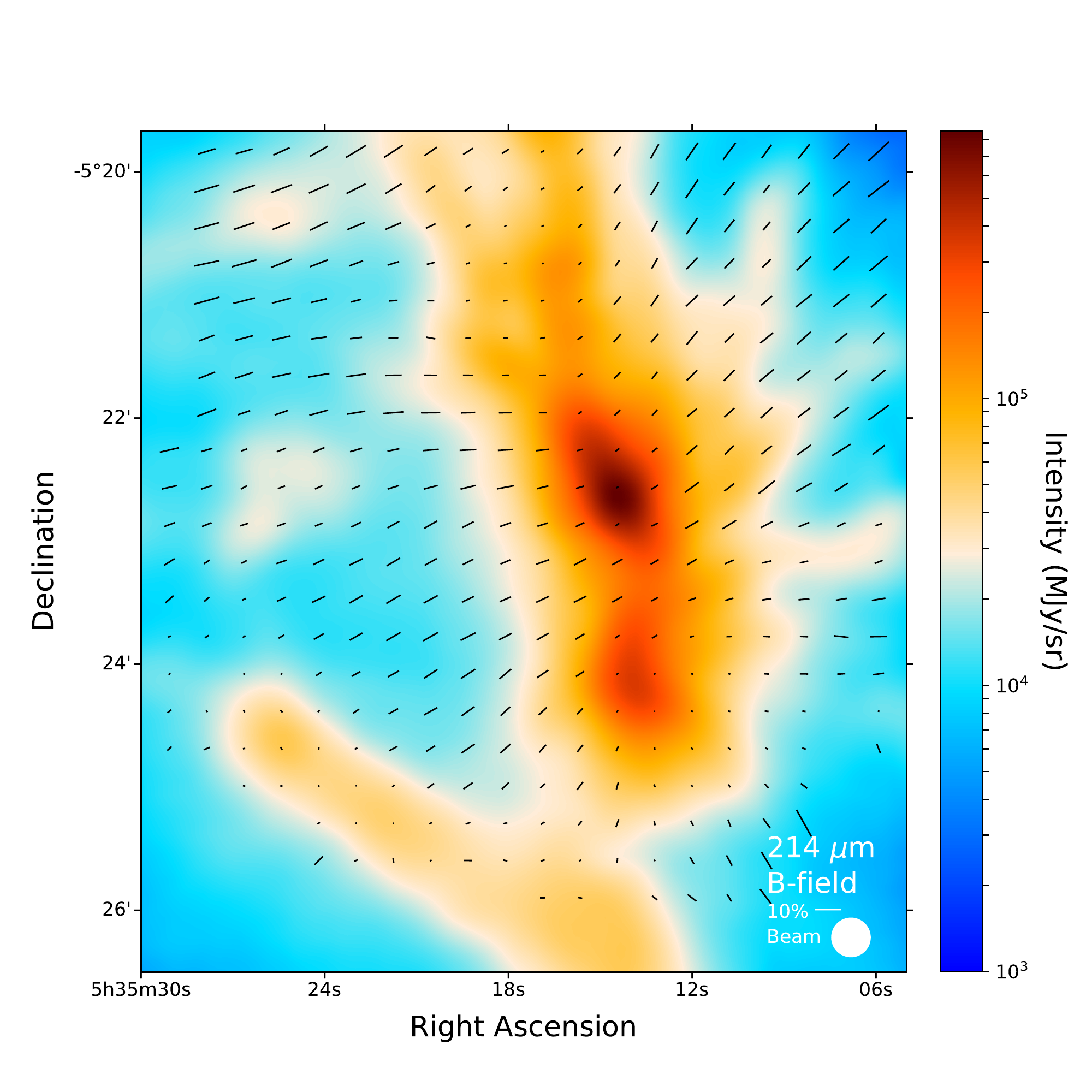}
    \caption{Polarization maps at 53, 89, 214, and 154 \micron, clockwise starting from the top left. The polarization vectors have been rotated by 90$^\circ$ to indicate the inferred direction of the magnetic field projected in the plane of the sky. Polarization data are from chop-nod HAWC+ observations. The background intensity images are obtained using scan data and reduced with the CRUSH analysis software. The vectors are plotted such that their spacing is equal to the beam size.} 
    \label{fig:maps}
\end{figure}

A signal-to-noise threshold of $p/\sigma_p>3$ was applied to the polarization maps and magnetic field analysis. This corresponds to a statistical uncertainty in position angle of $\sim10^\circ$. In all maps, polarization fractions have been debiased according to $p_{debias}=\sqrt{p^2-\sigma_p^2}$ \citep{Serkowski1974}. Table~\ref{tab:obs} shows a summary of the spectral and spatial resolution along with the number of Nyquist sampled detections above $3\sigma$ for the four polarimetry data sets.  The polarization maps are shown in Figure~\ref{fig:maps}. The vectors shown sample the maps with a spacing approximately equal to the beam size and have been rotated by 90$^\circ$ relative to the electric field orientation to represent the inferred magnetic field angle as projected onto the plane of the sky.

\begin{deluxetable}{cccccc}
\tablewidth{290pt}
\tabletypesize{\scriptsize}
\tablehead{\colhead{HAWC+ Band} & \colhead{Band Center} & \colhead{FWHM Bandwidth}  &  \colhead{Fractional Bandwidth} &\colhead{FWHM Beam Size} & \colhead{Number of Vectors} \\ 
\colhead{-} & \colhead{($\mu$m)} & \colhead{($\mu$m)} & \colhead{(-)} & \colhead{(\arcsec )} & \colhead{$>3\sigma$, Nyquist Sampled}} 
\startdata
A & 53 & 8.7 & 0.16 & 4.9 & 15,808 \\
C & 89 & 17 & 0.19 & 7.8 & 8,939 \\
D & 154& 34 & 0.22 & 13.6& 2,387\\
E & 214& 44 & 0.21 & 18.2& 1,880 
\enddata
\label{tab:obs}
\caption{Polarimetric data summary}
\end{deluxetable}

\section{Analysis}\label{sec:analysis}
\subsection{Spectral Energy Distributions}\label{sec:SED}
Obtaining a complete picture of OMC-1 requires understanding the environments in which dust grains reside. We have used new photometric measurements from HAWC+ along with archival multi-wavelength photometry data to produce improved spectral energy distributions (SEDs) for the OMC-1 region. The archival data used to constrain thermal emission include those from the Photodetector Array Camera and Spectrometer \citep[PACS][]{Poglitsch2010} and the Spectral and Photometric Imaging Receiver \citep[SPIRE][] {Griffin2010} instruments on the Herschel Space Observatory\footnote{Herschel is an ESA space observatory with science instruments provided by European-led Principal Investigator consortia and with important participation from NASA.} \citep{Pilbratt2010} and the Submillimetre Common-User Bolometer Array-2 \citep[SCUBA-2][]{Holland2013, Dempsey2013, Chapin2013} instrument operating on the James Clerk Maxwell Telescope (JCMT). Up to 30\% of the in-band radiation in the SCUBA-2 850 \micron\ data consists of free-free emission caused by UV from the Trapezium Cluster (S. Coud\'{e}, private communication). Therefore, we use also data from the MUltiplexed Squid TES Array at Ninety GHz \citep[MUSTANG][]{Dicker2008} on the Green Bank Telescope (GBT), and combined Very Large Array (VLA) and GBT X-band data to constrain the effects of free-free emission at longer wavelengths.

\subsubsection{Herschel Data}
Herschel data were obtained from the Herschel Science Archive (HSA\footnote{\url{http://archives.esac.esa.int/hsa/whsa/}}) for the 70, 100, and 160 \micron\ PACS photometry bands and the 250 \micron~SPIRE band. PACS observed the OMC-1 region with its 70 \micron~filter on February 23, 2010 (OBSID 1342191106, 1342191107) with a 20\arcsec~s$^{-1}$ scan speed and high-gain photometer setting. The PACS 100 and 160 \micron~filters observed OMC-1 on October 8, 2010 (OBSID 1342206052, 1342206053) with the same gain and scan speed settings used for the 70 \micron~data. We used the level 2.5 maps made with the \textit{scanamorphos} routine \citep{Roussel2013}. A SPIRE level 2.5 250 \micron~map was used for this region, which merged observations from OBSID 1342184386 and 1342239930, and utilized a cross-scan mode with a scanning velocity of 30\arcsec~s$^{-1}$. We do not utilize the SPIRE 350 and 500 \micron\ maps for OMC-1 due to their large beam size (25\arcsec\ and 36\arcsec, respectively). 

The angular resolution of each PACS band is dependent upon the scan speed of the individual observations and can be fitted as a two-dimensional Gaussian; the major and minor axes of this Gaussian are listed in \citet{Poglitsch2010}. We use the geometric mean of these axes as the resolution of the band. For PACS, we adopt resolutions of 5.6\arcsec, 6.8\arcsec, and 11.3\arcsec~for the 70, 100, and 160 \micron~bands, respectively. We follow \citet{Sadavoy2013} in adopting a SPIRE 250 \micron~resolution of 18.2\arcsec.

Due to the wide passbands of the PACS and SPIRE photometer filters and the assumption of a flat SED \citep{Muller2011, HESV_4}, small color corrections ($\lesssim 7\%$) are made to the data. \citet{Muller2011} and the HIPE program ; \citep[v15.0.1 with SPIRE calibration v.14\_3][]{Ott2010} list pre-tabulated color corrections for various modified SEDs. One point to note is that the values listed in \citet{Muller2011} are multiplicative-inverse corrections, while those in HIPE are multiplicative only. To use the pre-tabulated color corrections, we assume $\beta = 2$. This is close to that determined by \cite{Vaillancourt2002}, who finds a median temperature and dust emissivity index, $\beta$, in this region of $40\pm 10$~K and $1.8\pm 0.05$, respectively. For each PACS filter, we take the inverse of the color corrections listed for $\beta = 2$ and $T=30-50$~K to make them multiplicative factors, then adopt the root-mean-square (RMS) of these values as the factor; we take the error on this correction to be the root-mean-square-error (RMSE) of the values. For the 70, 100, and 160 \micron\ bands on PACS, these color corrections are 1.025, 1.004, and 0.929, with errors of 0.004, 0.018, and 0.027, respectively. For the SPIRE 250 \micron\ band, the process is the same without taking the inverse of the listed factors. Using the extended source corrections, we adopt a value of 0.970 with an error of 0.005 for the color correction.
Finally, we follow \citet{Arab2012} and  \citet{Sadavoy2013} in adopting calibration uncertainties of 20\% for the PACS data and 10\% for the SPIRE data, respectively.


\subsubsection{SCUBA-2 Data}
The SCUBA-2 instrument is able to simultaneously observe in two filters \citep{Holland2013}.  For this work, we choose to only use the 850 \micron~data, as the 450 \micron~band has large calibration uncertainties ($\sim$ 50\%) due to the high variabilty of the atmosphere at this wavelength \citep{Sadavoy2013} compared to 850 \micron.

In addition to possible free-free contamination, \citet{Coude2016} and \citet{Mairs2016} find molecular contamination in the 850 \micron\ data from the {$^{12}\text{CO}(\text{J}=3\xrightarrow{}2)$} rotational line of up to 20 percent. \citet{Coude2016} note that areas of lower column density will likely have a higher contamination level of this line. To correct for this, we utilized archival CO-corrected data\footnote{\url{http://www.cadc-ccda.hia-iha.nrc-cnrc.gc.ca/}} \citep{Mairs2016} that made use of data from the HARP instrument \citep{Buckle2009}. We adopt a 14.2\arcsec~resolution for these corrected data \citep{Mairs2016}. While \cite{Sadavoy2013} adopt a 10\% calibration uncertainty for the SCUBA-2 850 \micron~data, we adopt a 15\% error to account for any additional systematic effects in the HARP instrument \citep{Buckle2009}. 

\subsubsection{MUSTANG and X-Band Data}
The MUSTANG (90 GHz; 3.3 mm) and X-band (8.4 GHz; 3.5 cm) data used here were originally published in \citet{Dicker2009}, which describes the reduction process. For the MUSTANG data, we adopt a beam size of 9\arcsec, and for the X-band data, we use 8.4\arcsec~for the beam size. For both instruments, we follow \citet{Dicker2009} in adopting a calibration uncertainty of 15\%.

\begin{deluxetable}{ccccccc}
\tablewidth{290pt}
\tabletypesize{\scriptsize}
\tablehead{\colhead{Observatory/} & \colhead{Wavelength} &\colhead{Beam Size} & \colhead{Color} &\colhead{Color Correction}\vspace{-0.2cm} &\colhead{Calibration}\vspace{-0.2cm} & \colhead{Paper Reference}\\
\colhead{Instrument} & \colhead{} & \colhead{FWHM} & \colhead{Correction} & \colhead{Uncertainty} & \colhead{Uncertainty} & \colhead{}\\
\colhead{(-)} & \colhead{($\mu$m)} & \colhead{(\arcsec)} & \colhead{(-)} & \colhead{(-)} & \colhead{(\%)} & \colhead{(-)}}
\startdata
SOFIA/HAWC+ & 53 & 5.1 & -- & -- & 15 & This Paper\\
SOFIA/HAWC+ & 89 & 7.9 & -- & -- & 15 & This Paper\\
SOFIA/HAWC+ & 154& 14.0& -- & -- & 15 & This Paper\\
SOFIA/HAWC+ & 214& 18.7& -- & -- & 20 & This Paper\\
Herschel/PACS & 70 & 5.6 & 1.025&0.004 & 20 & \citet{Abergel2010}\\
Herschel/PACS & 100 & 6.8 & 1.004&0.018 & 20 & \citet{Andre2007}\\
Herschel/PACS & 160 & 11.3 & 0.929&0.027 & 20 & \citet{Andre2007}\\
Herschel/SPIRE & 250 & 18.2 & 0.970& 0.005& 10 & \citet{Andre2011, Bendo2013}\\
JCMT/SCUBA-2 & 850 & 14.2 & -- & -- & 15 & \citet{Mairs2016}\\
GBT/MUSTANG & 3500 & 9.0 & -- & -- & 15 & \citet{Dicker2009}\\
GBT and VLA & 35000 & 8.4 & -- & -- & 15& \citet{Dicker2009}\\
\enddata
\label{tab:corrections}
\caption{Adopted photometry calibration values.}
\end{deluxetable}

\subsubsection{Data Preparation}
\citet{Arab2012} correct for zero-point emission in the PACS data by subtracting the intensity in a region around $\alpha = 5^\text{h} 35^\text{m} 26.7^\text{s}$, $\delta = -5^\circ 26\arcmin 4.7\arcsec$ (J2000). We also correct for a zero-point; however, we set our zero value to that of the pixel located at this position. We propagate the error in quadrature to take this offset into account. The SPIRE data were zero-corrected in the default SPIRE HSA pipeline using Planck HFI/IRAS data as described in \citet{Bernard2010}.

We do not apply zero-point corrections to the HAWC+ and SCUBA-2 data due to their respective reduction methods. CRUSH removes residual DC offsets and systematic, correlated sky noise \citep{Kovacs2008} within a scan, and thus any arbitrary zero-point is already removed from the data. The SCUBA-2 reduction method (\textit{Starlink} SMURF software) similarly removes a zero-point \citep{Sadavoy2013}. 

The photometry and errors are color-corrected, if applicable, then are converted into common units of $\text{MJy}\, \text{sr}^{-1}$. \citet{Poglitsch2010} provide extended-source saturation levels for each PACS filter when using high-gain observations; any remaining pixels above these limits in the PACS images that were not flagged by the default PACS HSA pipeline were subsequently removed. To be more conservative on possible saturation, we enlarge the PACS 70~\micron\ map mask, which is then used to mask the PACS 100 and 160~\micron\ data near the BN/KL.  The masked regions near BN/KL cover 6.7, 6.0, and 6.0 square arcminutes for the 70, 100, and 160~\micron\ maps, respectively.

The photometry and error arrays were independently re-projected into a common WCS system with 3.7\arcsec\ square pixels using a flux-conserving algorithm. Finally, an error-weighted Gaussian convolution, with kernel size $\sqrt{\text{FWHM}_2^2 - \text{FWHM}_1^2}$, where $\text{FWHM}_1$ is the resolution of the each instrument and $\text{FWHM}_2$ is the target common resolution of 18.7\arcsec.

\subsubsection{Temperature, $\beta$ and Column Density Maps}
The data are fit in two steps. First, we use the MUSTANG and X-band data to fit the free-free emission at longer wavelengths assuming the form given in \citet{Hensley2015}
\begin{equation}
    I_{ff} = C\left(\dfrac{\nu}{30~\text{GHz}}\right)^{-0.12},
    \label{eq:ff}
\end{equation}
where C is a normalization constant. The free-free emission contamination at 850 \micron\ is extrapolated from the fit and removed. Then, using a single-temperature modified blackbody curve (equation \ref{eq:sed}) as defined in the Appendix of \citet{Vaillancourt2002}, the thermal component,
\begin{equation}
I_\nu = \left(1-e^{-\tau\left(\nu\right)}\right)B_\nu(T),
\label{eq:sed}
\end{equation} 
is fit.

We define the optical depth, $\tau\left(\nu\right) \equiv \varepsilon\left({\nu}/{\nu_0}\right)^\beta$, where $\varepsilon$ is a constant of proportionality directly related to the column density along the line of sight, $\beta$ is the dust emissivity index, and $B_\nu(T)$ is the Planck blackbody function at wavelength $\nu$ with temperature $T$. Following \cite{Sadavoy2013} (and references therein), we adopt $\nu_0 = 1000$ GHz. From the Python Scipy package, we use {\textit{curve\_fit}} to fit the function by minimizing the $\chi^2$ statistic using the Levenberg-Marquardt algorithm.
We limit the data used in the fits to those points where the signal-to-noise is greater than 3 and fit all pixels for which the number of degrees of freedom is greater than 1.
The vast majority of fitted pixels ($\sim87\%$) have a reduced $\chi^2$ ($\chi^2_r$) of 5 or less, with $\sim70\%$ having $\chi^2_r \leq 2$. Therefore, we apply a cut to the parameter maps to include only pixels for which $\chi^2<5$. We further apply a cut for $\beta<2.25$ as a proxy to eliminate edge pixels where data set limitations cause the fits to be suspect despite having reasonable $\chi^2_{r}$.

\begin{figure}
    \centering
    \includegraphics[width=3.5in]{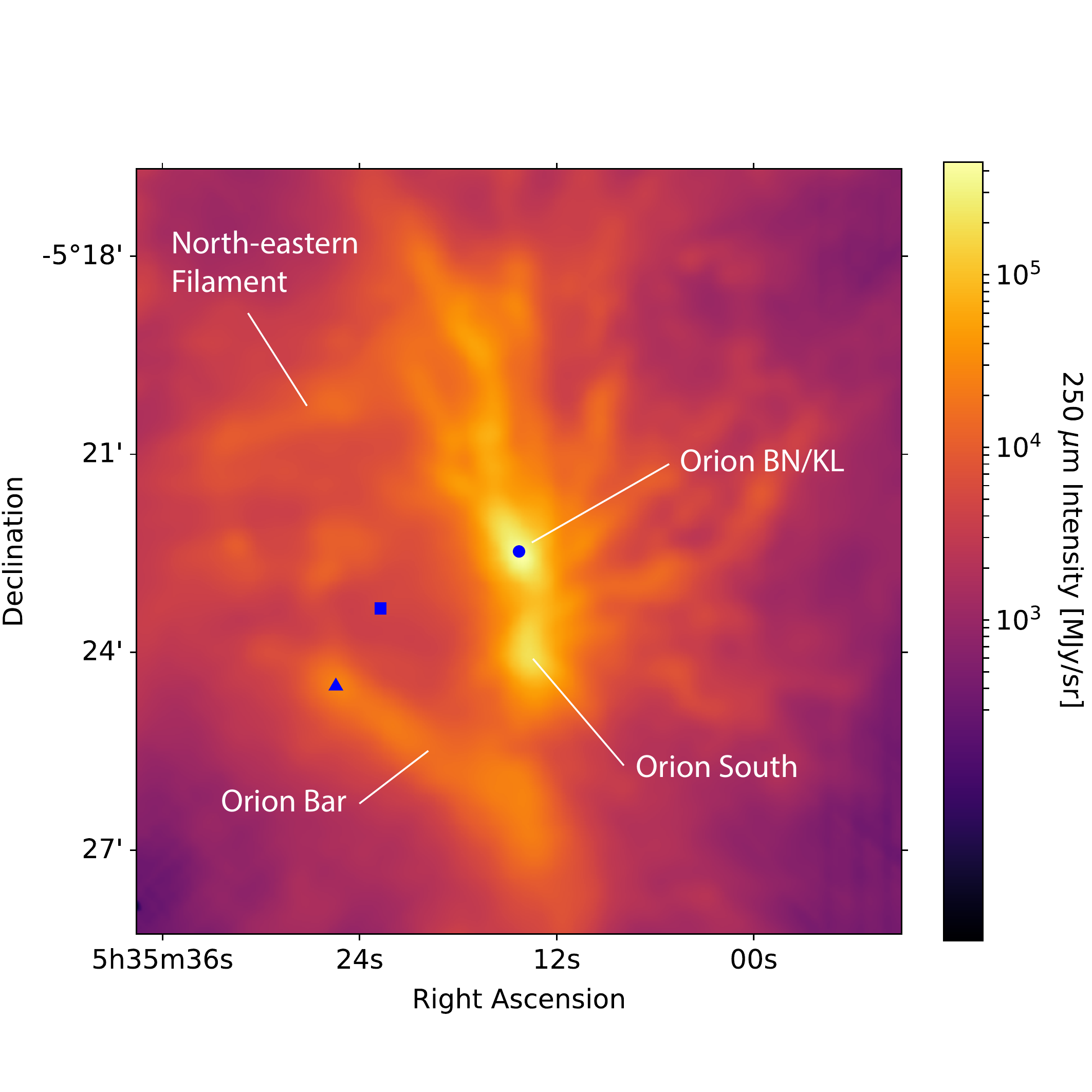}
    \includegraphics[width=3.5in]{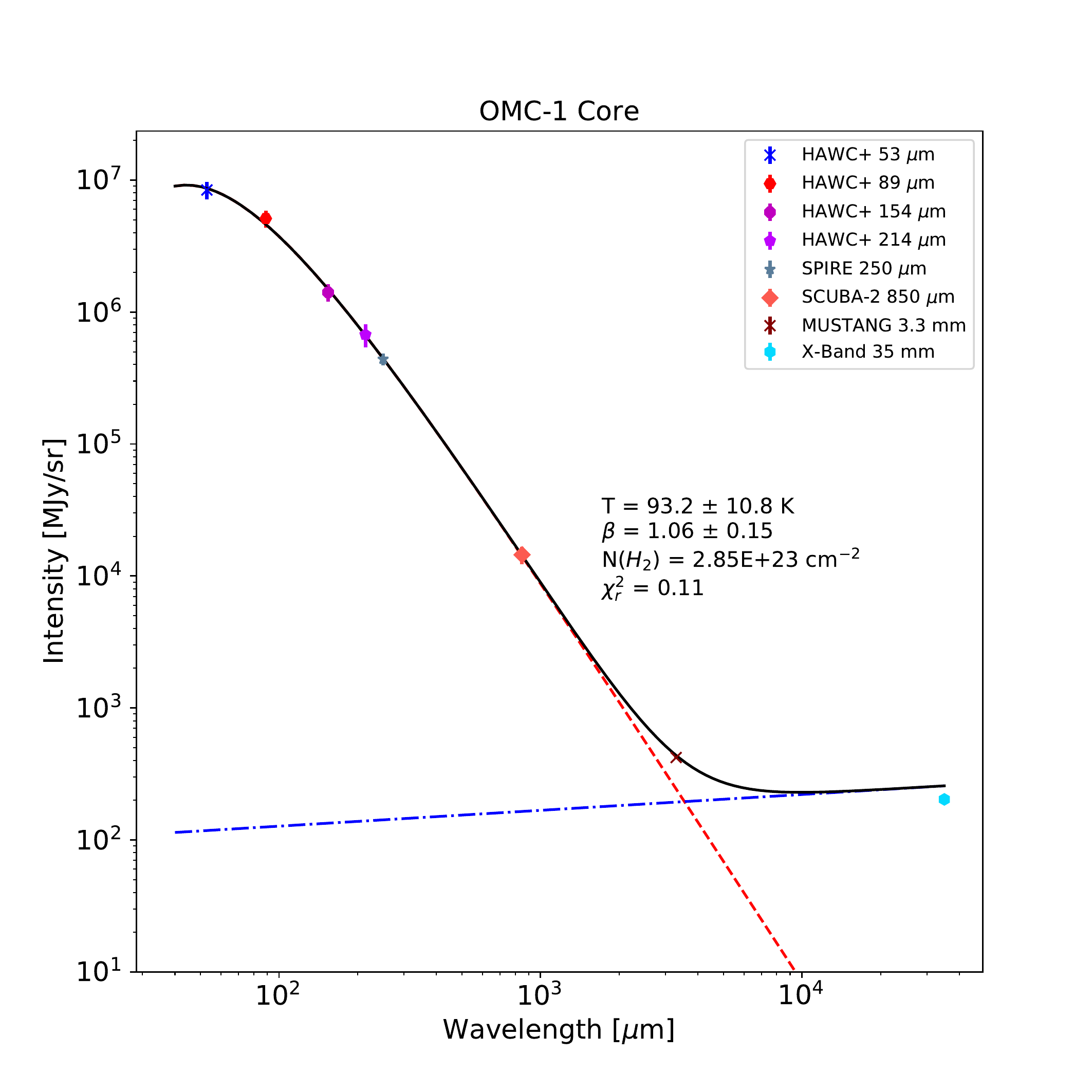}\\
    \includegraphics[width=3.5in]{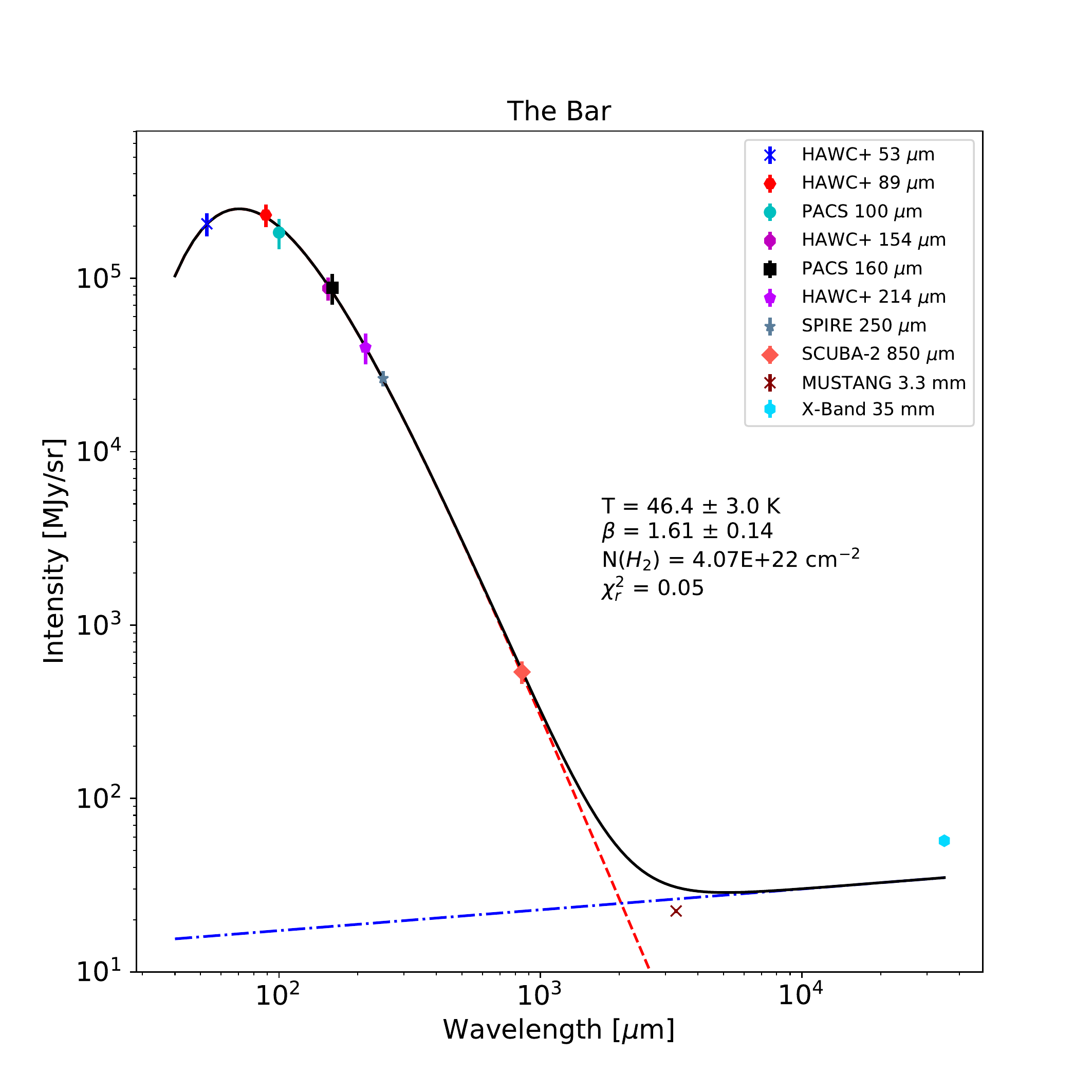}
    \includegraphics[width=3.5in]{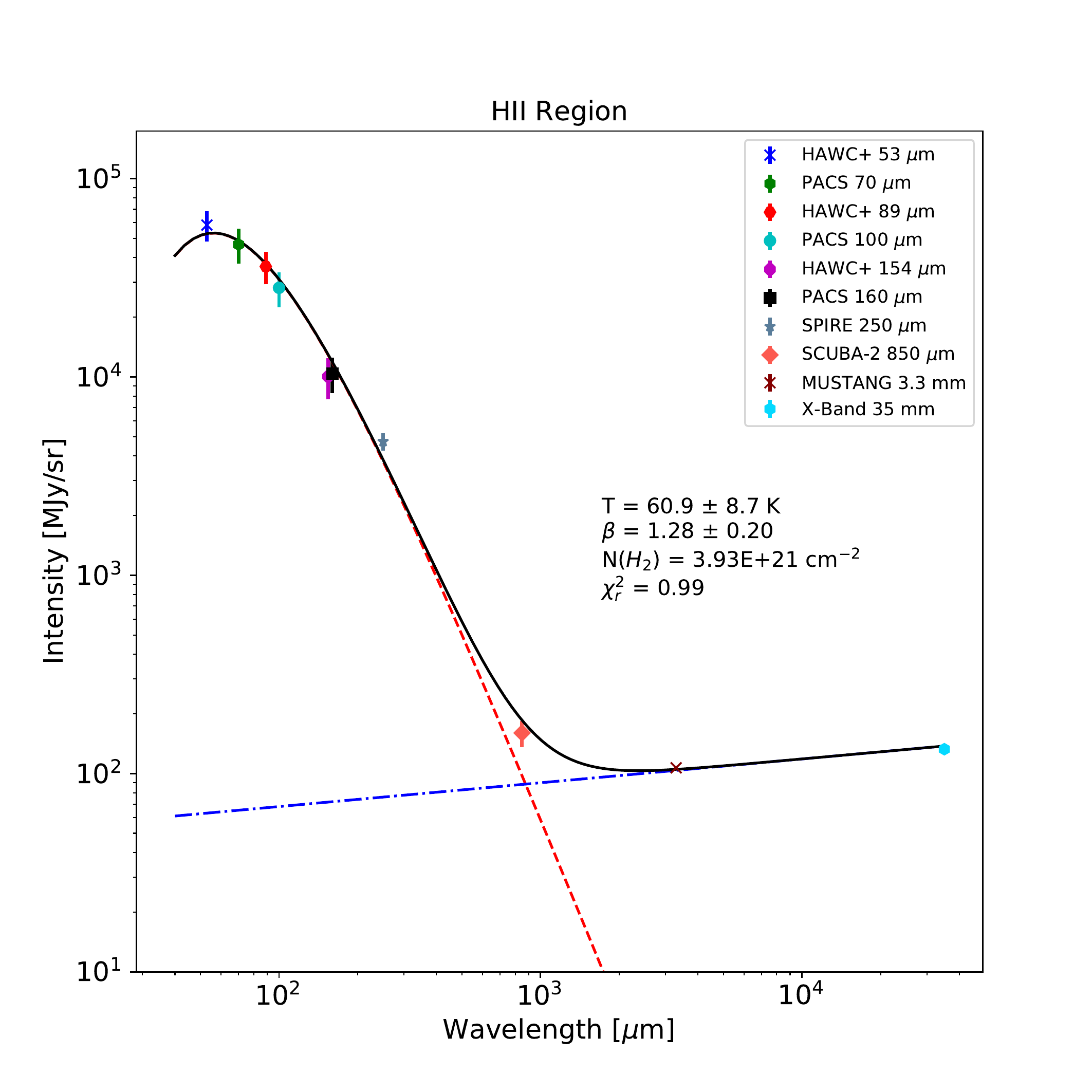}
    \caption{Representative SED fits are shown for three positions within OMC-1. The thermal model is represented by the red dashed line, the free-free model is shown by the blue dash-dot line. The black solid line represents the total SED. The top-left panel shows the locations of the individual fits for the OMC-1 Core (circle), \ion{H}{2} Region (square), and Bar (triangle) overplotted on the the SPIRE 250 \micron\ data smoothed to 18.7\arcsec\ resolution.}
    \label{fig:ind_SEDs}
\end{figure}

Comparing our definition of $\varepsilon$ with the modified blackbody function in \citet{Sadavoy2013}, we find: 
\begin{equation}
    \varepsilon = \kappa_{\nu_0} \mu m_H N(H_2)
    \label{eq:colden}
\end{equation}
Here, $\kappa_{\nu_0}$ is a reference dust opacity per unit mass at frequency $\nu_0$, $\mu$ is the mean molecular weight per hydrogen atom, $m_H$ is the atomic mass of hydrogen, and $N(H_2)$ is the gas column density (molecules per square centimeter) \citep{Sadavoy2013}. We adopt the values $\kappa_{\nu_0}(1000 \text{GHz}) = 0.1~\text{cm}^2~\text{g}^{-1}$ and $\mu = 2.8$ as in \citet{Sadavoy2013}. Example SEDs and fits for three regions of OMC-1 are shown in Figure~\ref{fig:ind_SEDs}.

\begin{figure}
    \centering
    \includegraphics[width=3.5in]{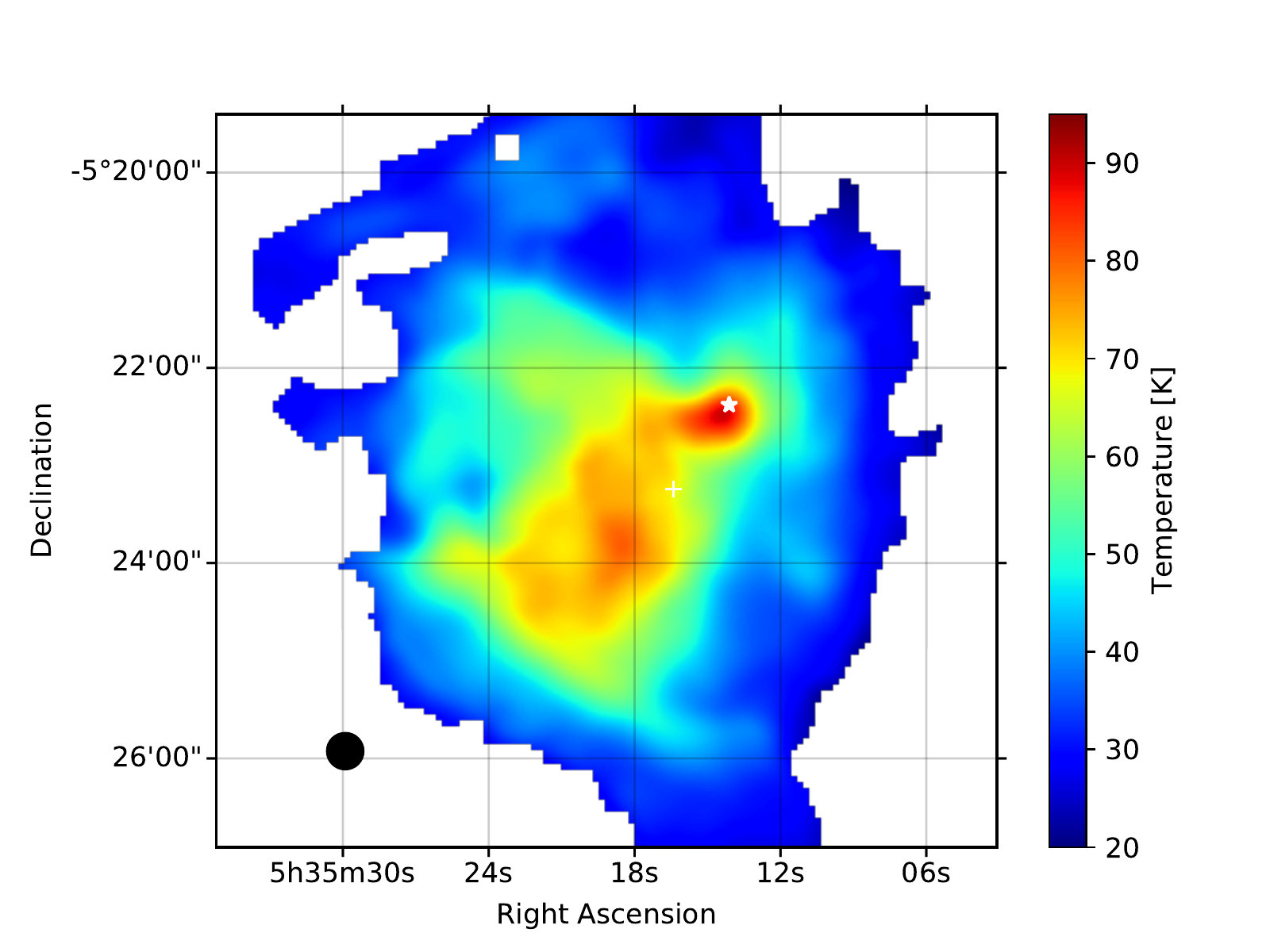}
    \includegraphics[width=3.5in]{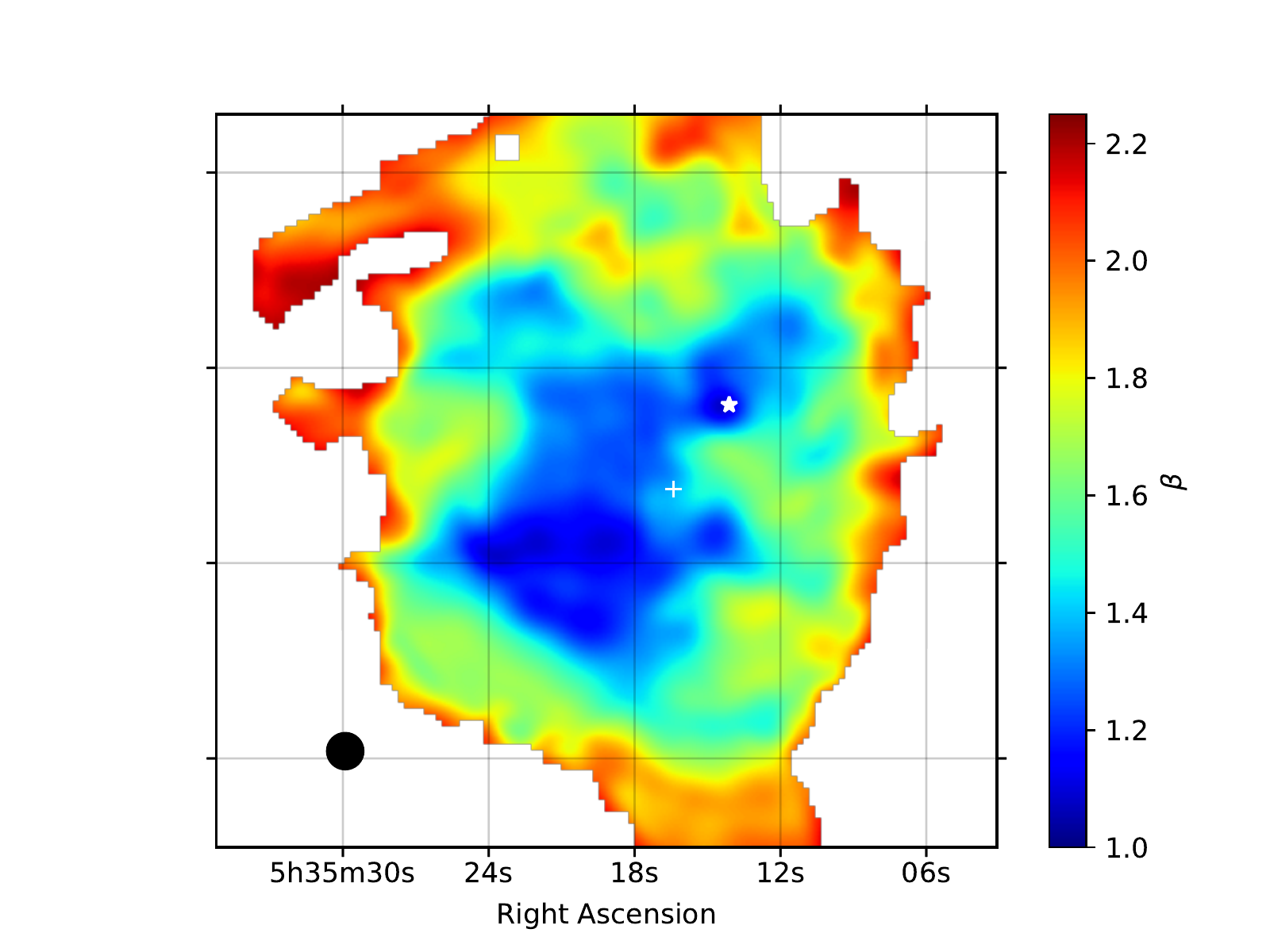}\\
    \includegraphics[width=3.5in]{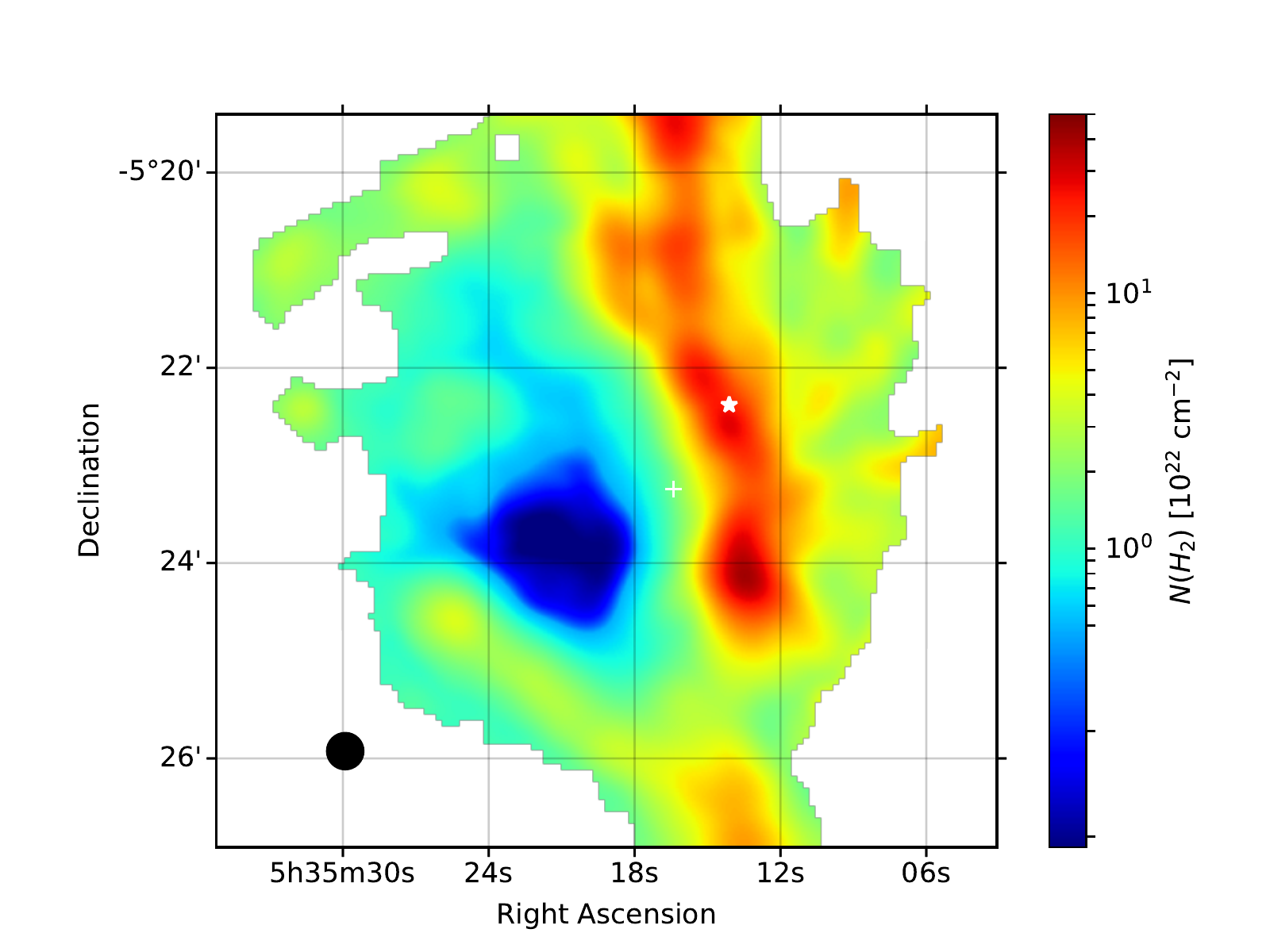}
    \includegraphics[width=3.5in]{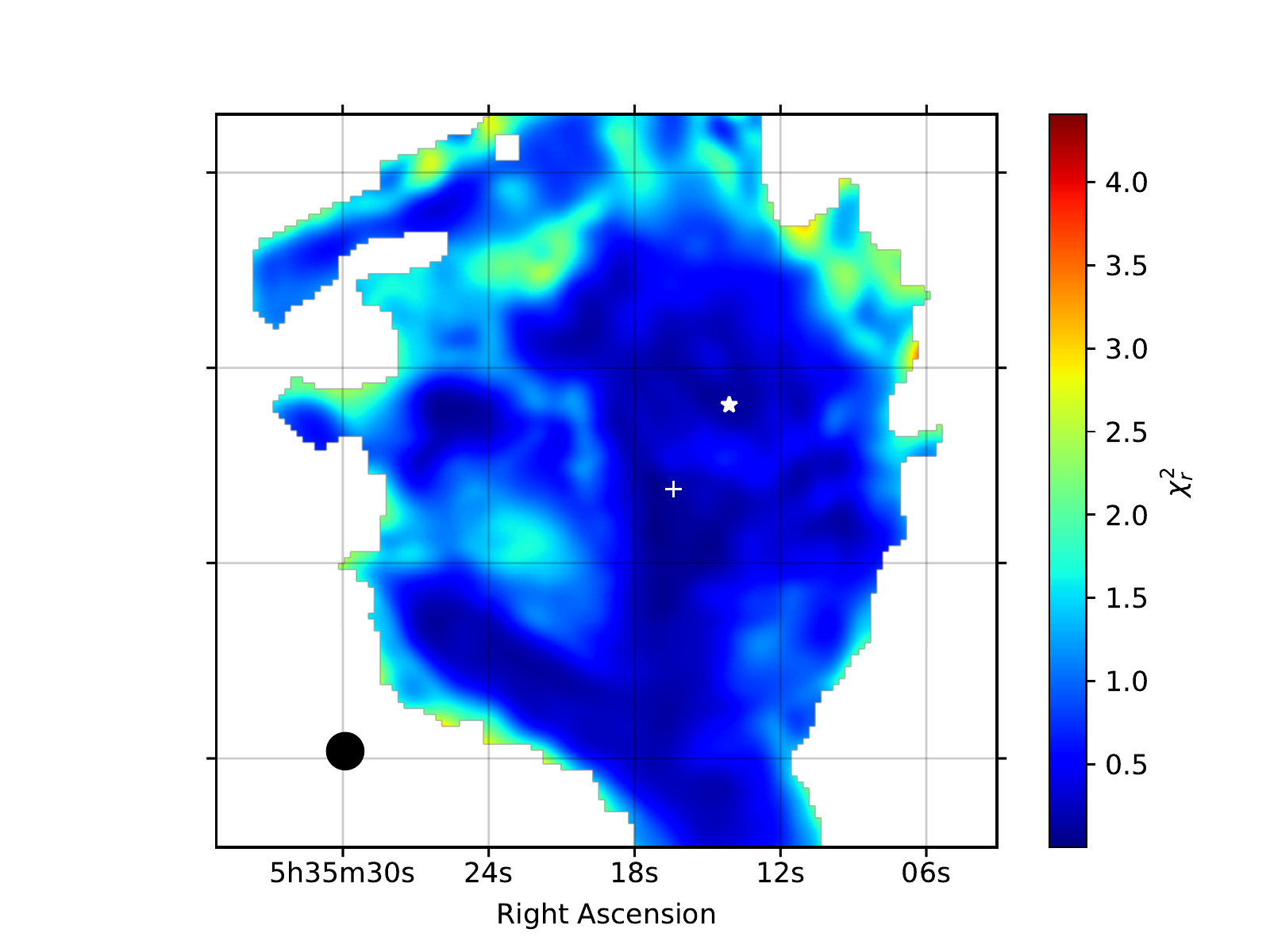}
    \caption{Clockwise from top left: The fitted temperatures, dust emissivity indices, $\chi^2_r$ values, and calculated column densities. All are smoothed to a 22\arcsec\ resolution. The effective beam size shown on the bottom left in each panel as a black filled circle. Points with only a $\chi^2_r < 5$ are shown. The white cross indicates the center of the Trapezium cluster, while the white star shows the location of the BN Object.}
    \label{fig:sedparams}
\end{figure}

Maps of the fitted parameters are shown in Figure~\ref{fig:sedparams}. 
The parameter maps have a median temperature and $\beta$ in this region of $36.3 \pm 2.5$ K and $1.7 \pm 0.15$, respectively, which are consistent with the results in \citet{Vaillancourt2002}. The warmest region outside of the BN/KL region ($83.9 \pm 19.3$~K) lies 1.9\arcmin\ southeast of the BN object or about 1\arcmin\ southeast of the Trapezium Cluster's center. Compared to \citet{Vaillancourt2002}, we find a temperature at this point that is $\sim$60\% higher at our 18.7\arcsec~ resolution, and 36\% higher when smoothing the temperature map to a 30\arcsec\ resolution to match the angular resolution of this previous work. Similarly, at the location of BN/KL, these authors find a temperature and $\beta$ of approximately 50 K and 1.5, respectively. We find $T \approx 92.3 \pm 11.5 \text{ K and } \beta \approx 1.02 \pm 0.15$ at our  18.7\arcsec\ resolution and $T \approx 80.4 \pm 1.0 \text{ K and } \beta \approx 1.20 \pm 0.02$ when smoothed to a 30\arcsec\ resolution. Thus, this disagreement in the fitted parameters cannot be described by a difference in resolution alone. 
 
\begin{figure}[h]
    \centering
    \includegraphics[width=4.0in]{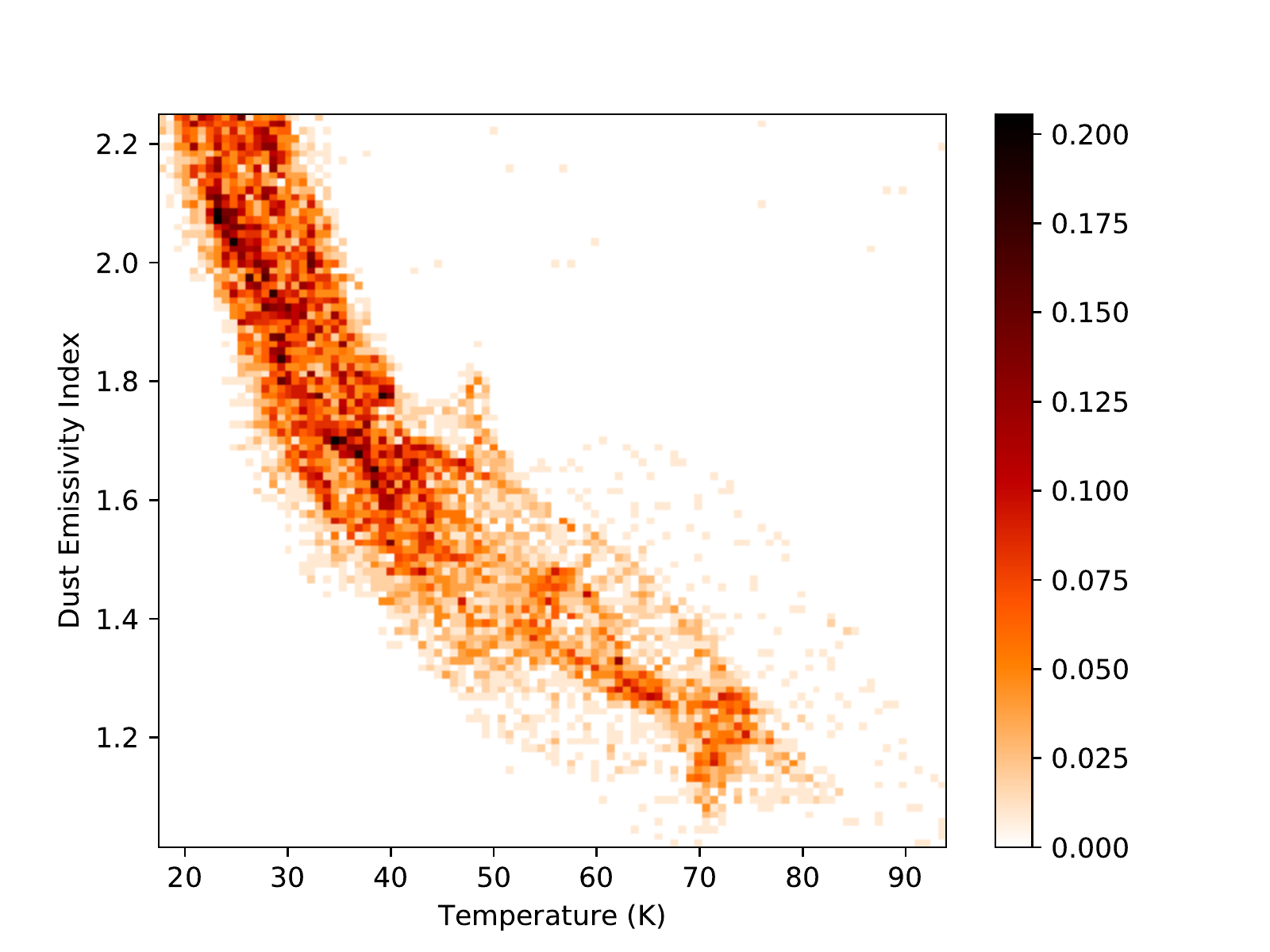}
    \caption{A two-dimensional histogram of dust emissivity index ($\beta$) versus temperature across the region shows an anticorrelation between the two quantities. }
    \label{fig:Tbeta}
\end{figure}

Across the mapped region, the value of $\beta$ is strongly and negatively correlated to the fitted temperature as shown in Figure~\ref{fig:Tbeta}.  This is similar to the trend found by \citet{Dupac2001, Dupac2003} for the OMC-1 region. \citet{Shetty2009} argues that this degeneracy could be due to line-of-sight variations. These authors also note that noise in the observations can cause such a degeneracy in the $T-\beta$ relation that is more apparent when using only data on the Rayleigh-Jeans tail of the SED than when the peak of the thermal spectrum is constrained. 

\begin{figure}[h]
    \centering
    \includegraphics[width=3.5in]{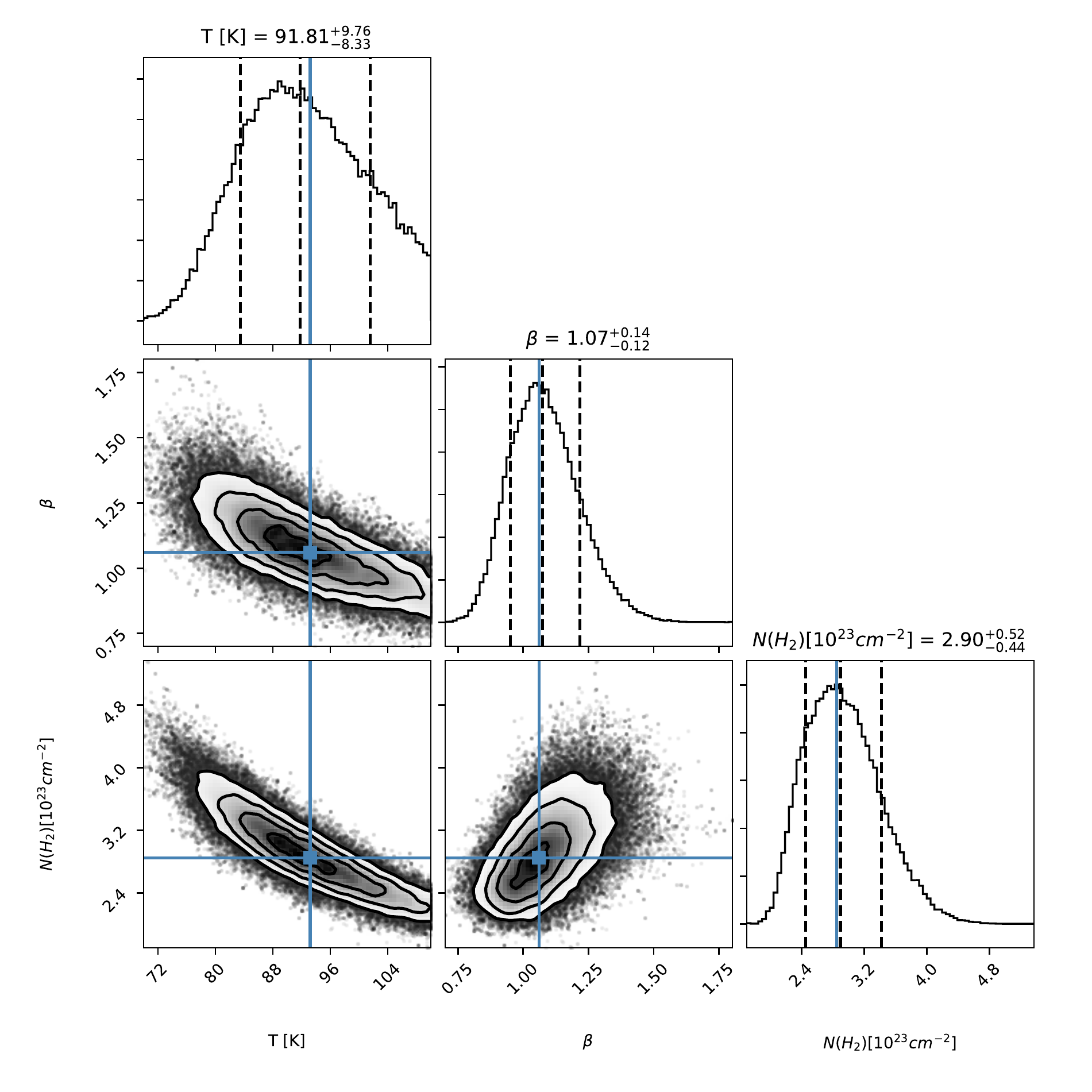}
    \includegraphics[width=3.5in]{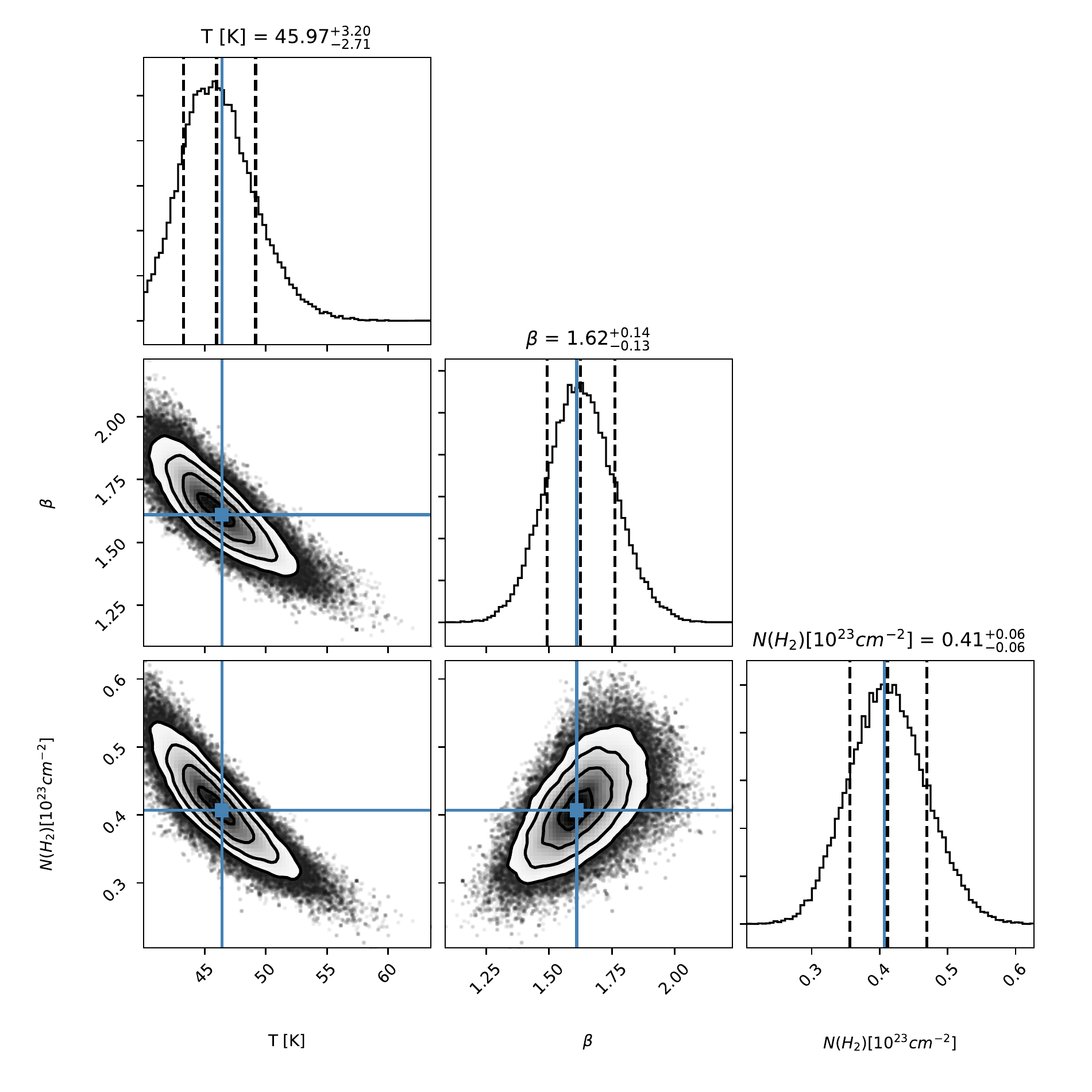}
    \includegraphics[width=3.5in]{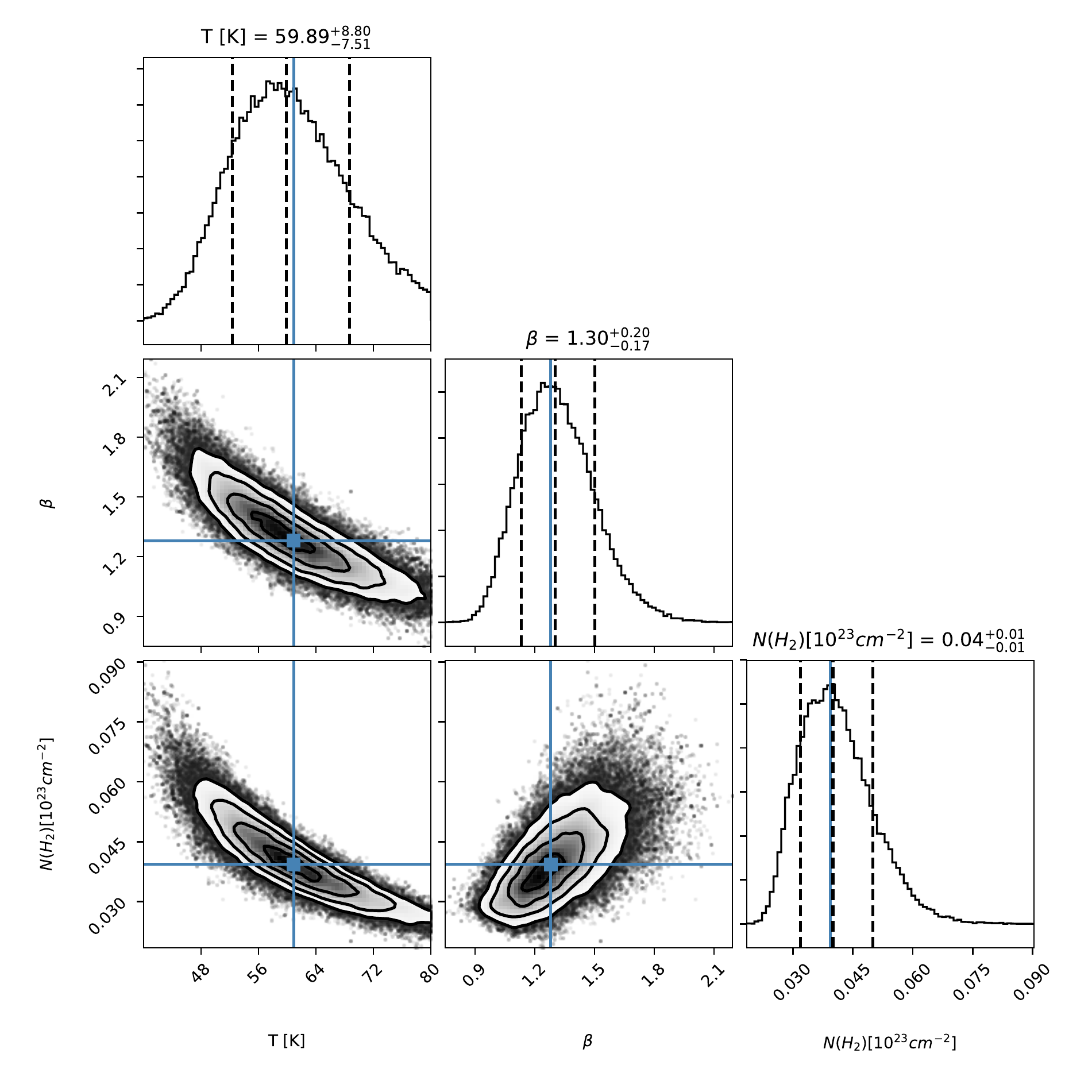}
    \includegraphics[width=3.5in]{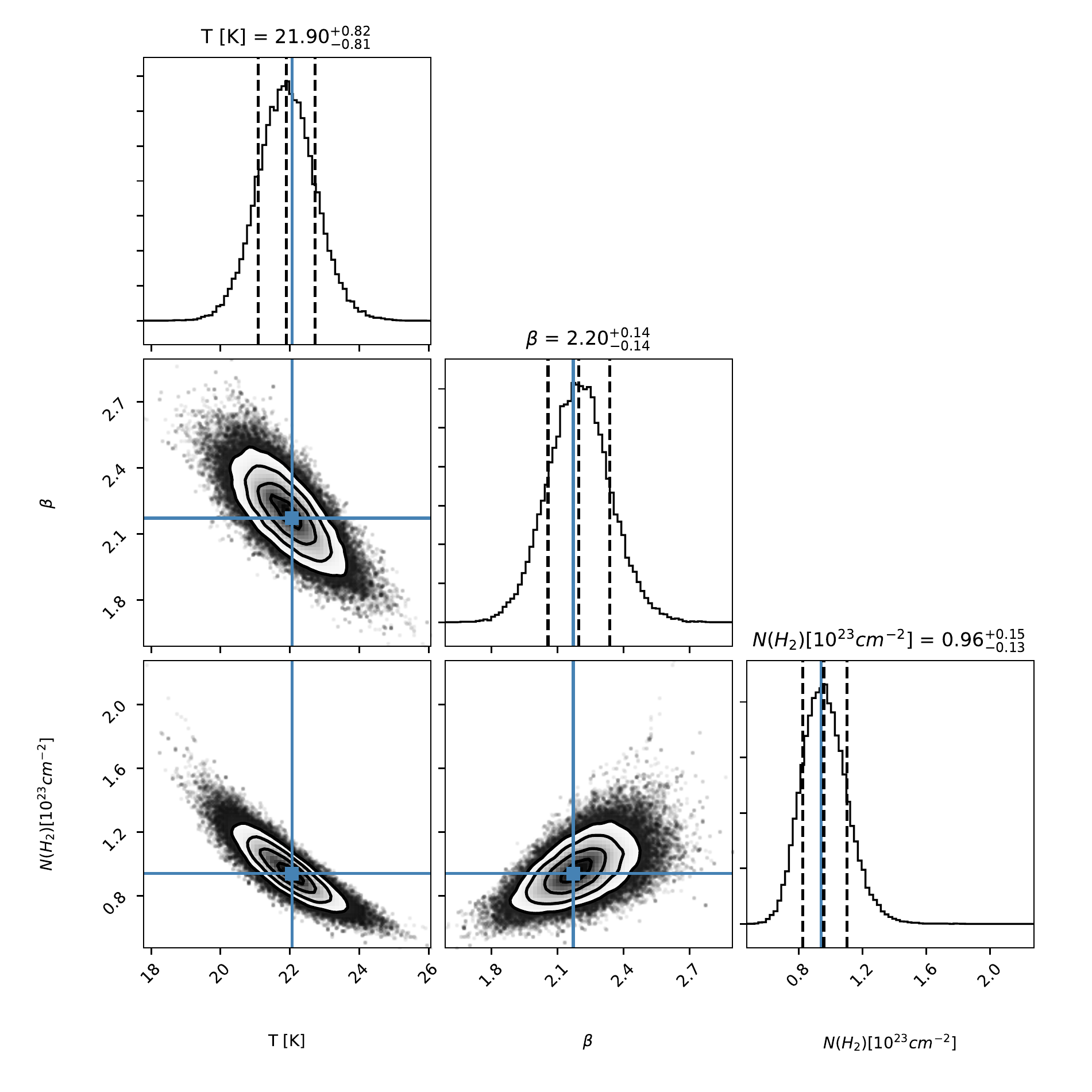}
    \caption{MCMC fits to four fiducial positions shown in Figure~\ref{fig:ind_SEDs}: Core (Top left), Bar (top right), \ion{H}{2} (bottom left), and an edge location of the map (bottom right, $\alpha = 5^\text{h} 35^\text{m} 5.62^\text{s}$, $\delta = -5^\circ 21\arcmin 14.35\arcsec$; J2000). Results from the initial fit described in the text are superposed on the likelihood distributions and are found to be in agreement with the MCMC results. Dashed lines on either side of the maximum likelihood value (reported uncertainties in the plot titles) correspond to the 68\% confidence limits (1~$\sigma$).}
    \label{fig:Tbetacheck}
\end{figure}

To check whether the correlation in Figure~\ref{fig:Tbeta} is an artifact of the fit (i.e. to explicitly search for systematic covariance between $T$ and $\beta$), we have re-fit the SED's for the three fiducial regions shown in Figure~\ref{fig:ind_SEDs} using a Markov Chain Monte Carlo technique \citep{Foreman-Mackey2013,Foreman-Mackey2016}. Figure~\ref{fig:Tbetacheck} shows the results from the MCMC for the three regions in Figure~\ref{fig:ind_SEDs} and one additional region at the edge of the map ($\alpha = 5^\text{h} 35^\text{m} 5.62^\text{s}$, $\delta = -5^\circ 21\arcmin 14.35\arcsec$; J2000) where $T$ is low and $\beta$ is high. There is some covariance between $T$ and $\beta$; however, the width of the likelihood function agrees with the uncertainties obtained from the initial fits, giving confidence that the reported uncertainties remain reasonable despite the underlying covariance. Our likelihoods are similar to the distributions shown in Figure~3 of \cite{Galametz2012} in which the variation in fit parameters resulting from Monte Carlo modifications to the spectral data points are explored. We conclude from these arguments that the $T$-$\beta$ correlation observed in our mapped region likely has a physical origin, as opposed to being entirely an artifact of the fitting process.

\subsection{General Features}\label{sec:Bgeom}
\begin{figure}
    \centering
    \includegraphics[width=3.5in]{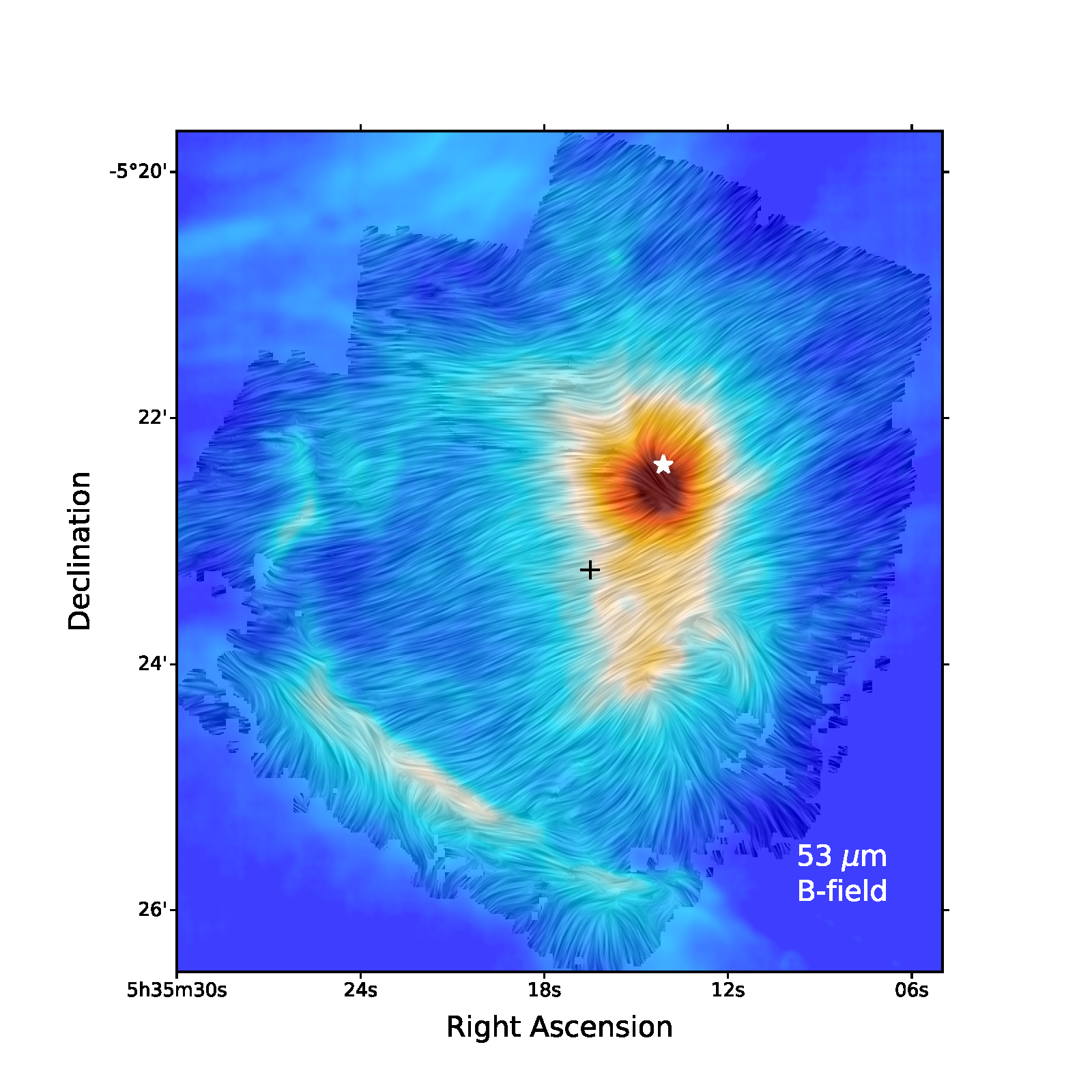}
    \includegraphics[width=3.5in]{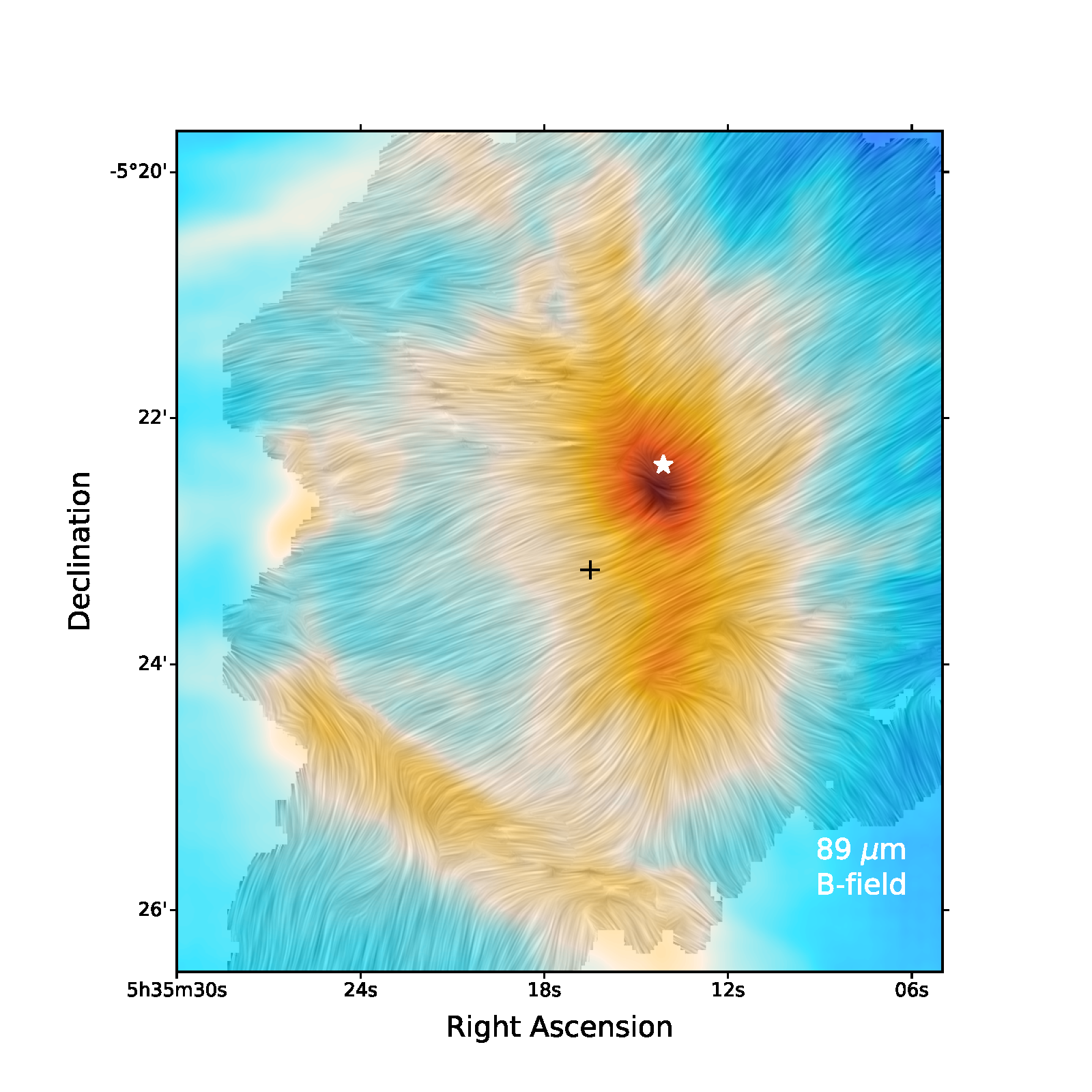}\\
    \includegraphics[width=3.5in]{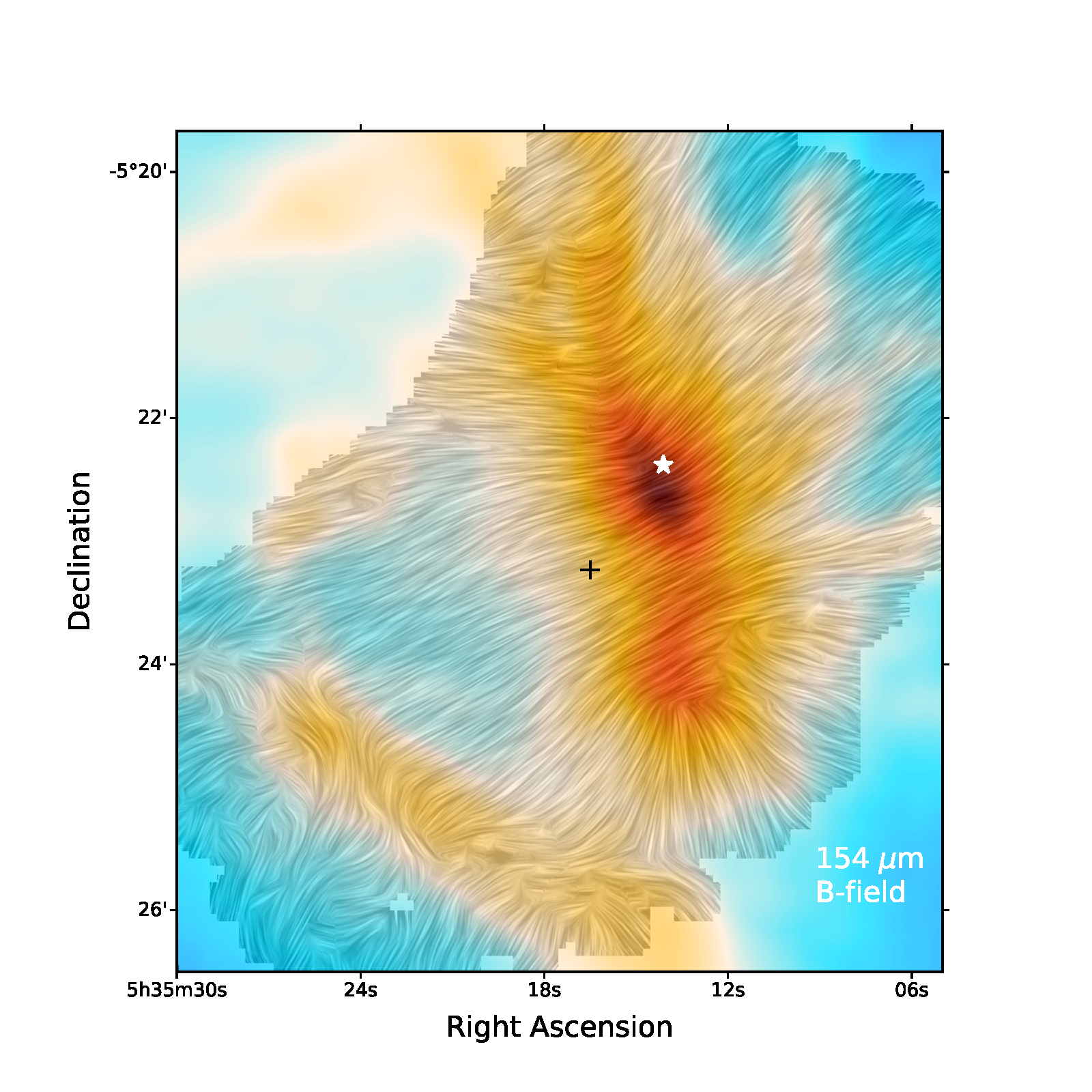}
    \includegraphics[width=3.5in]{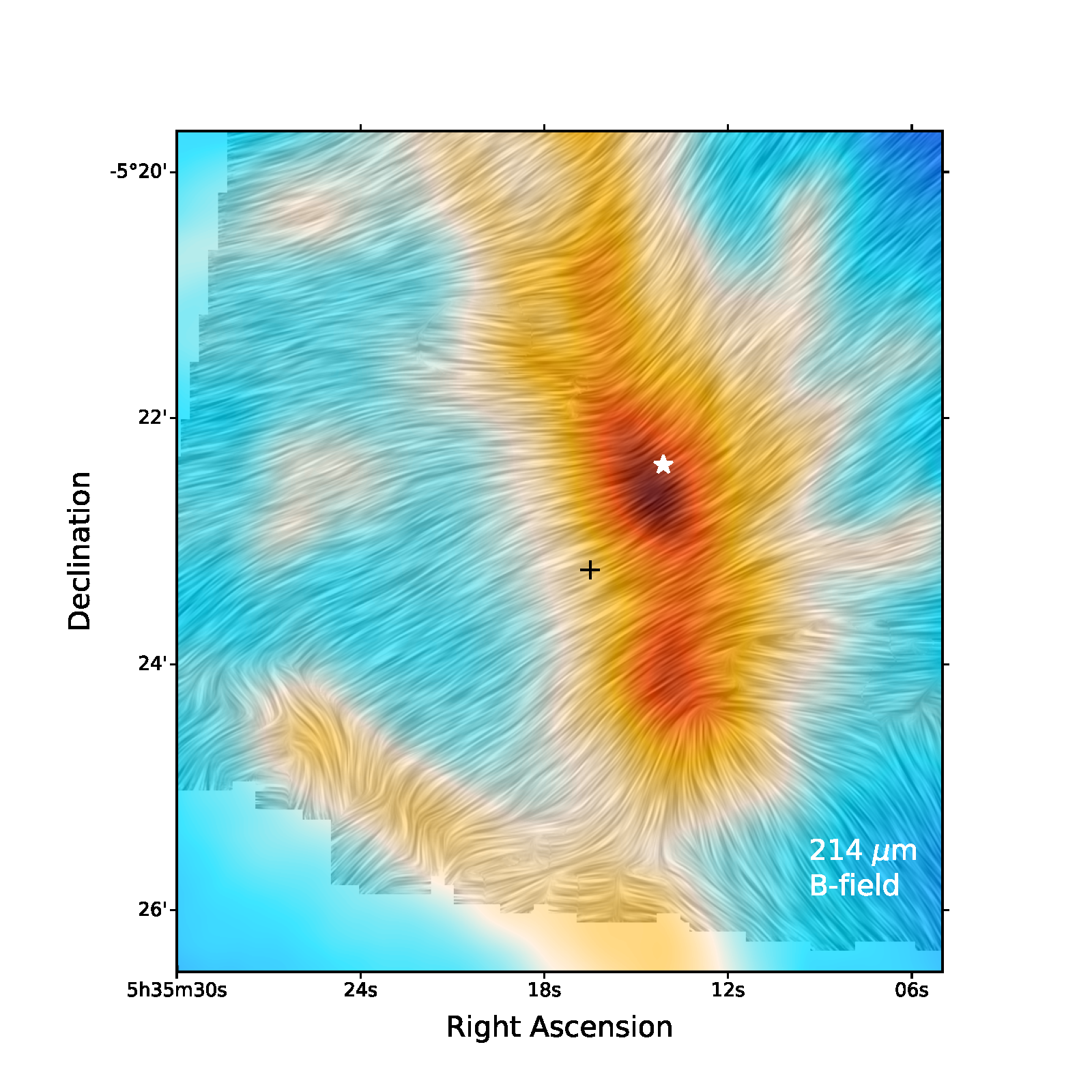}
    \caption{Line Integral Contour \citep{Cabral1993} maps for the 53, 89, 214, and 154 \micron\ inferred magnetic field directions, clockwise starting from the top left. Polarization data are from chop-nod HAWC+ observations. The background intensity images are the same as those used in Figure~\ref{fig:maps}. In each image, the star indicates the location of BN/KL, and the cross indicates the location of the Trapezium Cluster.}
    \label{fig:LIC}
\end{figure}

To compare the inferred magnetic field directions between the maps at different wavelengths, we construct Line Integral Contour (LIC) maps \citep{Cabral1993}, as shown in Figure~\ref{fig:LIC}. 
The general direction of the magnetic field is similar in all four bands and is oriented perpendicular to the Integral Shaped Filament (ISF). The hourglass pinching \citep{Schleuning1998,Houde2004, Ward-Thompson2017} of the magnetic field lines is apparent in all bands, albeit with less curvature than has been observed at 850 $\mu$m \citep{Ward-Thompson2017,Pattle2017}. This feature has been identified as evidence of magnetically-regulated collapse of the molecular cloud.

The inferred magnetic field direction near BN/KL differs between the short and long wavelength bands. At 53 and 89 \micron, there is a component of the field near the highest density region of the cloud that is parallel to the ISF.  We suggest that this may be a result of the BN/KL explosion compressing the fields perpendicular to the explosion direction. 
This feature is not seen at 154 or 214 \micron. The inferred field through the center of BN/KL in these bands is perpendicular to the ISF. This may be because the dust grains that dominate the emission at these wavelengths are cooler dust outside the region of the explosion. NIR (absorption) polarimetry of BN \citep{poid11}, which samples all of the dust along the line of sight to BN independent of dust temperature, yields a position angle of $115^\circ$ counterclockwise from north, in good agreement with our longer-wavelength results. Note that the NIR fractional polarization of BN is normal for the amount of extinction \citep{jone89}, indicating that the total column of dust in {\textit{front}} of BN does not show the effects of strong depolarization. 
\begin{figure}
    \centering
    \includegraphics[width=3.5in]{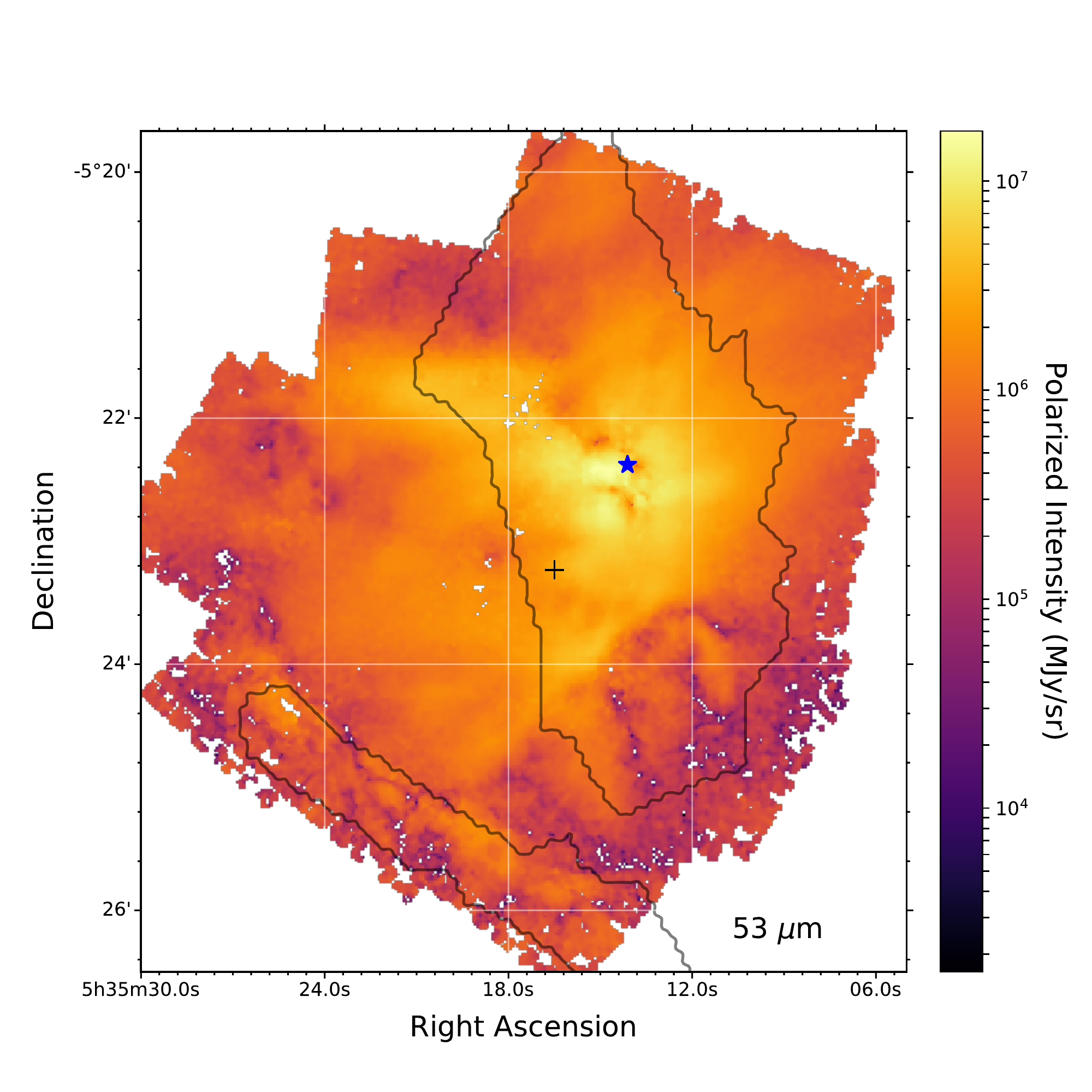}
    \includegraphics[width=3.5in]{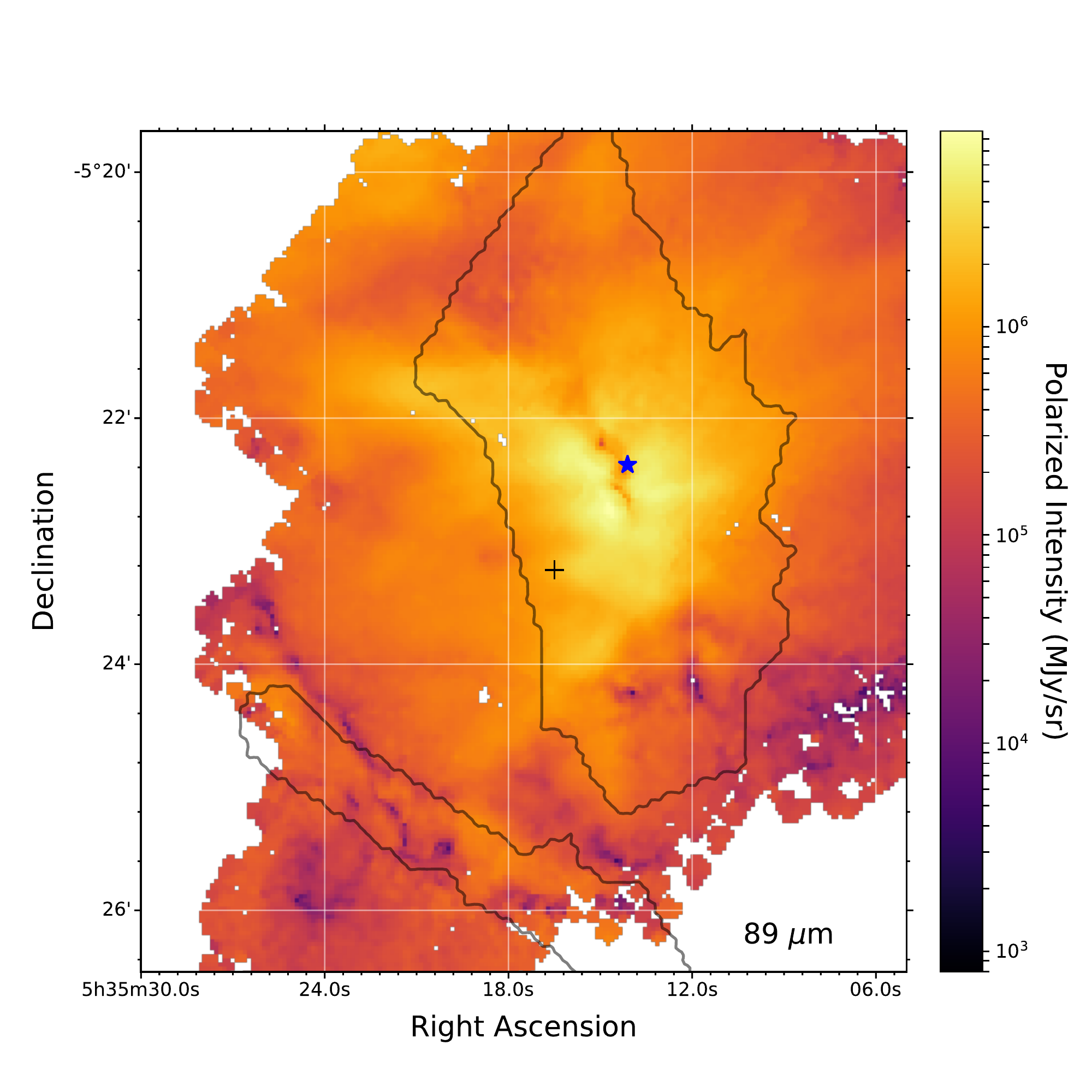}\\
    \includegraphics[width=3.5in]{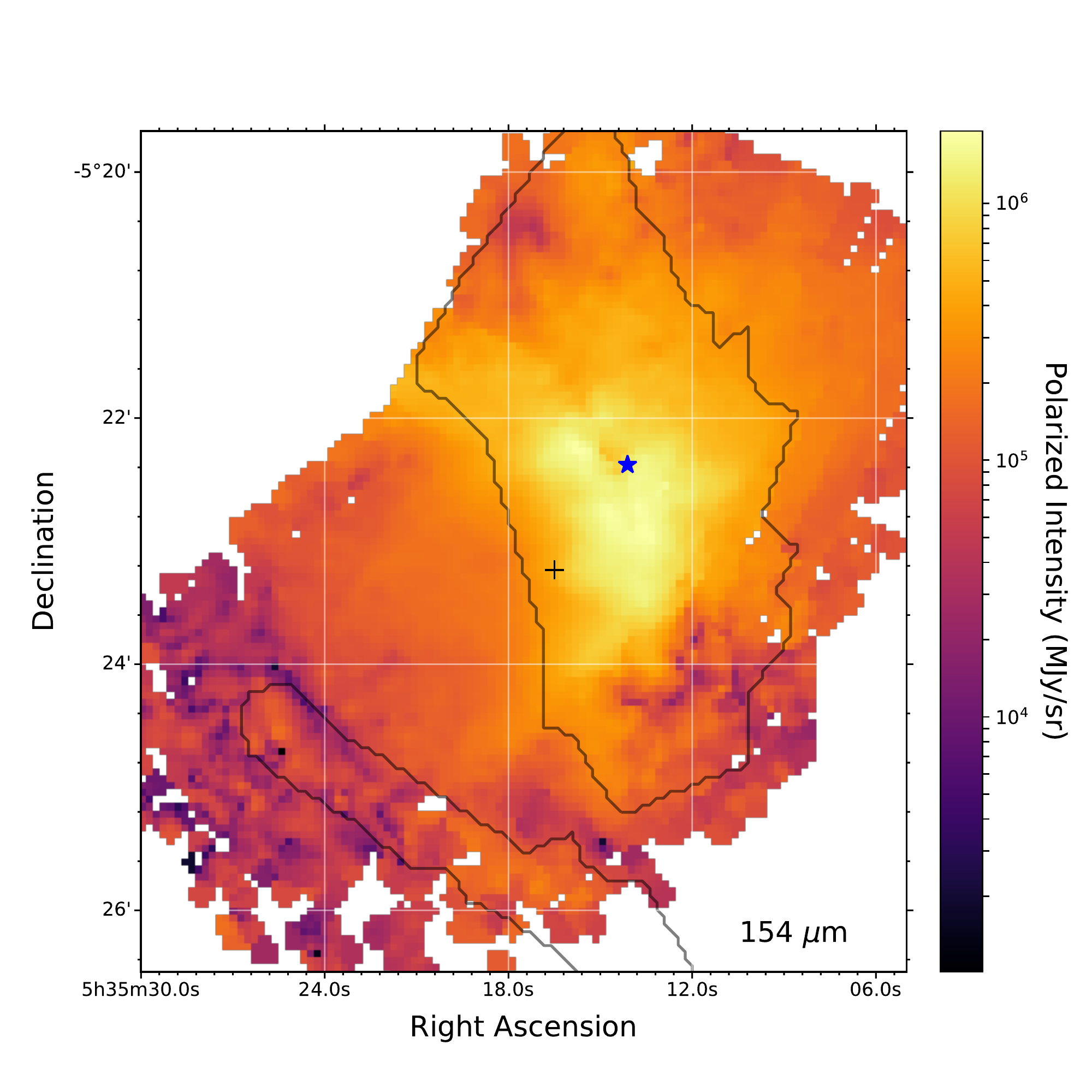}
    \includegraphics[width=3.5in]{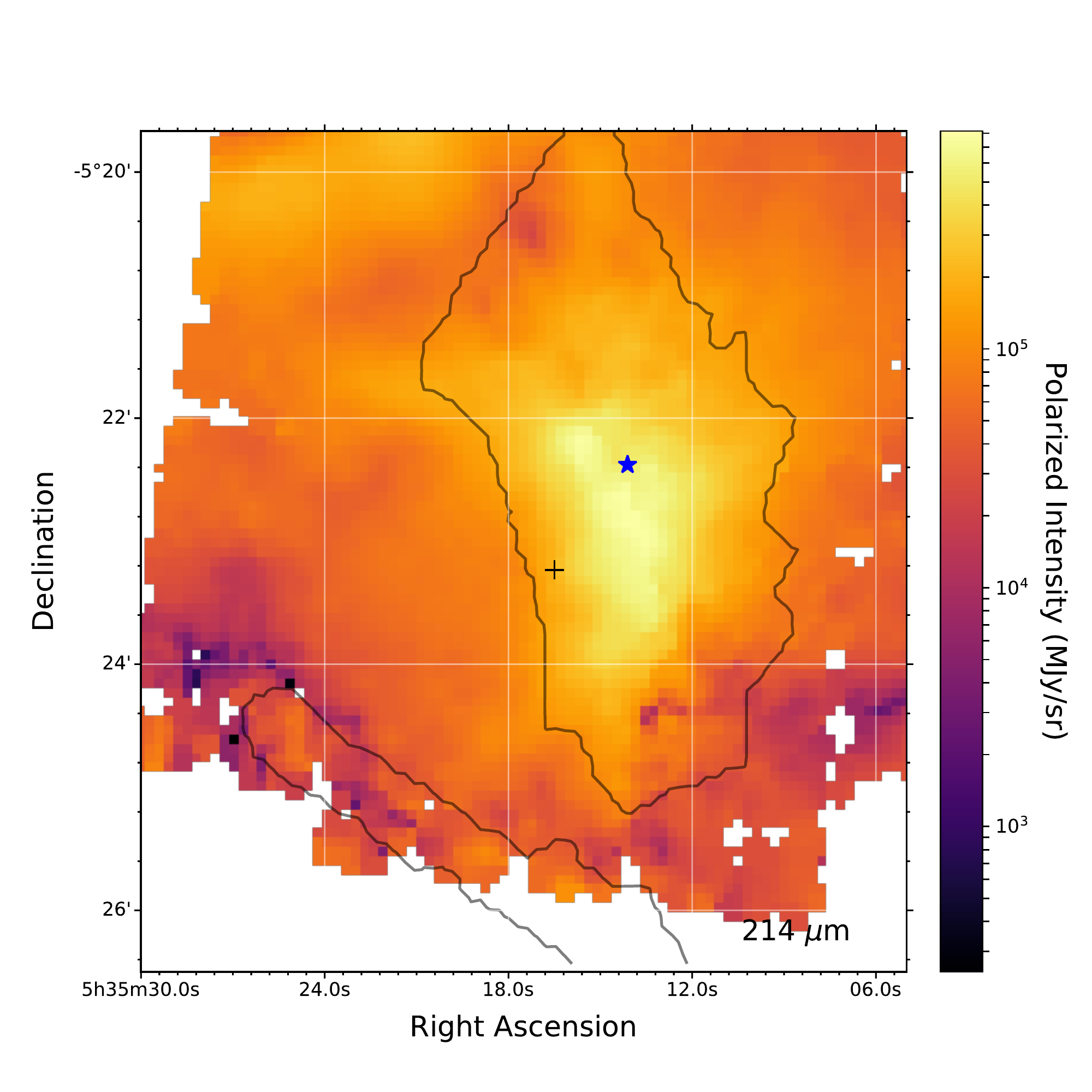}
    \caption{Polarized intensity ($p\times I$) plots for each of the four bands. The images are plotted such that extent and scale of the maps are the same for each band and are the same as the LIC images in Figure~\ref{fig:LIC}.  For reference, the blue star shows the position of the BN object and the black cross shows the location of the Trapezium cluster. Masks defining the BN/KL and Bar regions are shown in light gray contours for reference (see Section~\ref{sec:masks}). }
    \label{fig:PolFlux}
\end{figure}
The inferred magnetic field in the less dense material to the west of the Ridge is generally consistent with the northwest-southeast magnetic field structure (perpendicular to the ISF). 

The polarization measured in the Bar is considerably lower than in other regions in OMC-1, which is in agreement with longer-wavelength measurements \citep{Houde2004,Ward-Thompson2017}.  This may indicate either that the magnetic field is mostly projected along the line of sight, that the grain alignment efficiency is low, that there is significant variation in the direction of the polarization within the beam, or some combination of these effects. Specifically, the low polarization fraction may be a result of polarization cancellation due to multiple magnetic fields superposed along the line of sight.  For example, the ambient field of the cloud (running approximately northwest-southeast as observed in the \ion{H}{2} region) could be superposed on a field that is parallel to the geometry of the Bar, such as what one may expect from compression of the magnetic field if the Bar is indeed an edge-on region bounding a bubble created by the Trapezium Cluster \citep{Dotson1996, Novak2000}.

This interpretation assumes that the grains are magnetically aligned.  In regions where the grains are exposed to particularly high radiation fields, there is the possibility of radiative alignment \citep[$k$-RAT][]{Lazarian2007}. In this case, the angular momentum of the dust grains can become aligned with the $k$-vector direction of the radiation rather than with the magnetic field. Larger, cooler grains would be predominantly aligned by this process.  The longer wavelength observations would manifest this effect preferentially compared to shorter wavelengths, and the effect would be strongest closest to the radiation source.  The relationship between the Trapezium Cluster and the Orion Bar presents such a situation. In the \ion{H}{2} region between the Trapezium cluster and the Bar, the polarization direction (perpendicular to the $B$-field lines in Figures~\ref{fig:maps} and \ref{fig:LIC}) is such that the grains' angular momentum is roughly aligned with the $k$-vectors from the Trapezium Cluster. This is the direction expected for radiative alignment. In the center of the Bar (which is the region closest to the Trapezium cluster), the polarization direction changes between the 53~\micron\ data and the longer wavelengths. At 53~\micron, the polarization is perpendicular to the $k$-vectors from the Trapezium cluster. At longer wavelengths, the polarization direction in the center of the Bar is aligned with the polarization direction in the \ion{H}{2} region.  This observation is in tentative agreement with $k$-RAT theory. 

Alternative explanations for this wavelength-dependent polarization direction include the possibility that polarized reference beam intensity could be altering the direction of the polarization around the Bar. This effect is more significant at longer wavelengths (see Section~\ref{sec:refbeam}). In addition, different wavelengths can be preferentially probing different regions along the line of sight, which may have different field geometries. 

The polarized intensity ($p\times I$) maps are shown in Figure~\ref{fig:PolFlux}.  In all bands, the polarized intensity is concentrated toward the center of field, just as is total intensity, suggesting that aligned dust grains are present throughout the region.
Some features in polarized intensity correspond closely to features in the total intensity, but others do not.  In the 53 and 89 \micron\ maps, there are regions of low polarized intensity coincident with the center of the BN/KL explosion.  They are not seen in the longer wavelengths, possibly because the larger beams do not resolve them.  These low polarized flux features can be explained by either magnetic field spatial variations at scales below that of the beamsize or regions where the magnetic field is predominantly oriented along the line-of-sight or by the effects of optical depth. The 89 \micron\ image shows this ``depolarization'' as a line oriented approximately perpendicular to the BN/KL explosion axis (see Section~\ref{sec:bnkl}). This may indicate that the depolarization is related to the explosion. 

As is the case with the fractional polarization, the polarized intensity is also low near the Bar.  In this case, a thin line of low polarized intensity is observed in each of the 4 maps, which is located at the edge of the Bar that is closest to the trapezium cluster.  This supports the idea that the field is aligned with the edge of the bar and is predominantly oriented along the line-of-sight, perpendicular to the ambient field along the line-of-sight, or a combination of the two. 

\subsection{Object Masks}\label{sec:masks}

\begin{figure}
    \centering
    \includegraphics[width=3.0in]{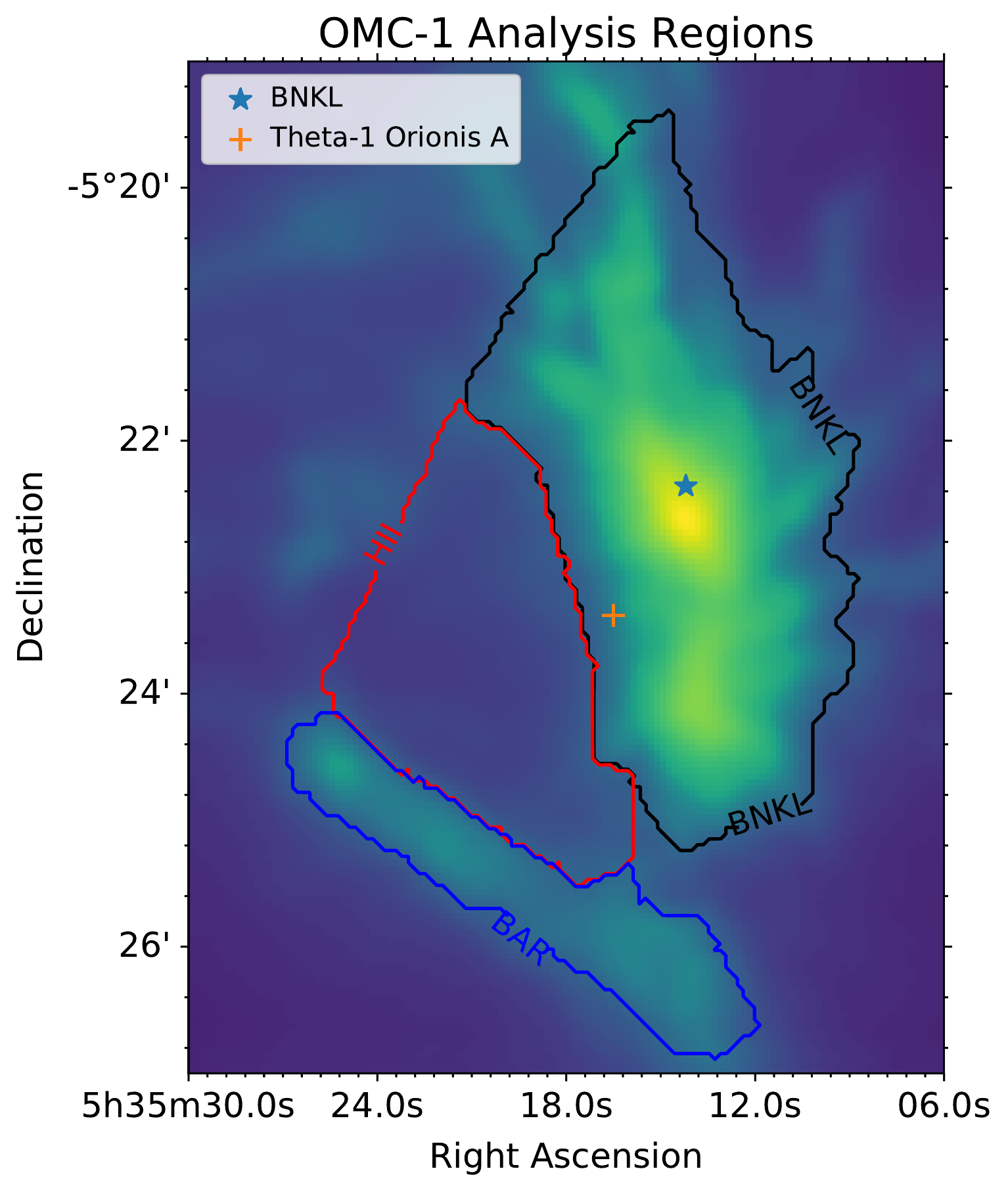}
    \caption{The three masked regions are indicated by the black (BNKL), red (HII), and blue (BAR) lines.}
    \label{fig:masks}
\end{figure}

In addition to the signal-to-noise threshold that is applied to the polarization maps, we apply additional cuts to our data for subsequent analysis. Temperature, density, and environmental conditions vary across the OMC-1 complex. Thus, we anticipate that the magnetic field strength will as well. Because of this and the relatively large number of vectors in the HAWC+ data, it is advantageous to apply analysis techniques to regions that are physically similar. Motivated by this, we have constructed a mask set that distinguishes three particular regions of interest. These regions correspond to 1. the BN/KL region and Molecular Ridge, 2. the Orion Bar, and 3. the less dense intercloud \ion{H}{2} region that is heated by the Trapezium cluster of O--B stars. These regions will be denoted ``BNKL,'' ``BAR,'' and ``HII,'' hereafter. The BNKL and BAR regions are defined by the 1,700 $MJy\, sr^{-1}$ contour in the HAWC+ 154 \micron\  chop-nod photometry map. The HII region is defined as the area between the two other masks. These masks are shown superposed on the 154 \micron\  HAWC+ scan map in Figure~\ref{fig:masks}. 

\subsection{Reference Beam Contamination}\label{sec:refbeam}
Because OMC-1 is located in a region of extended emission, it is necessary to quantify the effect of polarized emission in the reference beams of the observations. The measured normalized Stokes parameters, ($q_m, u_m$), are related to the source polarization ($q_s,u_s$), the reference polarization, ($q_r,u_r$), the source intensity, $I_s$, and the mean intensity in the two reference beams, $\bar{I}_r$.
\begin{align}
q_m&=\frac{q_sI_s-q_r\bar{I}_r}{I_s-\bar{I}_r}, \\
u_m&=\frac{u_sI_s-u_r\bar{I}_r}{I_s-\bar{I}_r}.
\end{align}
Here, we assume that the reference beam polarization fraction is the same in each reference beam because we will estimate a worst-case scenario by choosing a high but reasonable polarization fraction to make a pixel-by-pixel estimation of the maximum contamination. In this paper, we follow the formalism described in the Appendix of \cite{Novak1997} to estimate the maximum effect of reference beam contamination based on measured intensities and estimates of polarization of the reference beam contamination.  The estimates for minimum and maximum limits to the fractional polarization ($p_{sys}^+$ and $p_{sys}^-$, respectively) and the maximum angular error can be written in terms of the ratio of the reference beam intensity to the measured intensity, $w\equiv \bar{I}_r/I_m$, the measured polarization, $p_m$, and the assumed reference beam polarization, $p_r$.  
\begin{align}\label{eq:rbi}
p^{+}_{sys}&=\mathrm{max}\left[p_m,\left(\frac{p_m+p_rw}{1+w}\right)\right]\\
p^{-}_{sys}&=\frac{p_m-p_rw}{1+w}\\
\Delta \phi_{sys}&=\frac{1}{2}\mathrm{arctan}\left[\frac{p_rw}{(p_m^2-p_r^2w^2)^{1/2}}\right]
\end{align}
We use a subset of the Herschel maps described in Section~\ref{sec:SED} to estimate $I_r$ for each pixel in the HAWC+ chop-nod maps. We use the Herschel maps to fit a simple model for the emission,
\begin{align*}
    I=I_0\nu^2B_\nu(T),
\end{align*}
where $I_0$ is a fitted amplitude and $B_\nu(T)$ is the Planck function.
We then use this model to calculate the intensity for both the right and left reference beams for each observed sky pixel. $\overline{I}_r$ is then found by averaging these two quantities.  For the 53 and 89 \micron\ maps, we used the Herschel/PACS 100 and 160 \micron\ maps to model the intensity; for the 154 and 214 \micron\ maps, we added the Herschel/SPIRE 250 \micron\ maps when fitting for the intensity model. No smoothing was done for the Herschel maps. 

The measured intensity in each band, $I_m$, was taken from the calibrated Stokes I value of the HAWC+ chop-nod maps. From these, a map of $w\equiv\bar{I}_r/(I_s-\bar{I}_r)$ was produced for each band. 
We assumed that the polarization of the reference beams is $p_r=0.10$ in all HAWC+ bands. This represents one of the higher measurements of polarization observed and thus provides a conservative estimate. The values for the measured polarization, $p_m$, are the non-debiased polarization fractions from the HAWC+ polarimetry data sets.

\begin{figure}
    \centering
    \includegraphics[width=3.5in]{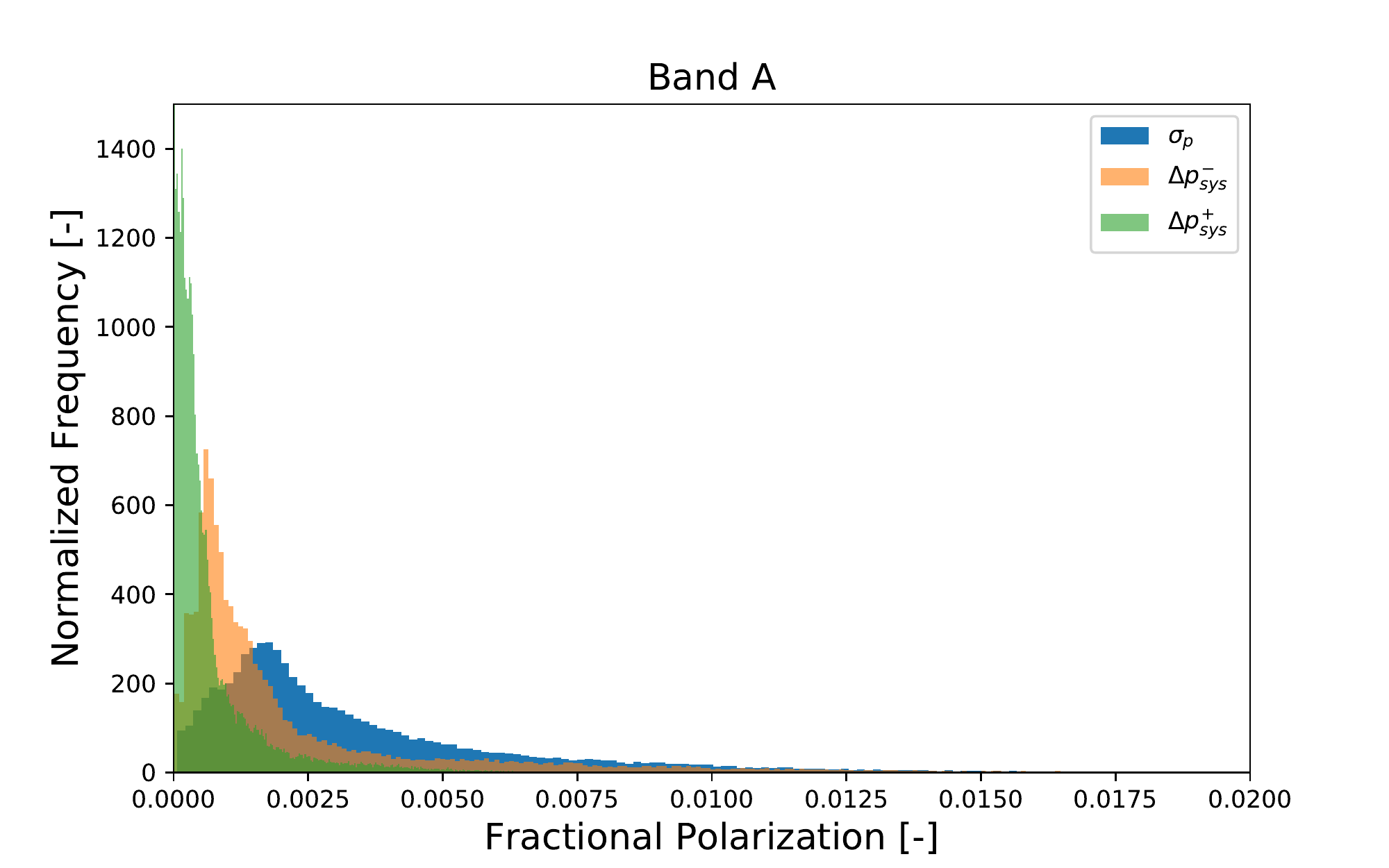}
    \includegraphics[width=3.5in]{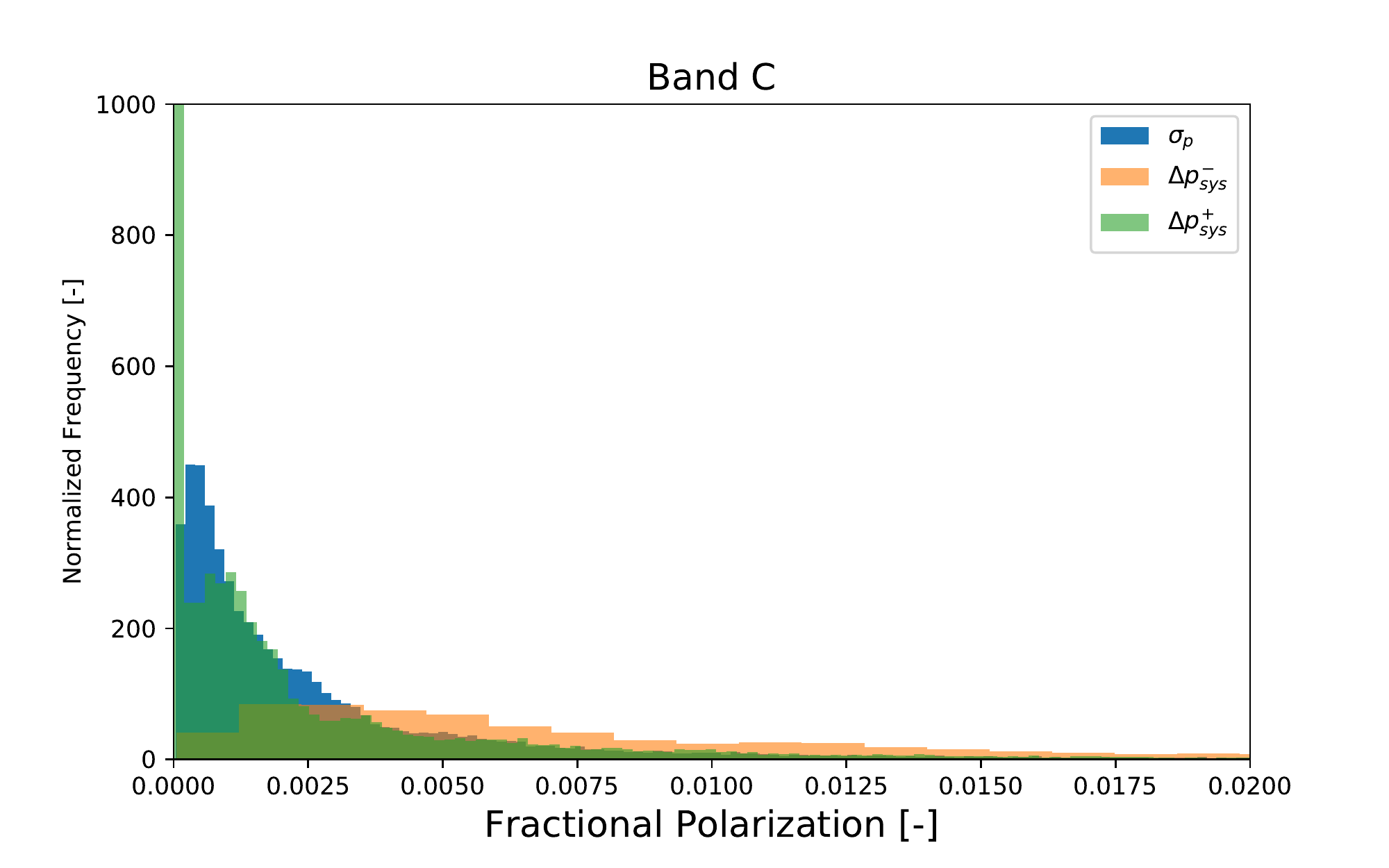}\\
    \includegraphics[width=3.5in]{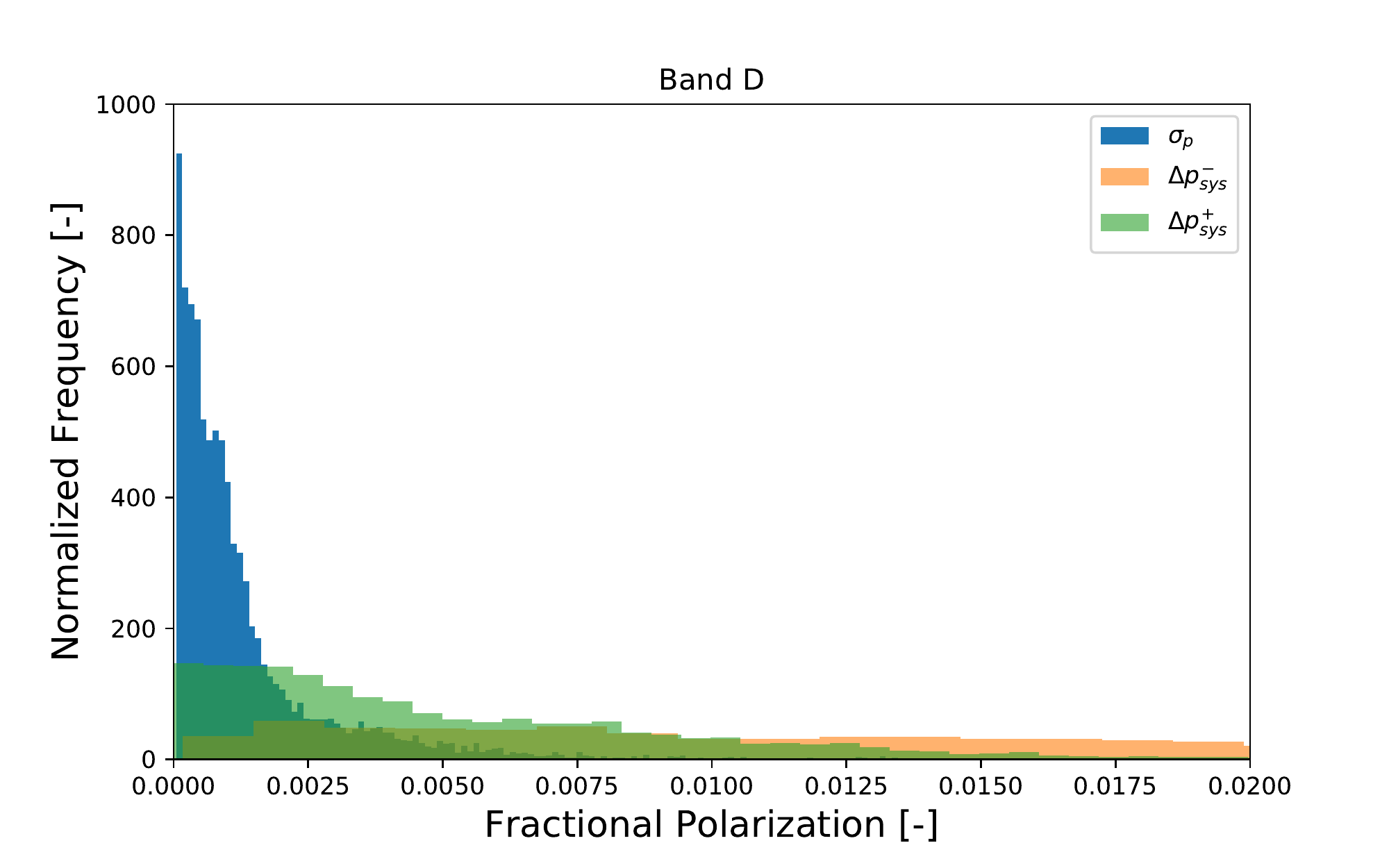}
    \includegraphics[width=3.5in]{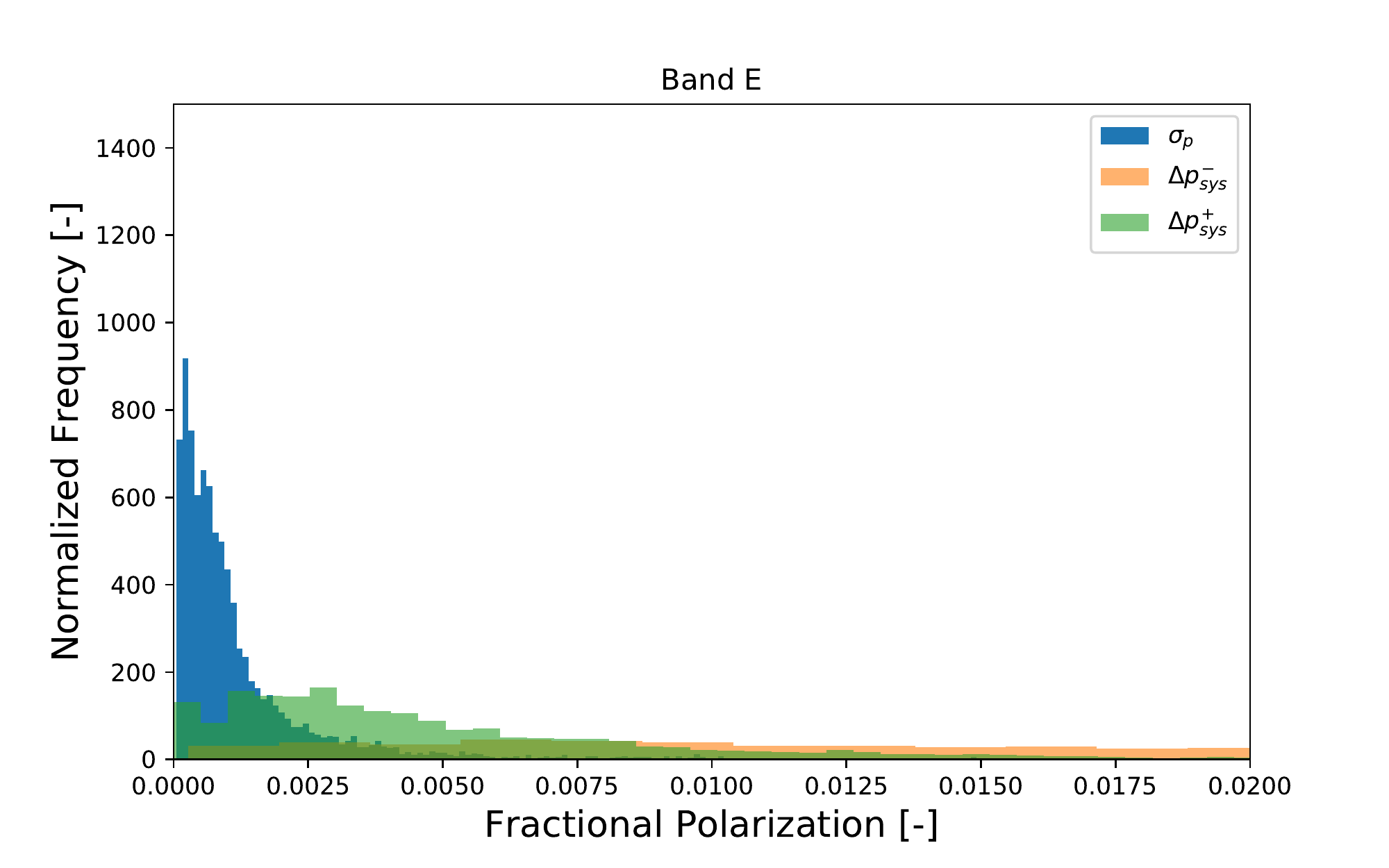}
    \caption{Histograms for the statistical errors on polarization fraction along with the upper and lower systematic errors for polarization fraction due to estimated reference beam contamination are shown for each of the four bands. For these plots, $\Delta p^+_{sys}\equiv p^+_{sys}-p_m$ and $\Delta p^-_{sys}\equiv p_m- p^-_{sys}$.}
    \label{fig:perr}
\end{figure}

From these estimates of $w$, $p_r$, $p_m$ and $\overline{I}_r$, maps of the quantities in Equations~\ref{eq:rbi}--8 were made. In all bands, maps of these relevant quantities were stored in FITS format and then applied for subsequent data cuts. 
Figure~\ref{fig:perr} shows histograms of the upper and lower uncertainties on polarization due to estimated reference beam contamination along with corresponding histograms for the statistical errors ($\sigma_p$) in each band.

\begin{figure}
    \centering
    \includegraphics[width=6in]{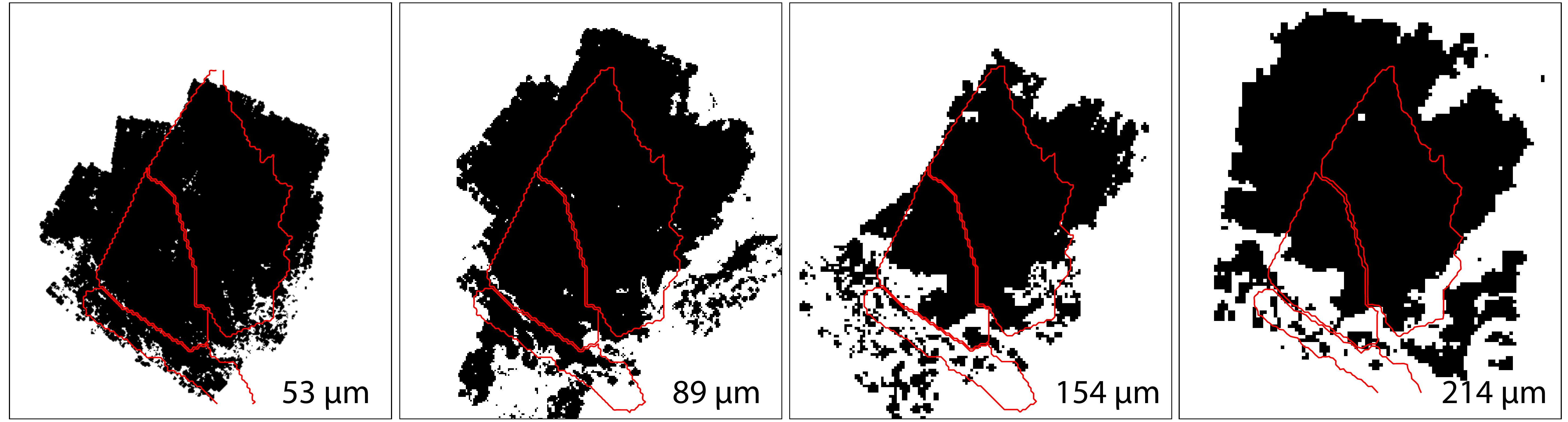}
    \caption{Pixel locations are shown (black) for the regions where the maximum effect of the reference beam affects the angles less than $10^\circ$. Region masks are overplotted for reference.The 3-$\sigma$ data cut is also included.}
    \label{fig:rbcuts}
\end{figure}

As an illustration of this method, Figure~\ref{fig:rbcuts} shows a mask set corresponding to the cut $\Delta\phi_{sys}<10^\circ$. This is the mask set utilized in Section~\ref{sec:PolAngDisp}. The 3-$\sigma$ data cut is also included. 
The 53 \micron\ map is mostly unaffected by the cuts for possible reference beam contamination due to the higher temperatures of the OMC-1 center relative to the surrounding cloud.  At longer wavelengths (89, 154, and 214 \micron), much of the Bar region is flagged by this algorithm due to its low polarization. As noted in Section~\ref{sec:Bgeom}, there are multiple possible causes for this.  

\subsection{Fractional Polarization vs. Intensity and Angle Dispersion}\label{sec:pvi}

The magnetic field in the ISM has both constant (threaded) and turbulent components \citep[see a recent treatment by][]{plan18}. The effects of a turbulent component can be seen in variations of the polarization angle with position on the sky, analyzed using angle dispersions \citep{Davis1951,Chandrasekhar1953,Myers1991,Pattle2017,plan18} or a type of structure function \citep{kobu94, hild99, Houde2016}, and in the trend of fractional polarization with column density \citep[e.g.][]{joba15, hild99}. We analyze the structure function in Section~\ref{sec:PolAngDisp}, and in this section we examine the trend of fractional polarization with column density and angle dispersion.

If the magnetic field geometry is perfectly constant with no bends or wiggles, the fractional polarization in emission will be constant \citep[][for a review]{jowh15} with column density.  (For most lines of sight in our OMC-1 map, the far-infrared optical depth is $\ll$ 1, so the effects of opacity on the polarization can be ignored.)  If the magnetic field varies in a purely stochastic way along the line of sight on scales comparable to the beam, the fractional polarization will decrease as the square root of the column density \citep{joba15}. A combination of a constant and a purely random component will cause the polarization to decrease with column density at a rate in between these two extremes. If there is a coherent departure from a purely constant component such as a spiral twist, regions of mutually perpendicular fields, or a simple large scale variation of the projected field along the line of sight, the fractional polarization can drop faster than the square root of the column density due to strong cancellation of the polarization. The \cite{plan18} analysis of the polarized foreground in the Milky Way using both the structure function and the observed fractional polarization suggests this is common. In addition to cancellation effects reducing the fractional polarization, loss of grain alignment can cause dilution of the fractional polarization by unpolarized intensity from regions with unaligned grains \citep{anla15, joba15}. 

\begin{figure}
    \centering
    \includegraphics[height=2.5in]{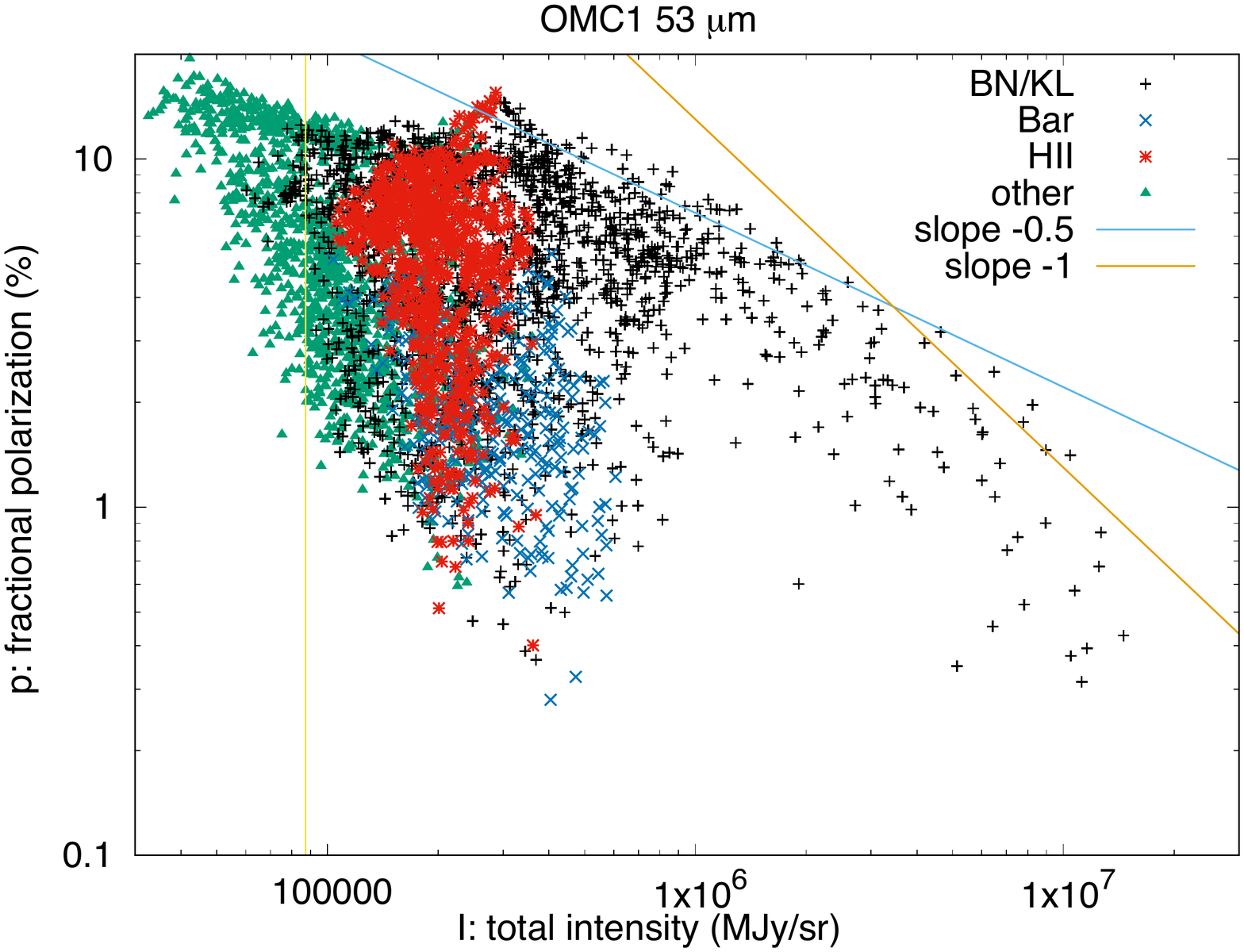}
    \includegraphics[height=2.5in]{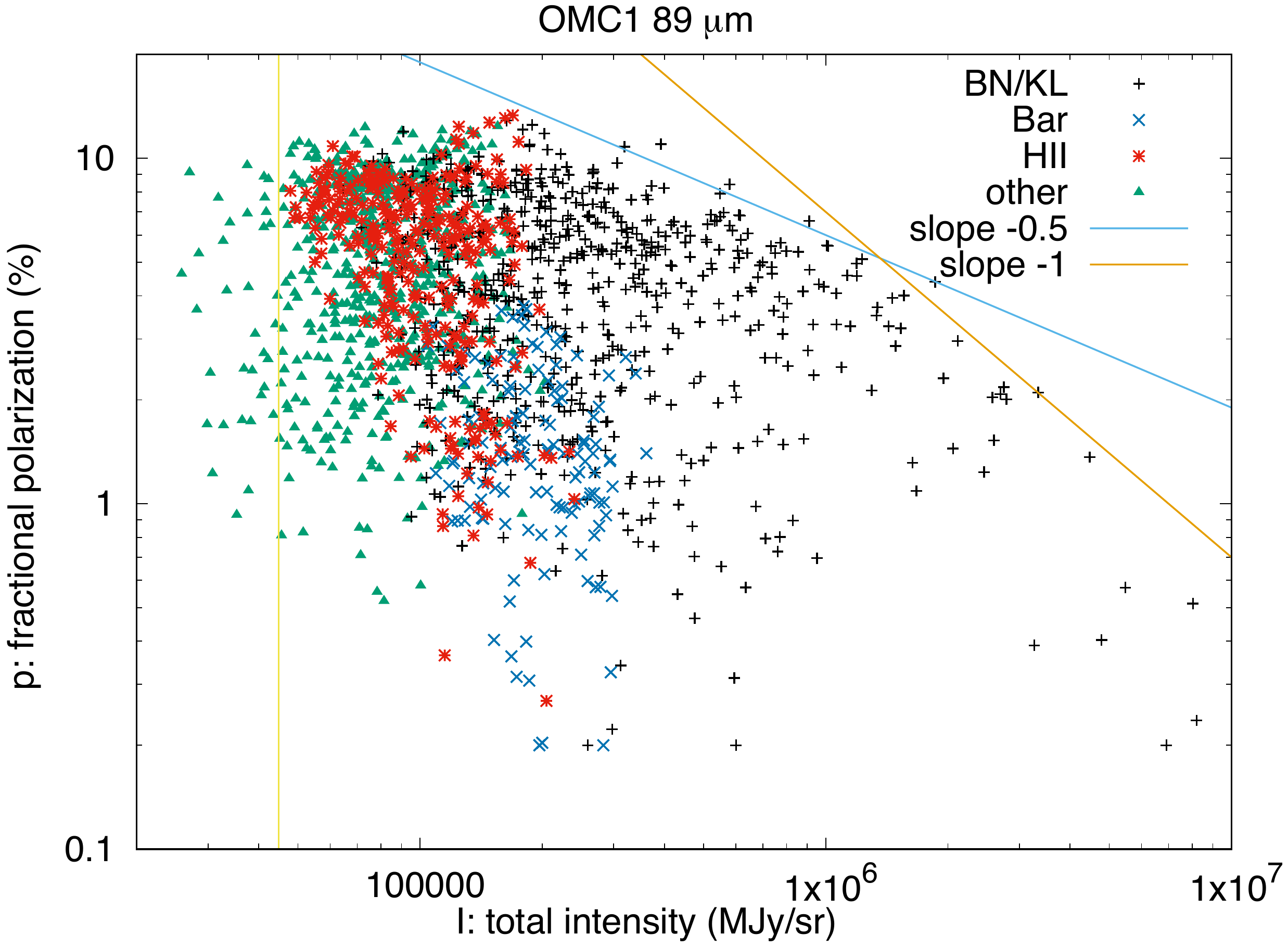}
    \includegraphics[height=2.5in]{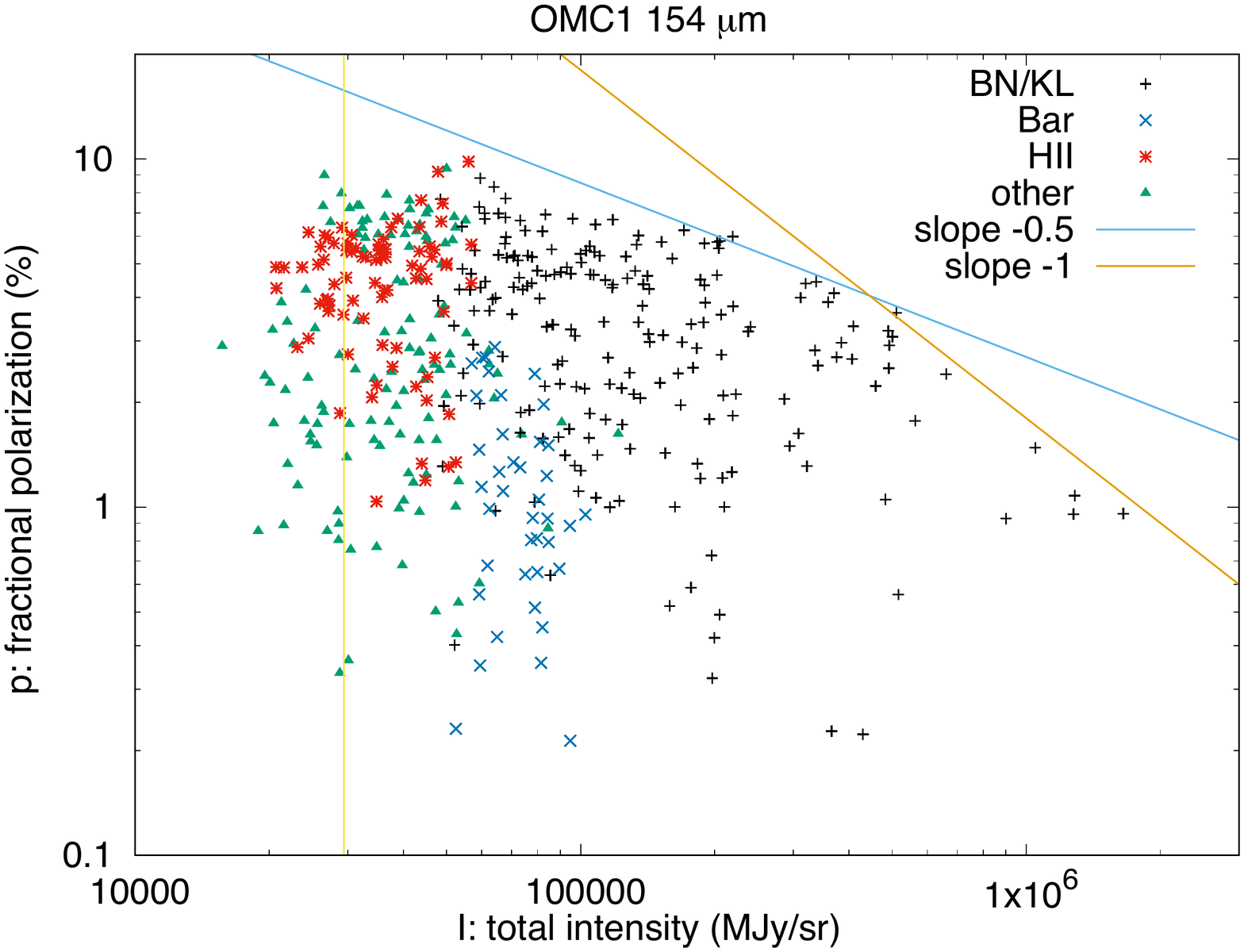}
    \includegraphics[height=2.5in]{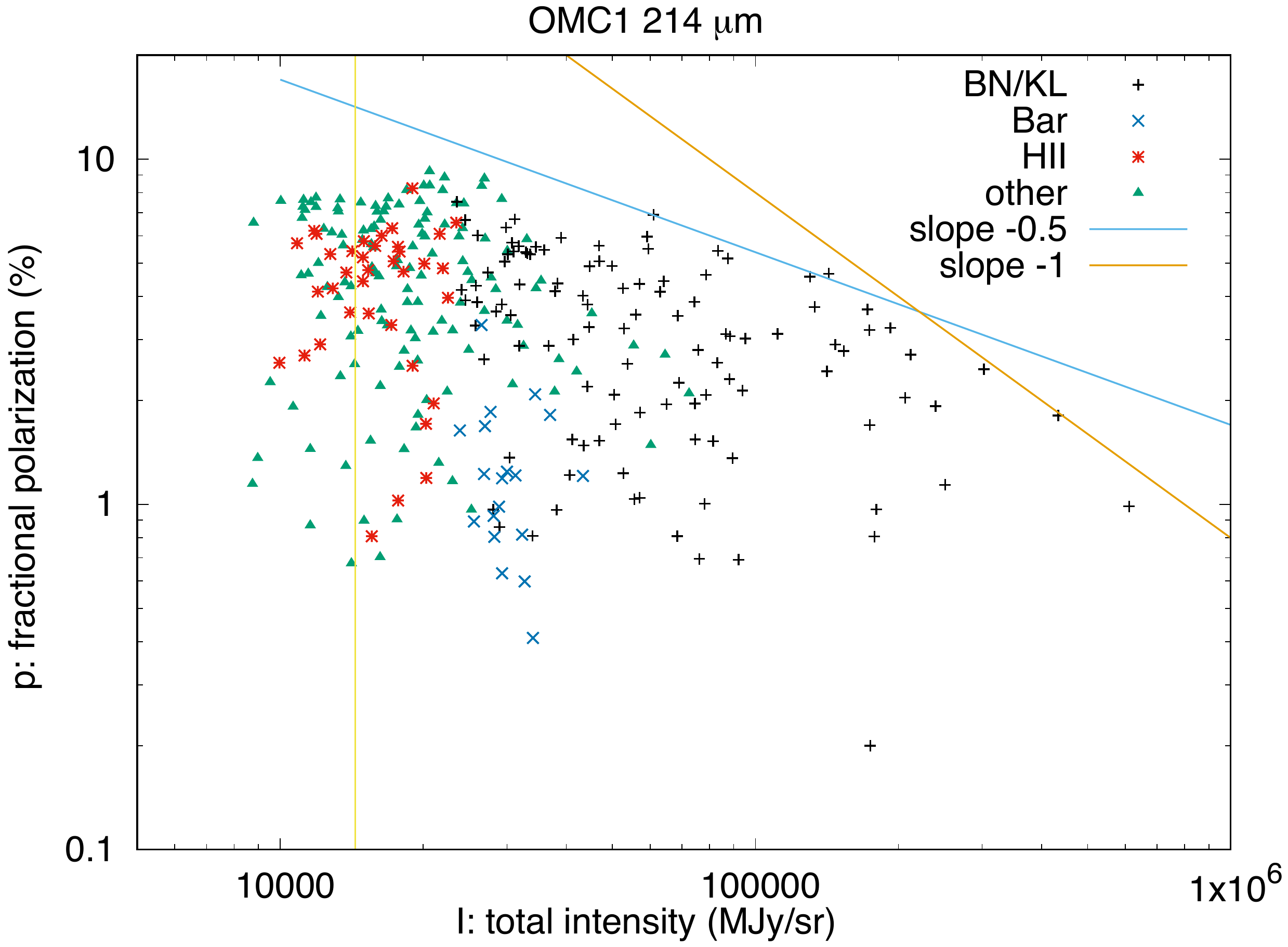}
    \caption{Plots of fractional polarization versus intensity for the four HAWC+ bands.  Points are sampled from the maps at an interval which is approximately the beam FWHM.  A cut has been made to exclude points with $|p_{sys} - p| > 2\%$, according to Section~\ref{sec:refbeam}.  This cut removes points at low total intensity, especially those below the flux indicated by the vertical yellow line.  Points with $p < 0.2\%$ are shown with $p = 0.2\%$.  Overall, the correlation between $p$ and $I$ is weak; however, the upper envelope of points defines a fairly clear trend, with $p$ decreasing with increasing $I$ with logarithmic slope between -0.5 and -1.  Lines with these slopes are indicated with diagonal lines in the plots.  The points at high intensity in all 4 bands correspond to the BN/KL core region.  See Section~\ref{sec:pvi} for discussion.}
    \label{fig:PVI}
\end{figure}

The trend in fractional polarization with surface brightness (intensity) for all four bands is illustrated in Figure~\ref{fig:PVI}. We are concentrating on the upper bound in these plots because that delineates lines of sight where the minimum depolarization effects are present.
If we roughly characterize the slope of the upper bound with a single power law $p \propto I^\alpha$ in each bandpass, we find $\alpha \sim$ -0.6 to -0.7. This is steeper than $p \propto {I^{ - 1/2}}$, indicating there must be large-scale variation of the projected field along the line of sight, loss of grain alignment for some fraction of the line of sight, or a combination of both. If the slope were $\alpha = -1$, often seen in dense protostellar cores \citep[e.g.][]{gala18}, then the denser regions would likely suffer loss of grain alignment \citep{joba15}. Our result of a shallower slope, and the fact that NIR polarization in extinction towards BN is the expected value for diffuse ISM extinction \citep{jone89, poid11}, suggests loss of grain alignment cannot be the sole explanation for the decrease in fractional polarization with intensity.  In fact, based on the evidence presented below, we conclude that the trend can be explained entirely by magnetic field structure, with no need to invoke variations in grain alignment.

As shown in Figure~\ref{fig:PVI}, the measured fractional polarization in OMC-1 ranges over a factor of 30 or more over the maps.  There is a tendency for the largest fractional polarization $p$ to be found where the total intensity $I$ is relatively low, and for the smallest $p$ to be found where $I$ is relatively high.  However, overall the correlation of $p$ and $I$ is weak.  A much better predictor of the fractional polarization along a particular line of sight is the dispersion of polarization angles in its vicinity, and in fact that is the strongest correlation we have found with $p$.  The correlation of $p$ with angle dispersion $S$ was explored previously for Planck and BLASTPol submillimeter polarimetry at $\geq$5\arcmin\ scales \citep{Planck2015_XIX,Fissel2016,plan18}.  Those authors found trends with $p \propto S^{-0.6}$ to $S^{-1}$.  \citet{plan18} developed a model for turbulent magnetic field structure predicting $p \propto S^{-1}$.

Here, we use a simple method of calculating the dispersion of polarization angles:  for a given line of sight, we compute the rms of the angles within a diameter of $\theta_S$ centered on that line of sight.  We exclude from the calculation angle measurements with statistical uncertainty $>10^\circ$.  We use Stokes parameters to avoid the complication of the branch of position angles.  In detail:
\begin{equation}
S = \sqrt{\langle(\phi - \overline{\phi})^2\rangle} \approx \sqrt{\langle\sin^2(\phi - \overline{\phi})\rangle} = \sqrt{(1-\langle\hat{q}\rangle\hat{q}(\overline{\phi}) - \langle\hat{u}\rangle\hat{u}(\overline{\phi}))/2} ,
\end{equation}
in radians, computed within a circular region centered on a given line of sight (associated with a particular measurement of $p$),
where:
\begin{equation}
    \hat{q_i} = \cos{2\phi_i}, \hat{u_i} = \sin{2\phi_i}
\end{equation}
To remove approximately the effect of noise bias on the measurements of $S$, we debias the results by subtracting the rms uncertainty in $\phi$ over the aperture:  $S_{\rm debiased} = \sqrt{S^2 - \sigma^2(\phi )}$. This is a small effect over most of these OMC-1 maps.  In the rest of the paper, we use $S$ as shorthand for $S_{\rm debiased}$.

\begin{figure}
    \centering
    \includegraphics[height=2.5in]{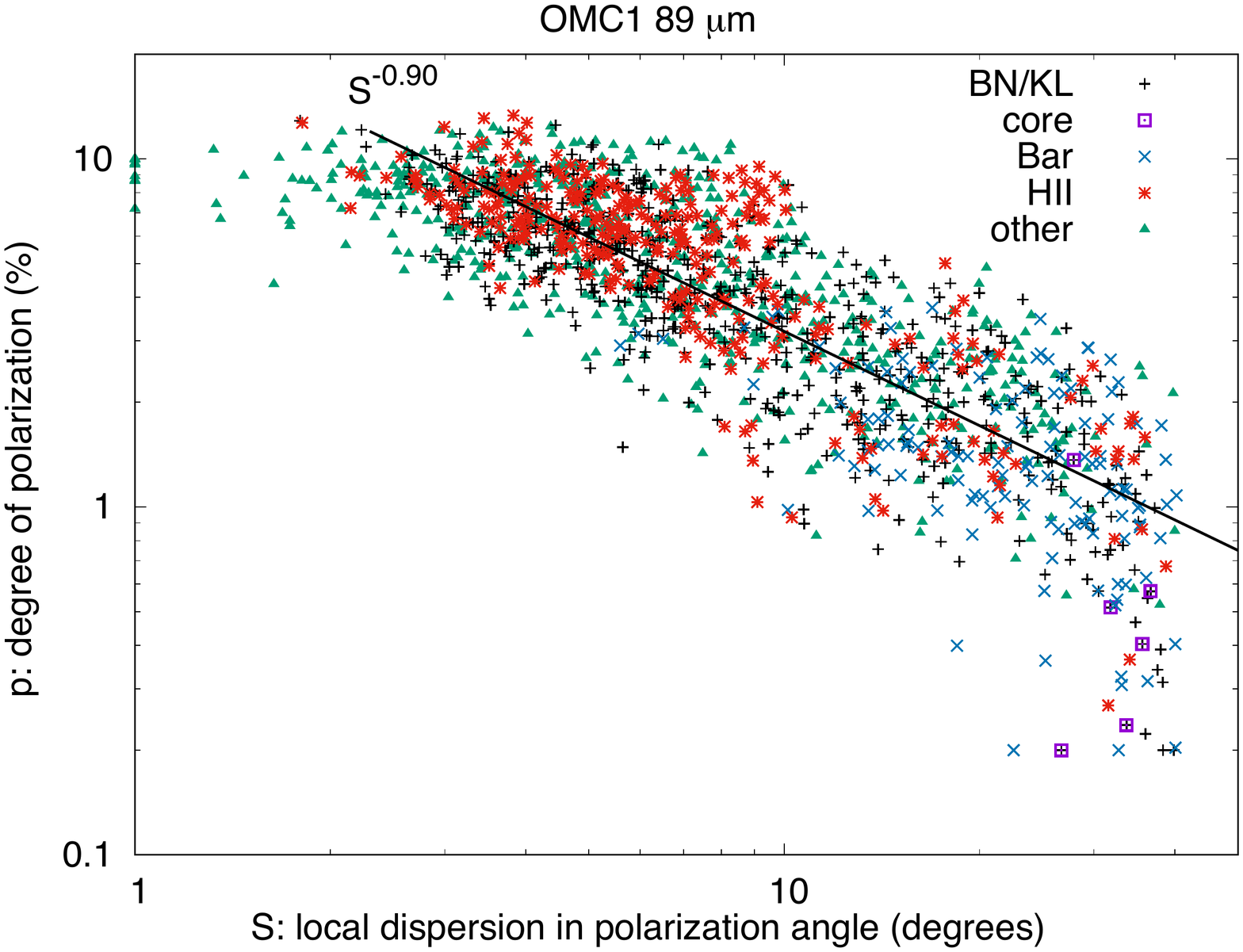}
    \includegraphics[height=2.5in]{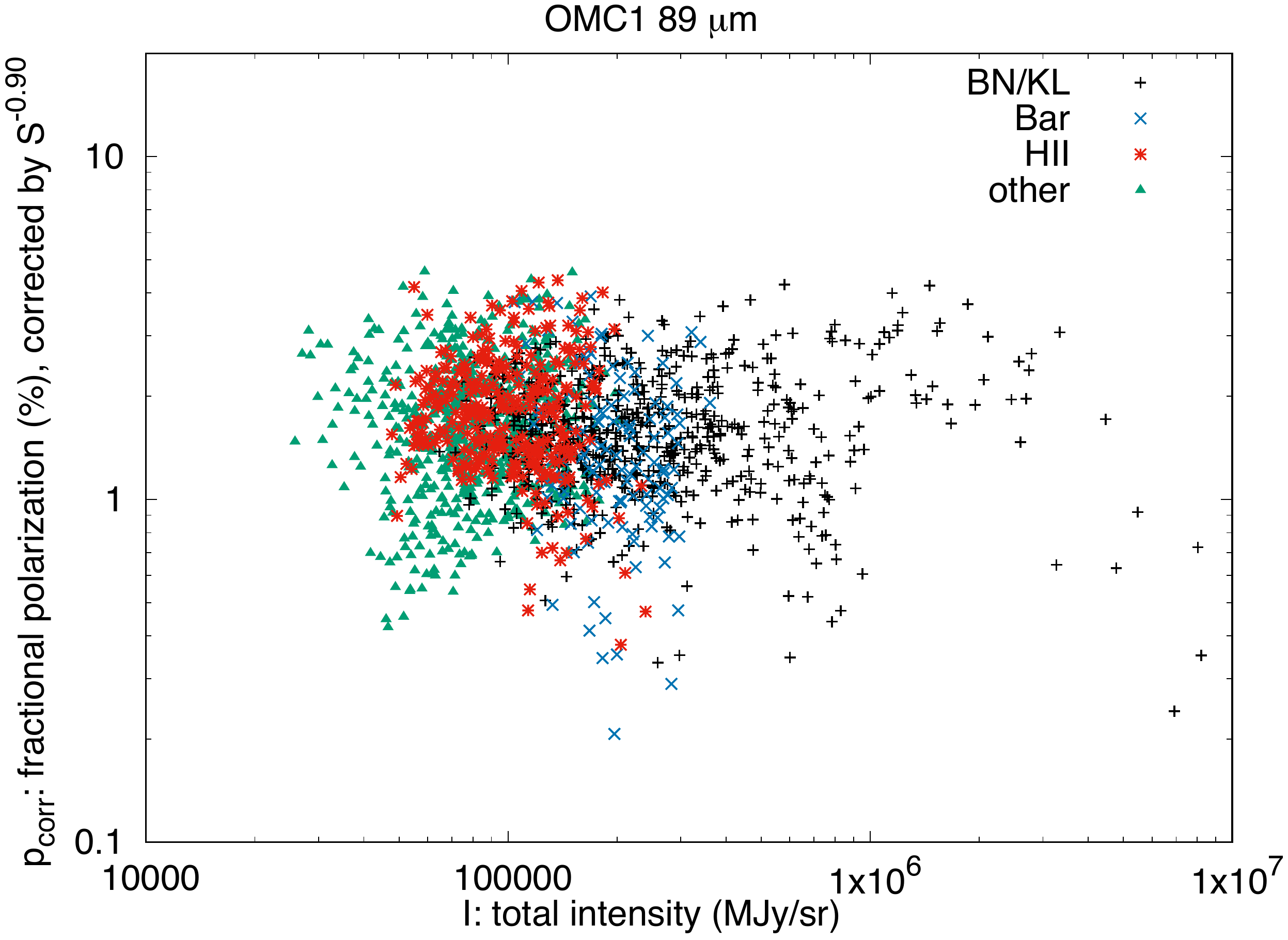}
    \caption{Left:  Fractional polarization vs. angle dispersion for OMC-1 at 89 \micron , using $\theta_S = 30\arcsec $.  In addition to the $|p_{sys} - p| > 2\%$ cut, points have been limited to $p \ge 0.2\%$ and $S \ge 1^\circ$.  Points within the FWHM of the BN/KL $I$ peak are highlighted (square "core" symbols); the polarization toward those lines of sight tends to fall below the trend.  Right: Fractional polarization vs. total intensity for OMC-1 at 89 \micron , now with $p$ corrected for the angle dispersion trend by dividing each point by the local value of $(S/12.5^\circ)^{-.90}$.  Again, the points at high intensity corresponding to the BN/KL core tend to fall below the trend.}
    \label{fig:PVS}
\end{figure}

Figure~\ref{fig:PVS} shows the observed relationship between $S$ and $p$ for the 89 \micron\ observations of OMC-1.  Following \citet{Fissel2016}, we fit the trend in $p$ as a function of two parameters, $S$ and total intensity $I$:
\begin{equation}
    p \approx p_0 (I/\overline{I})^{\alpha_I} (S/\overline{S})^{\alpha_S}
\end{equation}
The fit results are shown for all bands in Table~\ref{tab:p_vs_SI_results} and for 89 \micron\ in Figure~\ref{fig:PVS}.  For this initial look at the far-infrared angle dispersion, we used a common value of $\theta_S = 30\arcsec $ for all bands, just large enough to allow calculation of $S$ at 214 \micron .

\begin{table}[]
    \centering
    \begin{tabular}{ccc}
    Wavelength & $\theta_S$ & Best-Fit Trend \\
    (\micron ) & (arcsec) & \\
    \hline
    53 & 30 & $p \approx 3.1\% (I/3.8\times10^5 MJy/sr)^{-0.01} (S/14.2^\circ)^{-0.87}$ \\
    89  & 30 & $p \approx 2.6\% (I/2.4\times10^5 MJy/sr)^{-0.09} (S/12.5^\circ)^{-0.90}$ \\
    154 & 30 & $p \approx 1.9\% (I/1.0\times10^5 MJy/sr)^{-0.19} (S/12.5^\circ)^{-0.84}$ \\
    214 & 30 & $p \approx 2.3\% (I/0.42\times10^5 MJy/sr)^{-0.21} (S/8.6^\circ)^{-0.70}$ \\
    \hline
    \end{tabular}
    \caption{Fits to OMC-1 polarization trends.}
    \label{tab:p_vs_SI_results}
\end{table}

Our fit to the data finds a dependence $p \propto S^{-0.9}$ to $S^{-0.7}$ , close to the $S^{-1}$ form expected for a simple model of magnetic field structure \citep[Appendix E by][]{plan18}.  With that dependence removed, there is little further dependence on $I$ for OMC-1:  fits for the power law index $\alpha_I$ range from -0.01 to -0.21.  Standard error propagation indicates uncertainty $\sigma(\alpha_I)$ of $\sim0.01$ at 53 \micron\ increasing to $\sim0.03$ at 214 \micron\ due to the smaller number of measurements; however, the $\chi^2$ for the polarization model increases by only $\lesssim30\%$ if the $I^{\alpha_I}$ term is omitted.  On the other hand, \citet{Fissel2016} found a clear $I^{-0.45}$ dependence in their BLASTPol 500 \micron\ observations of Vela C (along with $S^{-0.60}$) with loss of grain alignment in denser regions offered as a possible explanation.  In OMC-1, we do not see clear evidence for poorer grain alignment in dense regions up to column densities of $N_H \approx 10^{23} cm^{-2}$, as further demonstrated qualitatively by Figure~\ref{fig:PolFlux}.  To first approximation, the observed distribution of fractional polarization can be explained by the magnetic field structure of the cloud.  OMC-1 has a stronger radiation field from its embedded stars than does Vela C, which in the context of radiative alignment torques ($B$-RAT, in this case) \citep{Lazarian2007} could explain the difference between our results and those of \citet{Fissel2016}.  To make a rough estimate of the difference in intensity of the radiation field, we use the dust temperature, which has a median value of 36 K for OMC-1 (Section~\ref{sec:SED}) and median value of $\sim$15 K for Vela C \citep{Hill2011}.  This corresponds to a ratio of $(36/15)^4 \approx 30$ in intensities.

Especially at 53 and 89 \micron , the fractional polarization toward the BN/KL core falls below the trend with $I$ and $S$ (Figures~\ref{fig:PVI} and \ref{fig:PVS}).  \citet{Schleuning1998} argued that low 100 \micron\ polarization toward BN/KL is due to optical depth $\approx$ 0.6. Such a value corresponds to $\sim4\times 10^{23}$ cm$^{-2}$ in Figure~\ref{fig:sedparams} and is localized to the two peaks along the Ridge. The suppression of polarization due to optical depth should be greater at shorter wavelengths, and in fact the HAWC+ data show a monotonic trend of fractional polarization decreasing with decreasing wavelength.  In a 30\arcsec\ diameter aperture centered on BN/KL \citep[matching][]{Schleuning1998}, the fractional polarization is 0.44\%, 0.71\%, 1.10\%, and 1.47 \% at 53, 89, 154, and 214 \micron , respectively.  Further supporting the hypothesis of optical depth significantly influencing the fractional polarization is the spectral energy distribution of BN/KL (Section~\ref{sec:SED}).  The calculated 53 \micron\ optical depth toward BN/KL, at 18.7\arcsec\ resolution, is 0.8; this optical depth reduces the emergent fractional polarization by a factor of $\sim 1.5$ at 53 \micron\ and less at longer wavelengths \citep{dowe97} -- insufficient to fully explain the trend with wavelength.  However, we note that there is clear wavelength-dependent polarization angle structure within the 30\arcsec\ aperture (Figures~\ref{fig:LIC} and \ref{fig:PolFlux}) which could also play a role in the variation of polarization fraction, and we also have not considered the clumpiness of the emitting medium.



The maximum fractional polarization, corresponding to favorable conditions of field orientation and order and of grain alignment, provides a lower limit to the elongation of dust grains \citep{Hildebrand1995,Draine2017,Guillet2018}.  In the HAWC+ maps of OMC-1, one local maximum in the fractional polarization is seen most clearly at 89 \micron , located at $\alpha_{J2000} = 5^\text{h} 35^\text{m} 21^\text{s}~\delta_{J2000} = -5^\circ 21\arcmin 50\arcsec$  in an elongated polarized flux feature which does not correspond closely to a feature in total intensity.  The maximum observed fractional polarization is 14.0\% , and the total intensity is 4 Jy/arcsec$^2$.  The Herschel 100 \micron\ map indicates a source to reference beam intensity ratio of approximately 25 for this line of sight and SOFIA chop (Section~\ref{sec:data}); the minimum intrinsic polarization that could produce 14.0\%\ observed polarization (via perpendicular source and reference beam polarization angles) is 12.9\% .  The same feature has an observed fractional polarization of 15.4\%, 9.3\%, and 8.2\% at 53, 154, and 214 \micron , respectively.  Other local maxima with higher fractional 89 \micron\ polarization are located where the intensity is far less, making those measurements vulnerable to significant reference beam effects, so they are not considered further.

\citet{Ward-Thompson2017} have noted a coherent magnetic field structure with high fractional 850 \micron\ polarization in a filament northeast of BN/KL.  Only our 214 \micron\ map has good coverage of this feature.  It is clearly defined in polarized intensity and has a maximum fractional polarization of 9.6\% toward $\alpha_{J2000} = 5^\text{h} 35^\text{m} 22^\text{s}~\delta_{J2000} = -5^\circ 19\arcmin 50\arcsec$, where the total intensity is 0.6 Jy/arcsec$^2$, approximately 10$\times$ the reference beam flux.  The minimum intrinsic polarization that could produce this is 7.9\%.

In summary, our observations of OMC-1 have identified lines of sight with fractional polarization $\geq$14\%, $\geq$13\%, and $\geq$8\% at 53, 89, and 214 \micron , respectively.  At the shorter wavelengths, this is somewhat higher than the maximum of 9\% found by \citet{Hildebrand1995} at 100 \micron ; with further HAWC+ observations of OMC or other fields, one may find still higher polarization in the far infrared.  At slightly longer wavelengths, a maximum fractional polarization of 8--13\% was observed in the Vela C cloud, mapped over a $\sim$1 degree area with 5\arcmin\ resolution at $\lambda$ = 250--500 \micron\ \citep{Gandilo2016}.  At 850 \micron , a maximum dust fractional polarization of $\sim$22\% has been measured elsewhere in the Galaxy \citep{Benoit2004,plan18}; the sensitivity to relatively diffuse clouds is a likely explanation for the higher value compared to the OMC-1 data we are reporting here.

\subsection{Magnetic Field Strength} \label{sec:PolAngDisp}

The Davis-Chandrasekhar-Fermi (DCF) method \citep{Davis1951,Chandrasekhar1953} can be used to obtain estimates of the plane-of-the-sky magnetic field strength by comparing the dispersion of polarization vectors to the velocity dispersion. One challenge with this technique is that the large scale field structure can contribute to the dispersion. In order to separate the dispersion due to the turbulent-field component from that of the large scale field, an isotropic two-point structure function, or dispersion function (DF), can be calculated to characterize the dispersion as a function of angular scale \citep{Hildebrand2009,Houde2009,Houde2011,Houde2016}. 
The dispersion function can be fit with a model that separates the large scale contribution from that of the turbulence \citep{Houde2016},

\begin{equation}
    1 - \langle\cos[\Delta \phi(l)]\rangle = \frac{1}{1+\mathcal{N} \left[\frac{\langle B_{t}^{2}\rangle}{\langle B_{0}^{2}\rangle}\right]^{-1}}\left\{1-\exp\left(-\frac{l^{2}}{2(\delta^{2} + 2W^{2})}\right)\right\} + a_{2}l^{2},
    \label{eq:disp_model}
\end{equation}

\noindent
where the first term accounts for the small-scale turbulent contribution to the dispersion (taking into account the correlations due to beam size), and the second term corresponds to the ordered, large-scale field contribution. In Eq. \ref{eq:disp_model}, $l$ is the distance between a pair of vectors with angle difference $\Delta \phi$, and $W$ corresponds to the beam radius. Angle brackets indicate average values. $\frac{\langle B_{t}^{2}\rangle}{\langle B_{0}^{2}\rangle}$ is the turbulent-to-ordered field ratio, and $\mathcal{N}$ is the number of turbulent cells in the gas column given by \citep{Houde2009}
\begin{equation}
\mathcal{N} = \frac{(\delta^{2}+2W^{2})\Delta'}{\sqrt{2\pi}\delta^{3}}.
\label{eq:N_turb}
\end{equation}

In Eqs. \ref{eq:disp_model} and \ref{eq:N_turb}, $\delta$ is the correlation length for the turbulent field, and $\Delta'$ is the effective thickness of the cloud. See \citet{Houde2009,Houde2011,Houde2016} for full details on the model above. In fitting the DFs using the equations above, there are four parameter to be determined: $\frac{\langle B_{t}^{2}\rangle}{\langle B_{0}^{2}\rangle}$, $\delta $, $a_{2}$, and $\Delta'$. The parameters $\frac{\langle B_{t}^{2}\rangle}{\langle B_{0}^{2}\rangle}$ and $\Delta'$ are highly degenerate, so we fit for $a_{2}$, $\delta$, and $\Delta'' \equiv \Delta'\left(\frac{\langle B_{t}^{2}\rangle}{\langle B_{0}^{2}\rangle}\right)^{-1}$.  Following \citet{Houde2009}, $\Delta'$ can be estimated by the FWHM value of the isotropic auto-correlation of the polarized intensity.

We implemented a Markov chain Monte Carlo (MCMC) solver for fitting the non-linear model of Eq. \ref{eq:disp_model} to the DFs and determining the optimal model parameters and their associated uncertainties. The model represented by  Eq.~\ref{eq:disp_model} is only valid at small values of $l$ ($\approx$ 0.1--0.5\arcmin\ up to $\sim$5--7 times the size of the beam; \citet{Houde2009,Houde2016}). First, a preliminary solution is found by running the MCMC algorithm using uncertainties calculated according to \citet{Houde2016}. Final solutions are found by repeating the MCMC process and inflating the errors by the square root of the reduced goodness-of-fit coefficient, $\chi^{2}_{r}$.

\begin{figure}
    \centering
    \includegraphics[width=3.0in]{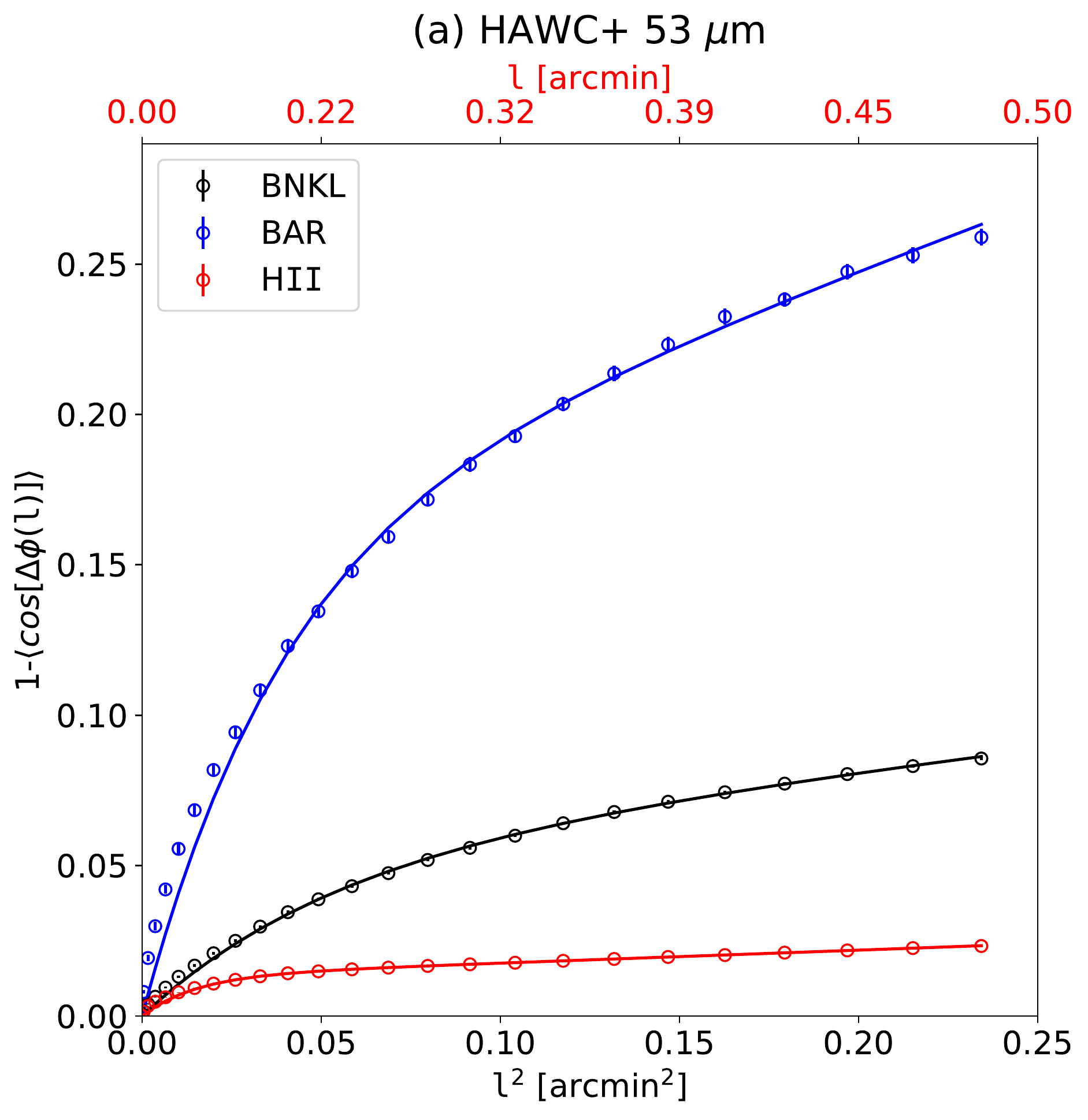}
    \includegraphics[width=3.0in]{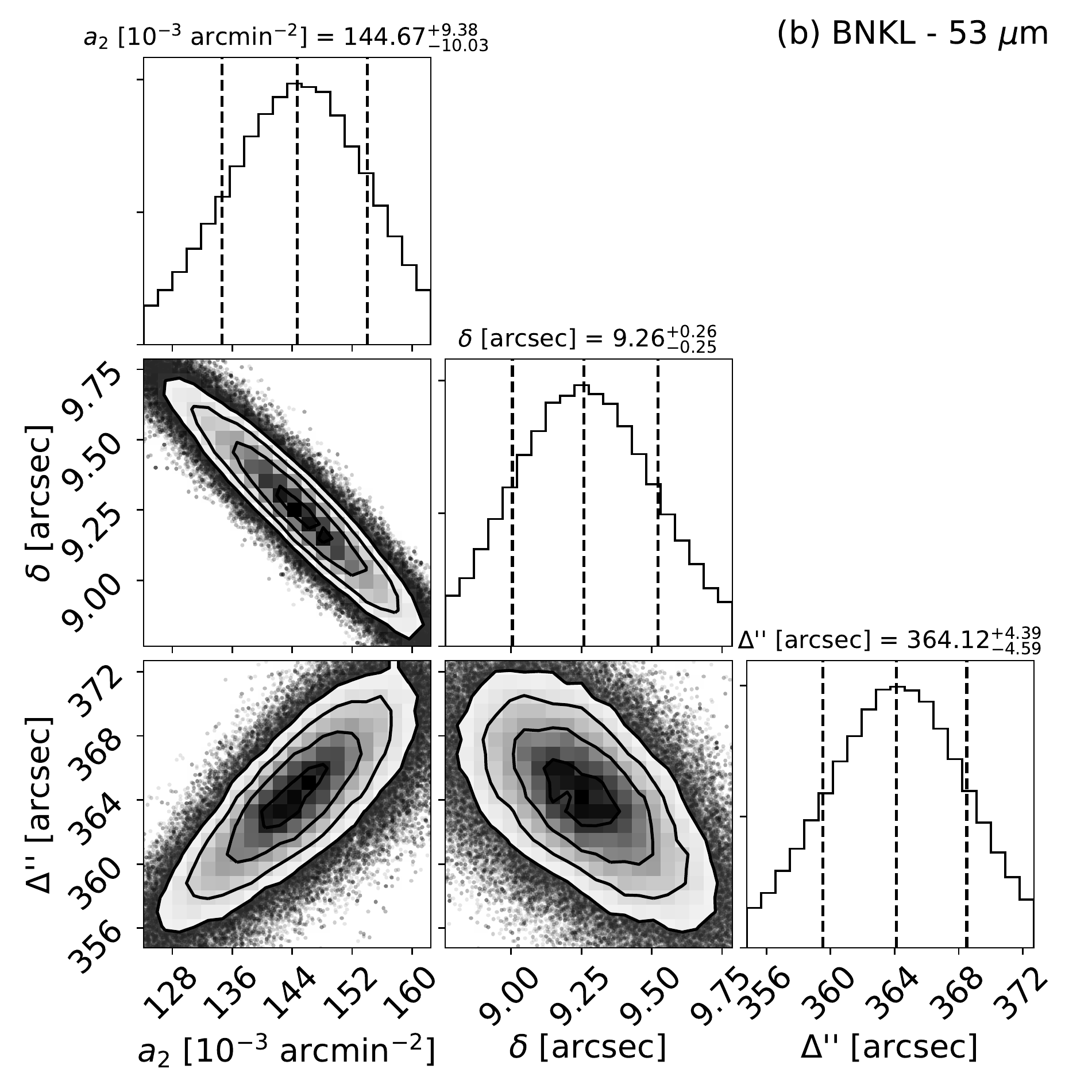}
    \includegraphics[width=3.0in]{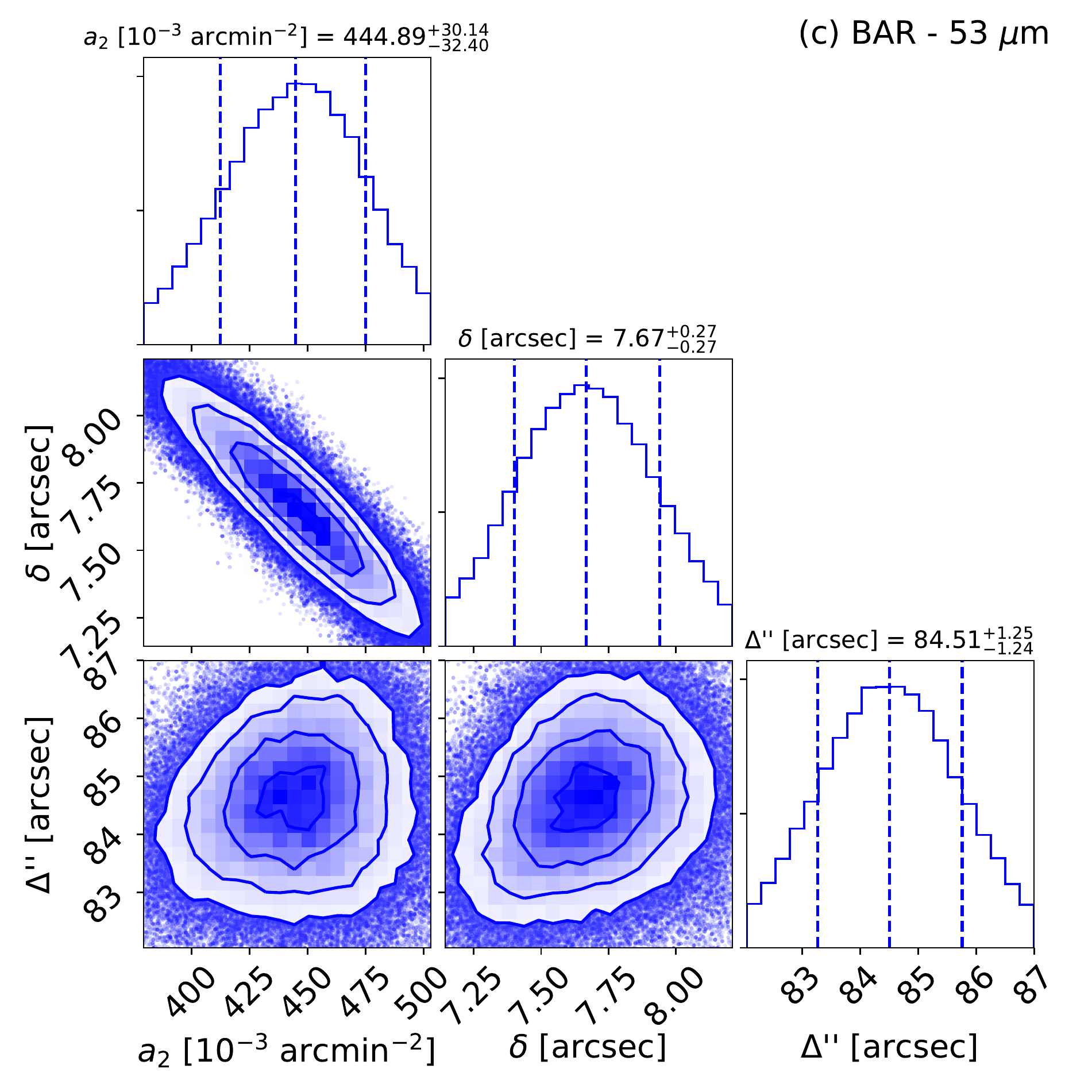}
    \includegraphics[width=3.0in]{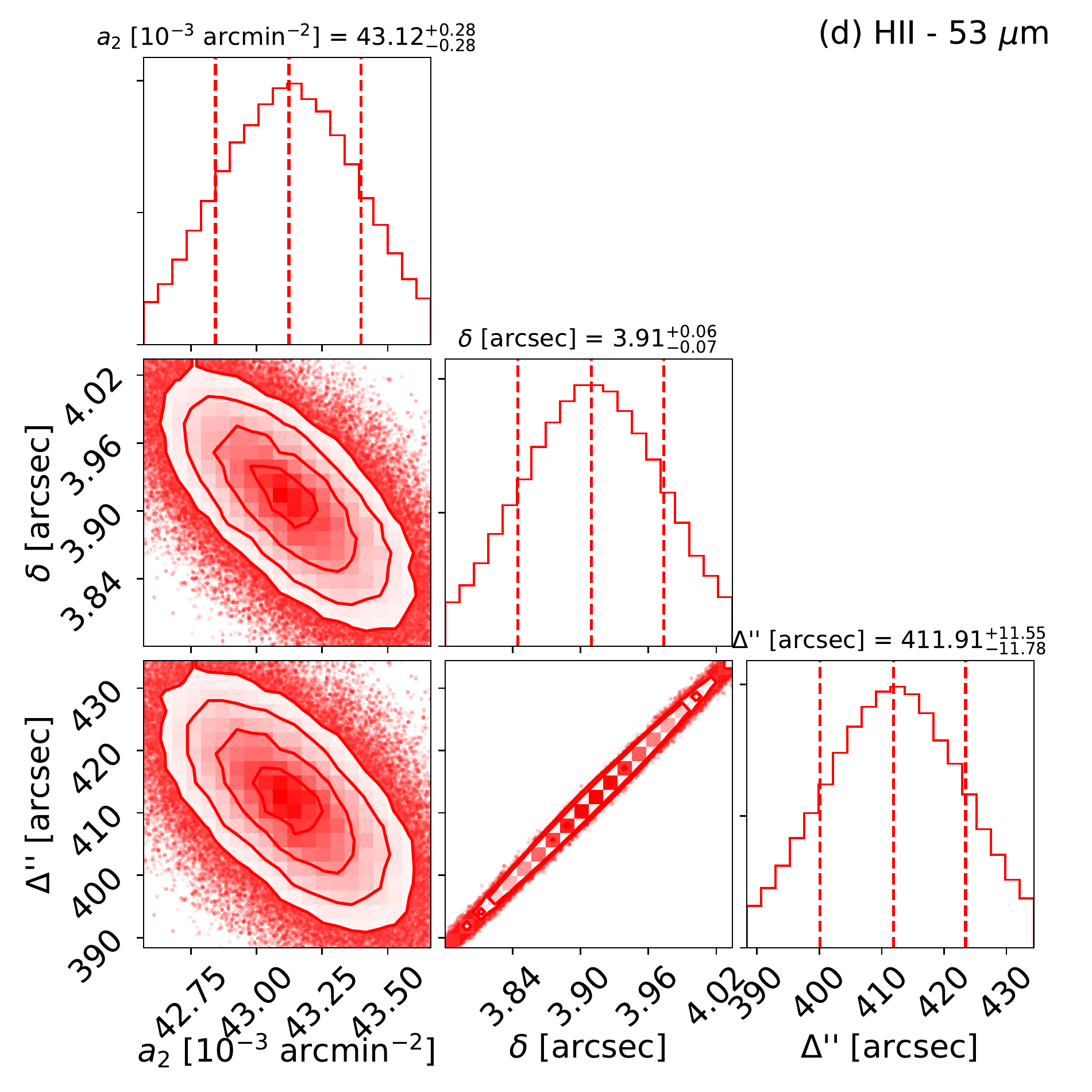}
    \caption{(a) Dispersion function for each OMC-1 region of interest using the 53 $\mu m$ data. Data points (circles) and fits (solid lines) are color-coded to match the region they represent: black, red, blue for BN/KL, the Bar, and the HII region, correspondingly. Solid lines correspond to the best fit of data using Eq. \ref{eq:disp_model} and the  parameters from Table \ref{tab:disp_results}. The model from Eq. \ref{eq:disp_model} only fits properly the dispersion functions at spatial scales smaller than the ordered (large scale) magnetic field, so the solid lines are shown only over the $l$-values used in each fit. (b), (c), and (d) show the results from the MCMC solver for each studied region.}
    \label{fig:DFsample}
\end{figure}

In Figure \ref{fig:DFsample}(a) we present the dispersion function (circles) for 53 $\mu$m data and best fit (solid lines) for small scales ($l\lesssim $ 0.5 arcmin). Data points and lines are color coded to match the regions in Figure \ref{fig:masks}: black, blue, and red for BNKL, BAR, and HII, respectively. As expected, all three curves show a dispersion that increases nonlinearly with angular distance $l$. As shown in Figure \ref{fig:DFsample}(a), the highest level of dispersion is present in the OMC-1 Bar. The BNKL region, in turn, shows a lower level of dispersion than BAR but higher than that of HII, which has the lowest level of dispersion of the three regions. This observation indicates the presence of a larger turbulent-field component in the BAR region than in the other two regions, which is in qualitative agreement with the visual inspection of the region in Figure \ref{fig:LIC}. However, if the explanation for the low observed fractional polarization is the superposition and cancellation of large scale fields, the dispersion, and ultimately $\frac{\langle B_{t}^{2}\rangle}{\langle B_{0}^{2}\rangle}$ may be overestimated by this technique. 

\begin{table}[]
    \centering
    \begin{tabular}{cccc}
    Wavelength &  $a_{2}$ & $\delta$ & $\Delta''$ \\
    $[\mu m]$ & [10$^{-3}$ arcmin$^{-2}$] & [arcsec] & [arcsec] \\
    \hline
    BNKL & & & \\
    \hline
    53 & 144.67$^{+ 9.38}_{- 10.03}$ & 9.26$^{+ 0.26}_{- 0.25}$ & 364.12$^{+ 4.39}_{- 4.59}$ \\
    89  & 74.34$^{+ 1.92}_{- 1.95}$ & 10.26$^{+ 0.30}_{- 0.30}$ & 387.36$^{+ 9.41}_{- 9.85}$ \\
    154 & 36.02$^{+ 3.34}_{- 3.70}$ & 21.69$^{+ 1.79}_{- 1.67}$ & 622.90$^{+ 23.16}_{- 22.55}$ \\
    214 & 7.15$^{+ 3.25}_{- 3.47}$ & 33.85$^{+ 2.68}_{- 2.61}$ & 707.10$^{+ 30.59}_{- 30.76}$ \\
    \hline
    BAR & & & \\
    \hline
    53 & 444.89$^{+ 30.14}_{- 32.40}$ & 7.67$^{+ 0.27}_{- 0.27}$ & 84.51$^{+ 1.25}_{- 1.24}$ \\
    89 & 158.65$^{+ 8.59}_{- 8.69}$ & 10.24$^{+ 0.21}_{- 0.20}$ & 94.74$^{+ 1.08}_{- 1.09}$ \\
    154 & -- & -- & -- \\
    214 & -- & -- & -- \\
    \hline
    HII & & & \\
    \hline
    53 & 43.12$^{+ 0.28}_{- 0.28}$ & 3.91$^{+ 0.06}_{- 0.07}$ & 411.91$^{+ 11.5}_{- 11.78}$ \\
    89 & 18.85$^{+ 1.48}_{- 1.55}$ & 9.29$^{+ 0.57}_{- 0.57}$ & 744.77$^{+ 46.93}_{- 50.14}$ \\
    154& 12.60$^{+ 0.73}_{- 0.73}$ & 9.37$^{+ 0.42}_{- 0.59}$ & 941.24$^{+ 43.67}_{- 95.68}$ \\
    214 & 14.78$^{+ 0.73}_{- 0.74}$ & 10.20$^{+ 0.61}_{- 1.09}$ & 888.12$^{+ 82.86}_{- 174.00}$ \\
    \hline
    \end{tabular}
    \caption{Parameters for the OMC-1 regions BN/KL, BAR, and HII, derived from the analysis of polarization vectors dispersion. These parameter were obtained by means of a Markov Chain Monte Carlo (MCMC) solver fitting the model in Eq. \ref{eq:disp_model} for the dispersion functions. Parameter values correspond to the quartile 0.5 (median) while errors correspond to the percentiles 0.16 and 0.84.}
    \label{tab:disp_results}
\end{table}

 Final best-fit parameters are summarized in Table \ref{tab:disp_results}. These results show clear differences among the three OMC-1 regions. Contributions in all regions to the dispersion from the ordered term, $a_{2}$, seem to decrease with increasing wavelength. When comparing $a_2$ in the different regions, the values for HII and BNKL are up to one order of magnitude lower than those in BAR. On the other hand, when examining the values of $\delta$ and $\Delta'$ we observe that in BNKL and BAR, these parameters increase with increasing wavelength, possibly in a non-linear way since the values for 53 and 89 $\micron$ are more similar to each other than to those of 154 and 214 $\micron$ values.  For the HII region, parameters $\delta$ and $\Delta'$ seem to increase with increasing wavelength (within their errors). This may be indicative of the presence of dust at different temperatures along the line-of-sight in BNKL and BAR.


Following \citet{Houde2009}, the parameters that characterize the turbulence in the studied regions, $\mathcal{N}$ and $\frac{\langle B_{t}^{2}\rangle}{\langle B_{0}^{2}\rangle}$, can be calculated as 

\begin{equation}
     \mathcal{N}(53~\mu m) = 6.67\left( \frac{\Delta'}{134~\mathrm{arcsec}}\right),
    \label{eq:N}
\end{equation}

\noindent
and 

\begin{equation}
    \frac{\langle B_{t}^{2}\rangle}{\langle B_{0}^{2}\rangle}(53~\micron) = \frac{\Delta'}{\Delta''} = 0.37 \left( \frac{\Delta'}{134~\mathrm{arcsec}}\right).
    \label{eq:Deltapp}
\end{equation}

In the equations above we have used $\delta$ and $W$ for BNKL 53 $\mu$m data from Tables \ref{tab:disp_results} and \ref{tab:obs}. As mentioned above, $\Delta'$ can be calculated as described in \citet{Houde2009}. We found $\Delta'$ = 2.27, 2.80, 3.80, and 4.97 arcmin for 53, 89, 154, and 214 $\micron$ data, respectively. These values, were obtained using the entire field of view in each band. Consequently, the strength of the large-scale magnetic field can be calculated as
 
\begin{equation}
    B_{0} \simeq \sqrt{4\pi\rho}\sigma(v)\!\!\left[\frac{\langle B_{t}^{2}\rangle}{\langle B_{0}^{2}\rangle}\right]^{-1/2} = \sqrt{4\pi\rho}\sigma(v)\!\!\left[\frac{\Delta'}{\Delta''}\right]^{-1/2},
    \label{eq:Bofit}
\end{equation}

\noindent
which is a modified version of the Davis-Chandrasekhar-Fermi relation. Using the fitted parameters for BNKL 53 $\mu$m, this leads to an estimate of the magnetic field strength in this region.

\begin{equation}
\begin{aligned}
     B_{0}(53~\mu m) & = 1002 \left(\frac{N(H_{2})}{9.85\times10^{22}\mathrm{cm^{-2}}}\right)^{1/2}\left(\frac{L}{4.34\times10^{17}\mathrm{cm}}\right)^{-1/2}\left(\frac{\sigma(v)}{1.85\times10^{5}\mathrm{cm/s}}\right)\left(\frac{\Delta'}{134~\mathrm{arsec}}\right)^{-1/2}~\mu G.
\end{aligned}
    \label{eq:B_0}
\end{equation}

\noindent
Here we have applied a nominal velocity dispersion value $\sigma(v) = $ 1.85 km/s \citep{Houde2009}, for all OMC-1 regions. Column densities correspond to average values for each region in the $N(H_{2})$ map of Figure \ref{fig:sedparams}. In order to transform column density to mass density we assume a uniform cloud depth $L = $4.34$\times$10$^{17}$ cm \citep{Pattle2017}. Resulting values for $\frac{\langle B_{t}^{2}\rangle}{\langle B_{0}^{2}\rangle}$, $B_{0}$, and $\mathcal{N}$ for all regions and bands are presented in Table \ref{tab:disp_results1}. Due to the potential contamination by reference intensity in some parts of the BAR region, only a very low number of pixels can be used for the analysis at 154 and 214 $\mu$m (Figure \ref{fig:rbcuts}), so dispersion functions for the BAR were not calculated in these bands. The results in Table \ref{tab:disp_results} show the BNKL region having the strongest magnetic field strength, $\sim$0.9 - 1.0 mG. The estimates for magnetic field strength in regions HII and BAR are approximately one third of the BNKL values, $\sim$300 $\mu$G. These values of plane-of-the-sky magnetic field strength are similar to the average value of 760 $\mu$G for the entire OMC-1 region estimated by \citet{Houde2009}, but significantly below the $6.6\pm4.7$ mG estimated by \cite{Pattle2017}. Our results indicate variation of magnetic field structure, not only in geometry but also in strength, across OMC-1. Consequently, it is possible that the magnetic field strength in OMC-1 displays significant spatial variations within each studied region, which, in the case of BNKL, can play a significant role in energy balance and magnetic dominance of the explosion observed in H$_{2}$ and CO emissions (See Section~\ref{sec:bnkl}).

\begin{table}[]
    \centering
    \begin{tabular}{ccccc}
    Wavelength & $N(H_{2})$ & $\frac{\langle B_{t}^{2}\rangle}{\langle B_{0}^{2}\rangle}$ & $B_{0}$  & $\mathcal{N}$ \\
    $[\mu m]$ & [cm$^{-2}$] &  & [$\mu$G] &  \\
    \hline
    BNKL & & & &   \\
    \hline
    53 & (9.85$\pm$8.96)$\times$10$^{22}$ & 0.37 & 1002 & 6.67 \\
    89 & \ldots & 0.43 & 931 & 8.42 \\
    154 & \ldots & 0.37 & 1013 & 5.02 \\
    214 & \ldots & 0.42 & 944 & 4.02 \\
    \hline
    BAR & & & &  \\
    \hline
    53 & (3.87$\pm$2.12)$\times$10$^{22}$ & 1.61 & 303 & 8.50 \\
    89 & \ldots & 1.77 & 289 & 8.44 \\
    154 & \ldots & -- & --  \\
    214 & \ldots & -- & -- \\
    \hline
    HII & & & \\
    \hline
    53 & (5.90$\pm$3.24)$\times$10$^{21}$ & 0.33 & 261 & 24.59\\
    89 & \ldots & 0.23 & 316 & 9.76 \\
    154 & \ldots & 0.24 & 305 & 19.32 \\
    214 & \ldots & 0.34 & 259 & 30.23 \\
    \hline
    \end{tabular}
    \caption{Physical parameters for OMC-1 regions derived from the results of the dispersion analysis. For each region/band the following parameters are reported: a) $N(H_{2})$, average column density of molecular hydrogen; b) turbulent-to-large-scale field ratio; c) $B_{0}$, plane-of-the-sky magnetic field intensity; d) $\mathcal{N}$, number of turbulent cells in the gas column. The uncertainties reported in the table are dominated by the uncertainty in the cloud depth (see Eqs.~\ref{eq:N}, \ref{eq:Deltapp}, and \ref{eq:Bofit}), which could be uncertain by a factor of $\sim$2.}
    \label{tab:disp_results1}
\end{table}

Our estimates of $\frac{\langle B_{t}^{2}\rangle}{\langle B_{0}^{2}\rangle}$ show that within the regions of the OMC-1 cloud studied here, HII and BAR are extreme cases in terms of turbulent states. The HII region seems to be a more ordered region with small turbulent components (turbulent field 0.23 - 0.34 times $\langle B_{0}^{2}\rangle$). The BAR, in contrast, appears as a highly turbulent region with components even greater than the large-scale field ($\frac{\langle B_{t}^{2}\rangle}{\langle B_{0}^{2}\rangle} >$1). However, as discussed in section~\ref{sec:Bgeom}, the low polarization in the Bar may indicate variations in grain alignment, a field predominantly oriented along the line of sight, a superposition of canceling (orthogonal) fields, or a combination of these effects. Such considerations would lower the value of the inferred large scale field and systematically inflate the dynamical importance of the turbulence. BNKL lies between these two regimes -- it shows between 0.37 and 0.43 times $\langle B_{0}^{2}\rangle$ for turbulent component. In terms of the number of turbulent cells present in the gas column, BNKL presents the lowest $\mathcal{N}$ ($\approx\ $5 - 8) while the HII region shows the highest $\mathcal{N}$ ($\approx\ $10 - 30).  

\subsection{The BN/KL Explosion}\label{sec:bnkl}
\begin{figure}
    \centering
    \includegraphics[width=7in]{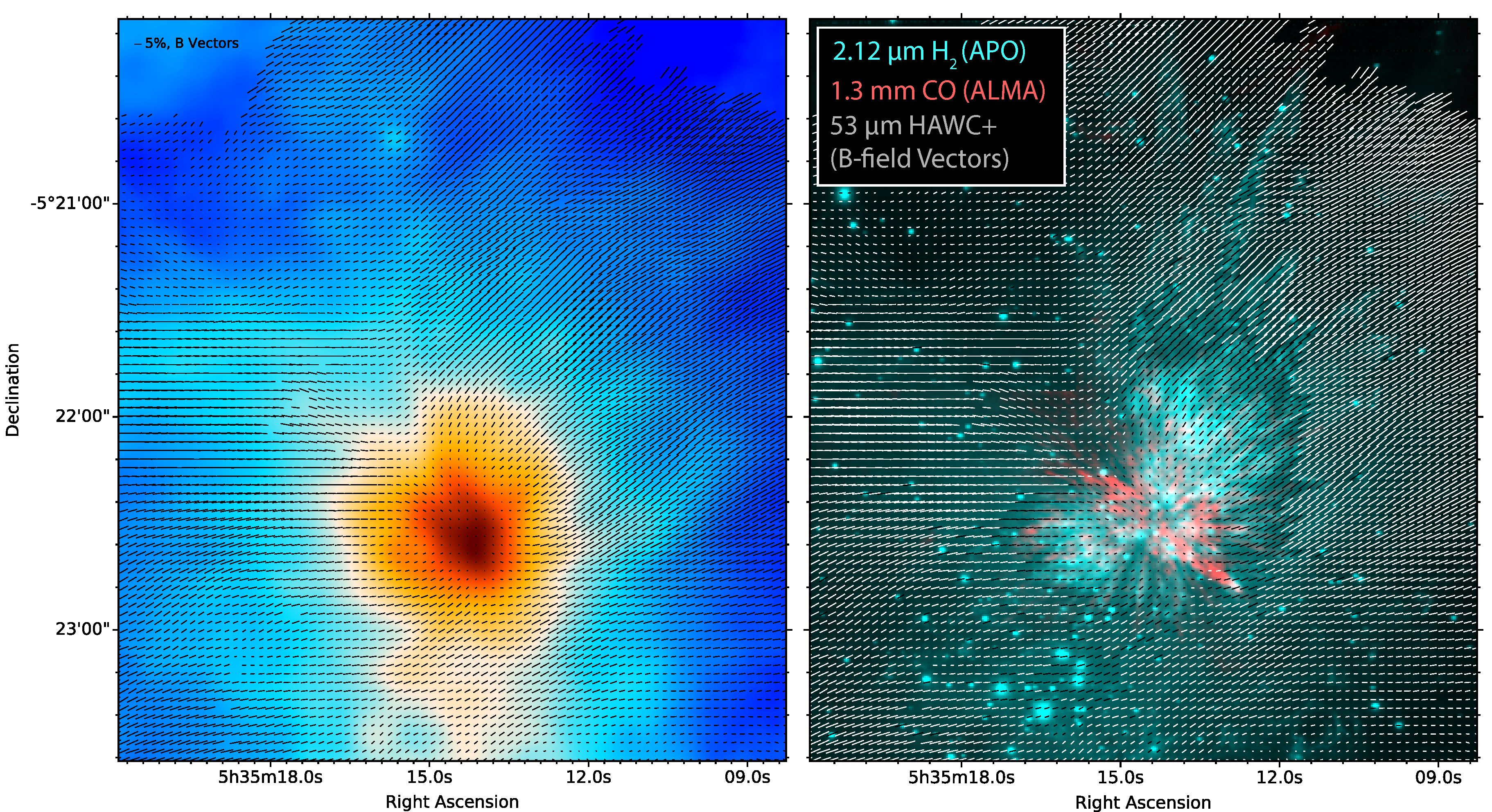}
    \caption{(left) The magnetic field vectors inferred from the 53 \micron\ polarimetry data are superposed on the 53 \micron\ intensity map. The polarization is sampled to Nyquist sample the HAWC+/SOFIA Band A beam.  (right) The same region is shown superposed on the 2.12 $\mu$m image from \cite{Bally2011} and the 1.3 mm ALMA CO map \citep{Bally2017}. The geometry of the magnetic field inferred by the band A polarimetry traces the bipolar outflow. }
    \label{fig:BNKL}
\end{figure}

The BNKL region has been identified as a site of a massive explosion possibly powered by stellar interactions \citep{Bally2005,Bally2011}. The energy associated with this explosion has been estimated to be of the order $\sim10^{47}$ ergs \citep{Snell1984}. The center of the explosion, which is roughly centered on the peak of the 53 $\mu$m intensity, is traced by high-velocity CO emission ``fingertips'' out to 30\arcsec--45\arcsec \citep{Bally2017}. These are distributed nearly isotropically in the plane of the sky around the center of the explosion.  Farther out, the explosion is traced by a bipolar outflow of H$_2$ ``fingers'' that extend 2\arcmin\ to 3\arcmin\ to the NNW and 2\arcmin\ to the SSE. The field lines inferred from the 53 $\mu$m HAWC+ observations appear to trace the H$_2$ fingers; however, they do not trace the isotropic pattern of the high-velocity CO, as shown in Figure~\ref{fig:BNKL}.  This suggests two possibilities for the magnetodynamics in the region.  First, the magnetic field could be confining the flow, shaping the bipolar feature by allowing the explosion to expand preferentially parallel to the large scale field. Second, the field could be being dragged by the explosion. 

It is possible to get a sense for the critical value of the field (i.e., that associated with the required energy needed to shape the outflow) based on energy considerations.  Because the field traces the bipolar pattern of the explosion in the larger volume defined by the H$_2$ fingers, but not in the smaller volume of the CO emission, we calculate the critical value of the field in each of these regions. For the smaller region, where the CO streams dominate, we assume a sphere of angular radius $\theta\sim30\arcsec$.  In this case, the mean energy density in the explosion can be approximated by 
\begin{equation}
    u_{\rm{explosion}}=8.3\times 10^{-6}\left(\frac{D}{400\, \mathrm{pc}}\right)^{-3}\left(\frac{\theta}{30\arcsec}\right)^{-3}\left(\frac{E}{2\times 10^{47}\,\mathrm{ergs}}\right)\mathrm{ergs}\,\mathrm{cm}^{-3}.
\end{equation}
Here, $D$ is the distance to BN/KL and $E$ is the total energy of the explosion. The magnetic field is given by $B=\sqrt{8\pi u_M}$, where $u_M$ is the magnetic energy density. We can define a mean critical field, $B_{\rm{crit}}$, as that required to produce magnetic energy density that is equal to the kinetic by setting $u_M=u_{\rm{explosion}}$,
\begin{equation}
    B_{\rm{crit}}=\sqrt{8\pi u_M}=14.4 \left(\frac{D}{400\, \mathrm{pc}}\right)^{-3/2}\left(\frac{\theta}{30\arcsec}\right)^{-3/2}\left(\frac{E}{2\times 10^{47}\,\mathrm{ergs}}\right)^{1/2}\, \mathrm{mG}.
\end{equation}
For the larger volume, we assume a cylindrical volume of radius $\theta_R\sim30$\arcsec and height $\theta_H\sim230$\arcsec.  We assume that the total energy in this volume is $\sim$1\% of that of the explosion itself. We calculate a similar critical field as above. 
\begin{equation}
    u_{\rm{explosion}}=1.4\times 10^{-8}\left(\frac{D}{400\, \mathrm{pc}}\right)^{-3}\left(\frac{\theta_R}{30\arcsec}\right)^{-2}\left(\frac{\theta_H}{230\arcsec}\right)^{-1}\left(\frac{E}{2\times 10^{45}\,\mathrm{ergs}}\right)\mathrm{ergs}\,\mathrm{cm}^{-3}.
\end{equation}
 The value of the mean critical magnetic field is then 
\begin{equation}
    B_{\rm{crit}}=0.6 \left(\frac{D}{400\, \mathrm{pc}}\right)^{-3/2}\left(\frac{\theta_R}{30\arcsec}\right)^{-1}\left(\frac{\theta_H}{230\arcsec}\right)^{-1/2}\left(\frac{E}{2\times 10^{45}\,\mathrm{ergs}}\right)^{1/2}\, \mathrm{mG}.
\end{equation}
Given the values for the BNKL region found in Section~\ref{sec:PolAngDisp} from the dispersion, we tentatively conclude that in the central 0.1 pc, the dynamics are dominated by the explosion as the magnetic field is much smaller than that required for energy balance with the explosion ($\sim$14 mG). This may explain why the high velocity CO gas shows an approximately isotropic distribution in the plane of the sky. On larger scales, the magnetic field strength estimates are $\sim$0.9-1.0 mG, which is in the range
of field strength for which the magnetic energy density to be of the same order of magnitude as that for the explosion  ($\sim$0.6 mG).  This could explain the common bipolar structure of the field and the H$_2$ gas in the outer regions of the explosion. 

This argument assumes that the characteristic field inferred from the DCF technique is approximately uniform over the BN/KL region.  Zeeman measurements of H$_2$O masers have indicated fields up to $\sim$40 mG near the infrared source IRc2 near the center of the BN/KL region \citep{Genzel1977,Garay1989,Fiebig1989} indicating that the details of the magnetic interaction are more complicated than the simple picture presented here. However, these large fields are likely confined to very small volumes of high density material and unlikely to significantly affect the dynamics in the volumes discussed here. Future polarimetric observations with ALMA may be able to shed more light on the details. 

\section{Summary} \label{sec:summary}
We have obtained new continuum far-infrared polarimetric and photometric maps of the OMC-1 region at 53, 89, 154, and 214 \micron\ using the HAWC+ instrument on SOFIA. 
\begin{enumerate}
    \item We have produced new maps of the temperatures and column densities of this region by combining HAWC+ photometry in four bands with other data sets. 
    \item The magnetic field geometry inferred from the polarization at these wavelengths indicate a similar large-scale field oriented roughly perpendicular to the BN/KL cloud as seen in previous studies. However, at the shorter (53, 89 \micron) wavelengths, the magnetic field structure around the BN Object shows a similar bipolar structure to molecular tracers of the BN/KL explosion.
    \item Analysis of the polarization fraction as a function of the local dispersion of the polarization vectors provides no evidence for loss of grain alignment within the cloud.  This could be due to stronger radiation fields in the OMC-1 region than in other Milky Way star formation regions.
    \item We estimate the magnetic field strength in the Bar and H II regions of the OMC-1 to be $\sim$250-300 $\mu$G. 
    \item Statistical estimates for the magnetic field indicate values of the field strength in th BN/KL region to be $\sim$1 mG, consistent with a picture in which the explosion dominates the magnetic field near the center, but the kinetic and magnetic energy densities are close to equipartition in the outer regions of the explosion.
    
\end{enumerate}

\section*{Acknowledgments}
Based on observations made with the NASA/DLR Stratospheric Observatory for Infrared Astronomy (SOFIA). SOFIA is jointly operated by the Universities Space Research Association, Inc. (USRA), under NASA contract NAS2-97001, and the Deutsches SOFIA Institut (DSI) under DLR contract 50 OK 0901 to the University of Stuttgart. Financial support for this work was provided by NASA through awards \#SOF 05-0038 and \#SOF 05-0018 issued by USRA.  
 
The authors would like to thank Joe Adams for skillful operation of HAWC+, SOFIA observatory personnel for the successful research flights leading to the OMC-1 results, Melanie Clarke for collaborative work on the HAWC+ data analysis pipeline, and Rahul Thapa and Lexi Tumblety for assistance with figure preparation and analysis. The authors would like to thank Simon Coud\'{e}, Steve Mairs, the JCMT Gould Belt Legacy Survey, and the East Asian Observatory (EAO) for their assistance in locating and using the CO-subtracted GBS data. The authors would like to thank Simon Dicker and Brian Mason for supplying the MUSTANG and X-band data. 

Parts of the analysis were performed using the Clusty Computing Facility in the Villanova Department of Astrophysics and Planetary Science. We thank Andrej Pr\v{s}a for his support in leading and maintaining this resource. 

PACS has been developed by a consortium of institutes led by MPE (Germany) and including UVIE (Austria); KU Leuven, CSL, IMEC (Belgium); CEA, LAM (France); MPIA (Germany); INAF-IFSI/OAA/OAP/OAT, LENS, SISSA (Italy); IAC (Spain). This development has been supported by the funding agencies BMVIT (Austria), ESA-PRODEX (Belgium), CEA/CNES (France), DLR (Germany), ASI/INAF (Italy), and CICYT/MCYT (Spain). SPIRE has been developed by a consortium of institutes led by Cardiff University (UK) and including Univ. Lethbridge (Canada); NAOC (China); CEA, LAM (France); IFSI, Univ. Padua (Italy); IAC (Spain); Stockholm Observatory (Sweden); Imperial College London, RAL, UCL-MSSL, UKATC, Univ. Sussex (UK); and Caltech, JPL, NHSC, Univ. Colorado (USA). This development has been supported by national funding agencies: CSA (Canada); NAOC (China); CEA, CNES, CNRS (France); ASI (Italy); MCINN (Spain); SNSB (Sweden); STFC, UKSA (UK); and NASA (USA). HCSS / HSpot / HIPE is a joint development (are joint developments) by the Herschel Science Ground Segment Consortium, consisting of ESA, the NASA Herschel Science Center, and the HIFI, PACS and SPIRE consortia. This work is based in part on observations made with Herschel, a European Space Agency Cornerstone Mission with significant participation by NASA. Portions of this work were carried out at the Jet propulsion Laboratory, operated by the California Institute of Technology under a contract with NASA. The James Clerk Maxwell Telescope has historically been operated by the Joint Astronomy Centre on behalf of the Science and Technology Facilities Council of the United Kingdom, the National Research Council of Canada and the Netherlands Organisation for Scientific Research. Additional funds for the construction of SCUBA-2 were provided by the Canada Foundation for Innovation. This paper made use of SCUBA-2 data taken as part of program ID MJLSG31.

We would like to thank the anonymous referee for their helpful comments. 

\software{ \texttt{python, Ipython} \citep{Perez2007}, \texttt{numpy} \citep{vanderWalt2011}, \texttt{scipy} \citep{Jones2001} \texttt{matplotlib} \citep{Hunter2007}, \texttt{emcee} \citep{Foreman-Mackey2013}, \texttt{corner} \citep{Foreman-Mackey2016}, \texttt{astropy} \citep{astropy:2013, astropy:2018}, LIC code (ported from publically-available IDL source by Diego Falceta-Gon\c{c}alves)} 

\bibliographystyle{aasjournal}
\bibliography{Chuss2018}

\begin{thebibliography}{}
\expandafter\ifx\csname natexlab\endcsname\relax\def\natexlab#1{#1}\fi
\providecommand{\url}[1]{\href{#1}{#1}}

\bibitem[{{Abergel}(2010)}]{Abergel2010}
{Abergel}, A. 2010, {SDP\_aabergel\_3: Evolution of interstellar dust},
  Herschel Space Observatory Proposal, ,

\bibitem[{{Allen} \& {Burton}(1993)}]{Allen1993}
{Allen}, D.~A., \& {Burton}, M.~G. 1993, \nat, 363, 54

\bibitem[{{Andersson} {et~al.}(2015){Andersson}, {Lazarian}, \&
  {Vaillancourt}}]{anla15}
{Andersson}, B.~G., {Lazarian}, A., \& {Vaillancourt}, J.~E. 2015, ARA\&A, 53,
  501

\bibitem[{{Andr{\'e}}(2007)}]{Andre2007}
{Andr{\'e}}, P. 2007, {KPGT\_pandre\_1: Probing the origin of the stellar
  initial mass function: A wide-field Herschel photometric survey of nearby
  star-forming cloud complexes}, Herschel Space Observatory Proposal, ,

\bibitem[{{Andr{\'e}}(2011)}]{Andre2011}
---. 2011, {GT2\_pandre\_5: Completion of the Gould Belt and HOBYS surveys},
  Herschel Space Observatory Proposal, ,

\bibitem[{{Arab} {et~al.}(2012){Arab}, {Abergel}, {Habart}, {Bernard-Salas},
  {Ayasso}, {Dassas}, {Martin}, \& {White}}]{Arab2012}
{Arab}, H., {Abergel}, A., {Habart}, E., {et~al.} 2012, \aap, 541, A19

\bibitem[{{Astropy Collaboration} {et~al.}(2013){Astropy Collaboration},
  {Robitaille}, {Tollerud}, {Greenfield}, {Droettboom}, {Bray}, {Aldcroft},
  {Davis}, {Ginsburg}, {Price-Whelan}, {Kerzendorf}, {Conley}, {Crighton},
  {Barbary}, {Muna}, {Ferguson}, {Grollier}, {Parikh}, {Nair}, {Unther},
  {Deil}, {Woillez}, {Conseil}, {Kramer}, {Turner}, {Singer}, {Fox}, {Weaver},
  {Zabalza}, {Edwards}, {Azalee Bostroem}, {Burke}, {Casey}, {Crawford},
  {Dencheva}, {Ely}, {Jenness}, {Labrie}, {Lim}, {Pierfederici}, {Pontzen},
  {Ptak}, {Refsdal}, {Servillat}, \& {Streicher}}]{astropy:2013}
{Astropy Collaboration}, {Robitaille}, T.~P., {Tollerud}, E.~J., {et~al.} 2013,
  \aap, 558, A33

\bibitem[{{Bally} {et~al.}(2011){Bally}, {Cunningham}, {Moeckel}, {Burton},
  {Smith}, {Frank}, \& {Nordlund}}]{Bally2011}
{Bally}, J., {Cunningham}, N.~J., {Moeckel}, N., {et~al.} 2011, \apj, 727, 113

\bibitem[{{Bally} {et~al.}(2017){Bally}, {Ginsburg}, {Arce}, {Eisner},
  {Youngblood}, {Zapata}, \& {Zinnecker}}]{Bally2017}
{Bally}, J., {Ginsburg}, A., {Arce}, H., {et~al.} 2017, \apj, 837, 60

\bibitem[{{Bally} \& {Zinnecker}(2005)}]{Bally2005}
{Bally}, J., \& {Zinnecker}, H. 2005, \aj, 129, 2281

\bibitem[{{Becklin} \& {Neugebauer}(1967)}]{Becklin1967}
{Becklin}, E.~E., \& {Neugebauer}, G. 1967, \apj, 147, 799

\bibitem[{{Bendo} {et~al.}(2013){Bendo}, {Griffin}, {Bock}, {Conversi},
  {Dowell}, {Lim}, {Lu}, {North}, {Papageorgiou}, {Pearson}, {Pohlen},
  {Polehampton}, {Schulz}, {Shupe}, {Sibthorpe}, {Spencer}, {Swinyard},
  {Valtchanov}, \& {Xu}}]{Bendo2013}
{Bendo}, G.~J., {Griffin}, M.~J., {Bock}, J.~J., {et~al.} 2013, \mnras, 433,
  3062

\bibitem[{{Beno{\^i}t} {et~al.}(2004){Beno{\^i}t}, {Ade}, {Amblard}, {Ansari},
  {Aubourg}, {Bargot}, {Bartlett}, {Bernard}, {Bhatia}, {Blanchard}, {Bock},
  {Boscaleri}, {Bouchet}, {Bourrachot}, {Camus}, {Couchot}, {de Bernardis},
  {Delabrouille}, {D{\'e}sert}, {Dor{\'e}}, {Douspis}, {Dumoulin}, {Dupac},
  {Filliatre}, {Fosalba}, {Ganga}, {Gannaway}, {Gautier}, {Giard},
  {Giraud-H{\'e}raud}, {Gispert}, {Guglielmi}, {Hamilton}, {Hanany},
  {Henrot-Versill{\'e}}, {Kaplan}, {Lagache}, {Lamarre}, {Lange},
  {Mac{\'{\i}}as-P{\'e}rez}, {Madet}, {Maffei}, {Magneville}, {Marrone},
  {Masi}, {Mayet}, {Murphy}, {Naraghi}, {Nati}, {Patanchon}, {Perrin}, {Piat},
  {Ponthieu}, {Prunet}, {Puget}, {Renault}, {Rosset}, {Santos}, {Starobinsky},
  {Strukov}, {Sudiwala}, {Teyssier}, {Tristram}, {Tucker}, {Vanel}, {Vibert},
  {Wakui}, \& {Yvon}}]{Benoit2004}
{Beno{\^i}t}, A., {Ade}, P., {Amblard}, A., {et~al.} 2004, \aap, 424, 571

\bibitem[{{Bernard} {et~al.}(2010){Bernard}, {Paradis}, {Marshall}, {Montier},
  {Lagache}, {Paladini}, {Veneziani}, {Brunt}, {Mottram}, {Martin},
  {Ristorcelli}, {Noriega-Crespo}, {Compi{\`e}gne}, {Flagey}, {Anderson},
  {Popescu}, {Tuffs}, {Reach}, {White}, {Benedettini}, {Calzoletti},
  {Digiorgio}, {Faustini}, {Juvela}, {Joblin}, {Joncas}, {Mivilles-Deschenes},
  {Olmi}, {Traficante}, {Piacentini}, {Zavagno}, \& {Molinari}}]{Bernard2010}
{Bernard}, J.~P., {Paradis}, D., {Marshall}, D.~J., {et~al.} 2010, \aap, 518,
  L88

\bibitem[{{Buckle} {et~al.}(2009){Buckle}, {Hills}, {Smith}, {Dent}, {Bell},
  {Curtis}, {Dace}, {Gibson}, {Graves}, {Leech}, {Richer}, {Williamson},
  {Withington}, {Yassin}, {Bennett}, {Hastings}, {Laidlaw}, {Lightfoot},
  {Burgess}, {Dewdney}, {Hovey}, {Willis}, {Redman}, {Wooff}, {Berry},
  {Cavanagh}, {Davis}, {Dempsey}, {Friberg}, {Jenness}, {Kackley}, {Rees},
  {Tilanus}, {Walther}, {Zwart}, {Klapwijk}, {Kroug}, \&
  {Zijlstra}}]{Buckle2009}
{Buckle}, J.~V., {Hills}, R.~E., {Smith}, H., {et~al.} 2009, \mnras, 399, 1026

\bibitem[{Cabral \& Leedom(1993)}]{Cabral1993}
Cabral, B., \& Leedom, L.~C. 1993, in Proceedings of the 20th annual conference
  on Computer graphics and interactive techniques, ACM, 263--270

\bibitem[{{Chandrasekhar} \& {Fermi}(1953)}]{Chandrasekhar1953}
{Chandrasekhar}, S., \& {Fermi}, E. 1953, \apj, 118, 113

\bibitem[{{Chapin} {et~al.}(2013){Chapin}, {Berry}, {Gibb}, {Jenness}, {Scott},
  {Tilanus}, {Economou}, \& {Holland}}]{Chapin2013}
{Chapin}, E.~L., {Berry}, D.~S., {Gibb}, A.~G., {et~al.} 2013, \mnras, 430,
  2545

\bibitem[{{Coud{\'e}} {et~al.}(2016){Coud{\'e}}, {Bastien}, {Kirk},
  {Johnstone}, {Drabek-Maunder}, {Graves}, {Hatchell}, {Chapin}, {Gibb},
  {Matthews}, \& {JCMT Gould Belt Survey Team}}]{Coude2016}
{Coud{\'e}}, S., {Bastien}, P., {Kirk}, H., {et~al.} 2016, \mnras, 457, 2139

\bibitem[{{Davis}(1951)}]{Davis1951}
{Davis}, L. 1951, PhRv, 81, 890

\bibitem[{{Dempsey} {et~al.}(2013){Dempsey}, {Friberg}, {Jenness}, {Tilanus},
  {Thomas}, {Holland}, {Bintley}, {Berry}, {Chapin}, {Chrysostomou}, {Davis},
  {Gibb}, {Parsons}, \& {Robson}}]{Dempsey2013}
{Dempsey}, J.~T., {Friberg}, P., {Jenness}, T., {et~al.} 2013, \mnras, 430,
  2534

\bibitem[{{Dicker} {et~al.}(2008){Dicker}, {Korngut}, {Mason}, {Ade},
  {Aguirre}, {Ames}, {Benford}, {Chen}, {Chervenak}, {Cotton}, {Devlin},
  {Figueroa-Feliciano}, {Irwin}, {Maher}, {Mello}, {Moseley}, {Tally},
  {Tucker}, \& {White}}]{Dicker2008}
{Dicker}, S.~R., {Korngut}, P.~M., {Mason}, B.~S., {et~al.} 2008, in Proc.
  SPIE, Vol. 7020, Millimeter and Submillimeter Detectors and Instrumentation
  for Astronomy IV, 702005

\bibitem[{{Dicker} {et~al.}(2009){Dicker}, {Mason}, {Korngut}, {Cotton},
  {Compi{\`e}gne}, {Devlin}, {Martin}, {Ade}, {Benford}, {Irwin}, {Maddalena},
  {McMullin}, {Shepherd}, {Sievers}, {Staguhn}, \& {Tucker}}]{Dicker2009}
{Dicker}, S.~R., {Mason}, B.~S., {Korngut}, P.~M., {et~al.} 2009, \apj, 705,
  226

\bibitem[{{Dolginov} \& {Mitrofanov}(1976)}]{dolginov1976}
{Dolginov}, A.~Z., \& {Mitrofanov}, I.~G. 1976, \apss, 43, 291

\bibitem[{{Dotson}(1996)}]{Dotson1996}
{Dotson}, J.~L. 1996, \apj, 470, 566

\bibitem[{Dowell(1997)}]{dowe97}
Dowell, C.~D. 1997, \apj, 487, 237

\bibitem[{{Draine} \& {Hensley}(2017)}]{Draine2017}
{Draine}, B.~T., \& {Hensley}, B.~S. 2017, ArXiv e-prints, arXiv:1710.08968

\bibitem[{{Draine} \& {Weingartner}(1997)}]{draine1997}
{Draine}, B.~T., \& {Weingartner}, J.~C. 1997, \apj, 480, 633

\bibitem[{{Dupac} {et~al.}(2001){Dupac}, {Giard}, {Bernard}, {Lamarre},
  {M{\'e}ny}, {Pajot}, {Ristorcelli}, {Serra}, \& {Torre}}]{Dupac2001}
{Dupac}, X., {Giard}, M., {Bernard}, J.~P., {et~al.} 2001, \apj, 553, 604

\bibitem[{{Dupac} {et~al.}(2003){Dupac}, {Bernard}, {Boudet}, {Giard},
  {Lamarre}, {M{\'e}ny}, {Pajot}, {Ristorcelli}, {Serra}, {Stepnik}, \&
  {Torre}}]{Dupac2003}
{Dupac}, X., {Bernard}, J.~P., {Boudet}, N., {et~al.} 2003, \aap, 404, L11

\bibitem[{{Fiebig} \& {Guesten}(1989)}]{Fiebig1989}
{Fiebig}, D., \& {Guesten}, R. 1989, \aap, 214, 333

\bibitem[{{Fissel} {et~al.}(2016){Fissel}, {Ade}, {Angil{\`e}}, {Ashton},
  {Benton}, {Devlin}, {Dober}, {Fukui}, {Galitzki}, {Gandilo}, {Klein},
  {Korotkov}, {Li}, {Martin}, {Matthews}, {Moncelsi}, {Nakamura},
  {Netterfield}, {Novak}, {Pascale}, {Poidevin}, {Santos}, {Savini}, {Scott},
  {Shariff}, {Diego Soler}, {Thomas}, {Tucker}, {Tucker}, \&
  {Ward-Thompson}}]{Fissel2016}
{Fissel}, L.~M., {Ade}, P.~A.~R., {Angil{\`e}}, F.~E., {et~al.} 2016, \apj,
  824, 134

\bibitem[{Foreman-Mackey(2016)}]{Foreman-Mackey2016}
Foreman-Mackey, D. 2016, JOSS, 24, doi:10.21105/joss.00024

\bibitem[{Foreman-Mackey {et~al.}(2013)Foreman-Mackey, Hogg, Lang, \&
  Goodman}]{Foreman-Mackey2013}
Foreman-Mackey, D., Hogg, D.~W., Lang, D., \& Goodman, J. 2013, PASP, 125, 306.
\newblock \url{http://stacks.iop.org/1538-3873/125/i=925/a=306}

\bibitem[{{Galametz} {et~al.}(2012){Galametz}, {Kennicutt}, {Albrecht},
  {Aniano}, {Armus}, {Bertoldi}, {Calzetti}, {Crocker}, {Croxall}, {Dale},
  {Donovan Meyer}, {Draine}, {Engelbracht}, {Hinz}, {Roussel}, {Skibba},
  {Tabatabaei}, {Walter}, {Weiss}, {Wilson}, \& {Wolfire}}]{Galametz2012}
{Galametz}, M., {Kennicutt}, R.~C., {Albrecht}, M., {et~al.} 2012, \mnras, 425,
  763

\bibitem[{{Galametz} {et~al.}(2018){Galametz}, {Maury}, {Girart}, {Rao},
  {Zhang}, {Gaudel}, {Valdivia}, {Keto}, \& {Lai}}]{gala18}
{Galametz}, M., {Maury}, A., {Girart}, J.~M., {et~al.} 2018, ArXiv e-prints,
  arXiv:1804.05801

\bibitem[{{Gandilo} {et~al.}(2016){Gandilo}, {Ade}, {Angil{\`e}}, {Ashton},
  {Benton}, {Devlin}, {Dober}, {Fissel}, {Fukui}, {Galitzki}, {Klein},
  {Korotkov}, {Li}, {Martin}, {Matthews}, {Moncelsi}, {Nakamura},
  {Netterfield}, {Novak}, {Pascale}, {Poidevin}, {Santos}, {Savini}, {Scott},
  {Shariff}, {Diego Soler}, {Thomas}, {Tucker}, {Tucker}, \&
  {Ward-Thompson}}]{Gandilo2016}
{Gandilo}, N.~N., {Ade}, P.~A.~R., {Angil{\`e}}, F.~E., {et~al.} 2016, \apj,
  824, 84

\bibitem[{{Garay} {et~al.}(1989){Garay}, {Moran}, \& {Haschick}}]{Garay1989}
{Garay}, G., {Moran}, J.~M., \& {Haschick}, A.~D. 1989, \apj, 338, 244

\bibitem[{{Genzel} \& {Downes}(1977)}]{Genzel1977}
{Genzel}, R., \& {Downes}, D. 1977, \aap, 61, 117

\bibitem[{{Griffin} {et~al.}(2010){Griffin}, {Abergel}, {Abreu}, {Ade},
  {Andr{\'e}}, {Augueres}, {Babbedge}, {Bae}, {Baillie}, {Baluteau}, {Barlow},
  {Bendo}, {Benielli}, {Bock}, {Bonhomme}, {Brisbin}, {Brockley-Blatt},
  {Caldwell}, {Cara}, {Castro-Rodriguez}, {Cerulli}, {Chanial}, {Chen},
  {Clark}, {Clements}, {Clerc}, {Coker}, {Communal}, {Conversi}, {Cox},
  {Crumb}, {Cunningham}, {Daly}, {Davis}, {de Antoni}, {Delderfield}, {Devin},
  {di Giorgio}, {Didschuns}, {Dohlen}, {Donati}, {Dowell}, {Dowell}, {Duband},
  {Dumaye}, {Emery}, {Ferlet}, {Ferrand}, {Fontignie}, {Fox}, {Franceschini},
  {Frerking}, {Fulton}, {Garcia}, {Gastaud}, {Gear}, {Glenn}, {Goizel},
  {Griffin}, {Grundy}, {Guest}, {Guillemet}, {Hargrave}, {Harwit}, {Hastings},
  {Hatziminaoglou}, {Herman}, {Hinde}, {Hristov}, {Huang}, {Imhof}, {Isaak},
  {Israelsson}, {Ivison}, {Jennings}, {Kiernan}, {King}, {Lange}, {Latter},
  {Laurent}, {Laurent}, {Leeks}, {Lellouch}, {Levenson}, {Li}, {Li},
  {Lilienthal}, {Lim}, {Liu}, {Lu}, {Madden}, {Mainetti}, {Marliani}, {McKay},
  {Mercier}, {Molinari}, {Morris}, {Moseley}, {Mulder}, {Mur}, {Naylor},
  {Nguyen}, {O'Halloran}, {Oliver}, {Olofsson}, {Olofsson}, {Orfei}, {Page},
  {Pain}, {Panuzzo}, {Papageorgiou}, {Parks}, {Parr-Burman}, {Pearce},
  {Pearson}, {P{\'e}rez-Fournon}, {Pinsard}, {Pisano}, {Podosek}, {Pohlen},
  {Polehampton}, {Pouliquen}, {Rigopoulou}, {Rizzo}, {Roseboom}, {Roussel},
  {Rowan-Robinson}, {Rownd}, {Saraceno}, {Sauvage}, {Savage}, {Savini},
  {Sawyer}, {Scharmberg}, {Schmitt}, {Schneider}, {Schulz}, {Schwartz},
  {Shafer}, {Shupe}, {Sibthorpe}, {Sidher}, {Smith}, {Smith}, {Smith},
  {Spencer}, {Stobie}, {Sudiwala}, {Sukhatme}, {Surace}, {Stevens}, {Swinyard},
  {Trichas}, {Tourette}, {Triou}, {Tseng}, {Tucker}, {Turner}, {Vaccari},
  {Valtchanov}, {Vigroux}, {Virique}, {Voellmer}, {Walker}, {Ward}, {Waskett},
  {Weilert}, {Wesson}, {White}, {Whitehouse}, {Wilson}, {Winter}, {Woodcraft},
  {Wright}, {Xu}, {Zavagno}, {Zemcov}, {Zhang}, \& {Zonca}}]{Griffin2010}
{Griffin}, M.~J., {Abergel}, A., {Abreu}, A., {et~al.} 2010, \aap, 518, L3

\bibitem[{{Guillet} {et~al.}(2018){Guillet}, {Fanciullo}, {Verstraete},
  {Boulanger}, {Jones}, {Miville-Desch{\^e}nes}, {Ysard}, {Levrier}, \&
  {Alves}}]{Guillet2018}
{Guillet}, V., {Fanciullo}, L., {Verstraete}, L., {et~al.} 2018, \aap, 610, A16

\bibitem[{Harper {et~al.}(2018)Harper, Runyan, Dowell, Wirth, Amato, Ames,
  Amiri, Banks, Bartels, Benford, Berthoud, Buchanan, Casey, Chapman, Chuss,
  Derro, Dotson, Evans, Fixsen, Gatley, Guerra, Halpern, Hamilton, Hamlin,
  Hansen, Heimsath, Hermida, Hilton, Hirsch, Hollister, Hostetter, Irwin,
  Jhabvala, Jhabvala, Kastner, Kovács, Loewenstein, Looney, Lopez-Rodriguez,
  Maher, Michail, Miller, Moseley, Novak, Pernic, Rennick, Rhody, Sandberg,
  Sandford, Santos, Shafer, Sharp, Shirron, Siah, Silverberg, Sparr, Spotz,
  Staguhn, Toorian, Towey, Tuttle, Vaillancourt, Voellmer, Volpert, i~Wang, \&
  Wollack}]{Harper2018}
Harper, D.~A., Runyan, M.~C., Dowell, C.~D., {et~al.} 2018, JAI, 7, 1840008

\bibitem[{{Heiles} {et~al.}(1993){Heiles}, {Goodman}, {McKee}, \&
  {Zweibel}}]{Heiles1993}
{Heiles}, C., {Goodman}, A.~A., {McKee}, C.~F., \& {Zweibel}, E.~G. 1993, in
  Protostars and Planets III, ed. E.~H. {Levy} \& J.~I. {Lunine}, 279

\bibitem[{Hensley {et~al.}(2015)Hensley, Murphy, \& Staguhn}]{Hensley2015}
Hensley, B., Murphy, E., \& Staguhn, J. 2015, MNRAS, 449, 809.
\newblock \url{http://dx.doi.org/10.1093/mnras/stv287}

\bibitem[{Hildebrand {et~al.}(2000)Hildebrand, Davidson, Dotson, Dowell, Novak,
  \& Vaillancourt}]{Hildebrand2000}
Hildebrand, R.~H., Davidson, J.~A., Dotson, J.~L., {et~al.} 2000, PASP, 112,
  1215.
\newblock \url{http://stacks.iop.org/1538-3873/112/i=775/a=1215}

\bibitem[{{Hildebrand} {et~al.}(1999){Hildebrand}, {Dotson}, {Dowell},
  {Schleuning}, \& {Vaillancourt}}]{hild99}
{Hildebrand}, R.~H., {Dotson}, J.~L., {Dowell}, C.~D., {Schleuning}, D.~A., \&
  {Vaillancourt}, J.~E. 1999, \apj, 516, 834

\bibitem[{{Hildebrand} \& {Dragovan}(1995)}]{Hildebrand1995}
{Hildebrand}, R.~H., \& {Dragovan}, M. 1995, \apj, 450, 663

\bibitem[{{Hildebrand} {et~al.}(2009){Hildebrand}, {Kirby}, {Dotson}, {Houde},
  \& {Vaillancourt}}]{Hildebrand2009}
{Hildebrand}, R.~H., {Kirby}, L., {Dotson}, J.~L., {Houde}, M., \&
  {Vaillancourt}, J.~E. 2009, \apj, 696, 567

\bibitem[{{Hill} {et~al.}(2011){Hill}, {Motte}, {Didelon}, {Bontemps},
  {Minier}, {Hennemann}, {Schneider}, {Andr{\'e}}, {Men'shchikov}, {Anderson},
  {Arzoumanian}, {Bernard}, {di Francesco}, {Elia}, {Giannini}, {Griffin},
  {K{\"o}nyves}, {Kirk}, {Marston}, {Martin}, {Molinari}, {Nguyen Luong},
  {Peretto}, {Pezzuto}, {Roussel}, {Sauvage}, {Sousbie}, {Testi},
  {Ward-Thompson}, {White}, {Wilson}, \& {Zavagno}}]{Hill2011}
{Hill}, T., {Motte}, F., {Didelon}, P., {et~al.} 2011, \aap, 533, A94

\bibitem[{{Holland} {et~al.}(2013){Holland}, {Bintley}, {Chapin},
  {Chrysostomou}, {Davis}, {Dempsey}, {Duncan}, {Fich}, {Friberg}, {Halpern},
  {Irwin}, {Jenness}, {Kelly}, {MacIntosh}, {Robson}, {Scott}, {Ade},
  {Atad-Ettedgui}, {Berry}, {Craig}, {Gao}, {Gibb}, {Hilton}, {Hollister},
  {Kycia}, {Lunney}, {McGregor}, {Montgomery}, {Parkes}, {Tilanus}, {Ullom},
  {Walther}, {Walton}, {Woodcraft}, {Amiri}, {Atkinson}, {Burger}, {Chuter},
  {Coulson}, {Doriese}, {Dunare}, {Economou}, {Niemack}, {Parsons},
  {Reintsema}, {Sibthorpe}, {Smail}, {Sudiwala}, \& {Thomas}}]{Holland2013}
{Holland}, W.~S., {Bintley}, D., {Chapin}, E.~L., {et~al.} 2013, \mnras, 430,
  2513

\bibitem[{{Houde} {et~al.}(2004){Houde}, {Dowell}, {Hildebrand}, {Dotson},
  {Vaillancourt}, {Phillips}, {Peng}, \& {Bastien}}]{Houde2004}
{Houde}, M., {Dowell}, C.~D., {Hildebrand}, R.~H., {et~al.} 2004, \apj, 604,
  717

\bibitem[{{Houde} {et~al.}(2016){Houde}, {Hull}, {Plambeck}, {Vaillancourt}, \&
  {Hildebrand}}]{Houde2016}
{Houde}, M., {Hull}, C. L.~H., {Plambeck}, R.~L., {Vaillancourt}, J.~E., \&
  {Hildebrand}, R.~H. 2016, \apj, 820, 38

\bibitem[{{Houde} {et~al.}(2011){Houde}, {Rao}, {Vaillancourt}, \&
  {Hildebrand}}]{Houde2011}
{Houde}, M., {Rao}, R., {Vaillancourt}, J.~E., \& {Hildebrand}, R.~H. 2011,
  \apj, 733, 109

\bibitem[{{Houde} {et~al.}(2009){Houde}, {Vaillancourt}, {Hildebrand},
  {Chitsazzadeh}, \& {Kirby}}]{Houde2009}
{Houde}, M., {Vaillancourt}, J.~E., {Hildebrand}, R.~H., {Chitsazzadeh}, S., \&
  {Kirby}, L. 2009, \apj, 706, 1504

\bibitem[{Hunter(2007)}]{Hunter2007}
Hunter, J.~D. 2007, CSE, 9

\bibitem[{{Johnston} {et~al.}(1989){Johnston}, {Migenes}, \&
  {Norris}}]{Johnston1989}
{Johnston}, K.~J., {Migenes}, V., \& {Norris}, R.~P. 1989, \apj, 341, 847

\bibitem[{Jones {et~al.}(2001)Jones, Oliphant, \& et~al.}]{Jones2001}
Jones, E., Oliphant, T., \& et~al., P.~P. 2001, SciPy: Open Source Scientific
  Tools for Python, , .
\newblock \url{http://www.scipy.org/}

\bibitem[{{Jones}(1989)}]{jone89}
{Jones}, T.~J. 1989, \apj, 346, 728

\bibitem[{{Jones} {et~al.}(2015){Jones}, {Bagley}, {Krejny}, {Andersson}, \&
  {Bastien}}]{joba15}
{Jones}, T.~J., {Bagley}, M., {Krejny}, M., {Andersson}, B.~G., \& {Bastien},
  P. 2015, \aj, 149, 31

\bibitem[{{Jones} \& {Whittet}(2015)}]{jowh15}
{Jones}, T.~J., \& {Whittet}, Douglas, C.~B. 2015, {Interstellar Polarization},
  147

\bibitem[{{Kleinmann} \& {Low}(1967)}]{Kleinmann1967}
{Kleinmann}, D.~E., \& {Low}, F.~J. 1967, \apj, 149, L1

\bibitem[{{Kobulnicky} {et~al.}(1994){Kobulnicky}, {Molnar}, \&
  {Jones}}]{kobu94}
{Kobulnicky}, H.~A., {Molnar}, L.~A., \& {Jones}, T.~J. 1994, \aj, 107, 1433

\bibitem[{Kounkel {et~al.}(2017)Kounkel, Hartmann, Loinard, Ortiz-LeÃ³n,
  Mioduszewski, RodrÃ­guez, Dzib, Torres, Pech, Galli, Rivera, Boden, II,
  BriceÃ±o, \& Tobin}]{Kounkel2017}
Kounkel, M., Hartmann, L., Loinard, L., {et~al.} 2017, ApJ, 834, 142.
\newblock \url{http://stacks.iop.org/0004-637X/834/i=2/a=142}

\bibitem[{{Kov{\'a}cs}(2006)}]{Kovacs2006}
{Kov{\'a}cs}, A. 2006, PhD thesis, Caltech

\bibitem[{{Kov{\'a}cs}(2008)}]{Kovacs2008}
{Kov{\'a}cs}, A. 2008, in Millimeter and Submillimeter Detectors and
  Instrumentation for Astronomy IV, Vol. 7020, 70201S

\bibitem[{{Lazarian} \& {Hoang}(2007)}]{Lazarian2007}
{Lazarian}, A., \& {Hoang}, T. 2007, \mnras, 378, 910

\bibitem[{{Mairs} {et~al.}(2016){Mairs}, {Johnstone}, {Kirk}, {Buckle},
  {Berry}, {Broekhoven-Fiene}, {Currie}, {Fich}, {Graves}, {Hatchell},
  {Jenness}, {Mottram}, {Nutter}, {Pattle}, {Pineda}, {Salji}, {Di Francesco},
  {Hogerheijde}, {Ward-Thompson}, {Bastien}, {Bresnahan}, {Butner}, {Chen},
  {Chrysostomou}, {Coud{\'e}}, {Davis}, {Drabek-Maunder}, {Duarte-Cabral},
  {Fiege}, {Friberg}, {Friesen}, {Fuller}, {Greaves}, {Gregson}, {Holland},
  {Joncas}, {Kirk}, {Knee}, {Marsh}, {Matthews}, {Moriarty- Schieven}, {Mowat},
  {Rawlings}, {Richer}, {Robertson}, {Rosolowsky}, {Rumble}, {Sadavoy},
  {Thomas}, {Tothill}, {Viti}, {White}, {Wouterloot}, {Yates}, \&
  {Zhu}}]{Mairs2016}
{Mairs}, S., {Johnstone}, D., {Kirk}, H., {et~al.} 2016, \mnras, 461, 4022

\bibitem[{{M\"{u}ller} {et~al.}(2011){M\"{u}ller}, {Okumura}, \&
  {Klaas}}]{Muller2011}
{M\"{u}ller}, T., {Okumura}, K., \& {Klaas}, U. 2011, {PACS Photometer
  Passbands and Colour Correction Factors for Various Source SEDs}.
\newblock
  \url{https://www.cosmos.esa.int/documents/12133/996891/PACS+Photometer+Passbands+and+Colour+Correction+Factors+for+Various+Source+SEDs}

\bibitem[{{Myers} \& {Goodman}(1991)}]{Myers1991}
{Myers}, P.~C., \& {Goodman}, A.~A. 1991, \apj, 373, 509

\bibitem[{{Novak}(2011)}]{Novak2011}
{Novak}, G. 2011, in ASP Conference Series, Vol. 449, Astronomical Polarimetry
  2008: Science from Small to Large Telescopes, ed. P.~{Bastien}, N.~{Manset},
  D.~P. {Clemens}, \& N.~{St- Louis}, 50

\bibitem[{Novak {et~al.}(1997)Novak, Dotson, Dowell, Goldsmith, Hildebrand,
  Platt, \& Schleuning}]{Novak1997}
Novak, G., Dotson, J.~L., Dowell, C.~D., {et~al.} 1997, ApJ, 487, 320.
\newblock \url{http://stacks.iop.org/0004-637X/487/i=1/a=320}

\bibitem[{{Novak} {et~al.}(2000){Novak}, {Dotson}, {Dowell}, {Hildebrand},
  {Renbarger}, \& {Schleuning}}]{Novak2000}
{Novak}, G., {Dotson}, J.~L., {Dowell}, C.~D., {et~al.} 2000, \apj, 529, 241

\bibitem[{{Ott}(2010)}]{Ott2010}
{Ott}, S. 2010, in Astronomical Data Analysis Software and Systems XIX, Vol.
  434, 139

\bibitem[{{Pattle} {et~al.}(2017){Pattle}, {Ward-Thompson}, {Berry},
  {Hatchell}, {Chen}, {Pon}, {Koch}, {Kwon}, {Kim}, {Bastien}, {Cho},
  {Coud{\'e}}, {Di Francesco}, {Fuller}, {Furuya}, {Graves}, {Johnstone},
  {Kirk}, {Kwon}, {Lee}, {Matthews}, {Mottram}, {Parsons}, {Sadavoy},
  {Shinnaga}, {Soam}, {Hasegawa}, {Lai}, {Qiu}, \& {Friberg}}]{Pattle2017}
{Pattle}, K., {Ward-Thompson}, D., {Berry}, D., {et~al.} 2017, \apj, 846, 122

\bibitem[{P\'{e}rez \& Granger(2007)}]{Perez2007}
P\'{e}rez, F., \& Granger, B.~E. 2007, CSE, 9, doi:10.1109/MCSE.2007.53

\bibitem[{{Pilbratt} {et~al.}(2010){Pilbratt}, {Riedinger}, {Passvogel},
  {Crone}, {Doyle}, {Gageur}, {Heras}, {Jewell}, {Metcalfe}, {Ott}, \&
  {Schmidt}}]{Pilbratt2010}
{Pilbratt}, G.~L., {Riedinger}, J.~R., {Passvogel}, T., {et~al.} 2010, \aap,
  518, L1

\bibitem[{{Planck Collaboration} {et~al.}(2015){Planck Collaboration}, {Ade},
  {Aghanim}, {Alina}, {Alves}, {Armitage-Caplan}, {Arnaud}, {Arzoumanian},
  {Ashdown}, {Atrio-Barandela}, \& et~al.}]{Planck2015_XIX}
{Planck Collaboration}, {Ade}, P.~A.~R., {Aghanim}, N., {et~al.} 2015, \aap,
  576, A104

\bibitem[{{Planck Collaboration} {et~al.}(2018){Planck Collaboration},
  {Aghanim}, {Akrami}, {Alves}, {Ashdown}, {Aumont}, {Baccigalupi},
  {Ballardini}, {Banday}, {Barreiro}, {Bartolo}, {Basak}, {Benabed}, {Bernard},
  {Bersanelli}, {Bielewicz}, {Bock}, {Bond}, {Borrill}, {Bouchet}, {Boulanger},
  {Bracco}, {Bucher}, {Burigana}, {Calabrese}, {Cardoso}, {Carron}, {Chary},
  {Chiang}, {Colombo}, {Combet}, {Crill}, {Cuttaia}, {de Bernardis}, {de
  Zotti}, {Delabrouille}, {Delouis}, {Di Valentino}, {Dickinson}, {Diego},
  {Dor{\'e}}, {Douspis}, {Ducout}, {Dupac}, {Efstathiou}, {Elsner},
  {En{\ss}lin}, {Eriksen}, {Fantaye}, {Fernandez-Cobos}, {Ferri{\`e}re},
  {Forastieri}, {Frailis}, {Fraisse}, {Franceschi}, {Frolov}, {Galeotta},
  {Galli}, {Ganga}, {G{\'e}nova-Santos}, {Gerbino}, {Ghosh},
  {Gonz{\'a}lez-Nuevo}, {G{\'o}rski}, {Gratton}, {Green}, {Gruppuso},
  {Gudmundsson}, {Guillet}, {Handley}, {Hansen}, {Helou}, {Herranz}, {Hivon},
  {Huang}, {Jaffe}, {Jones}, {Keih{\"a}nen}, {Keskitalo}, {Kiiveri}, {Kim},
  {Krachmalnicoff}, {Kunz}, {Kurki-Suonio}, {Lagache}, {Lamarre}, {Lasenby},
  {Lattanzi}, {Lawrence}, {Le Jeune}, {Levrier}, {Liguori}, {Lilje},
  {Lindholm}, {L{\'o}pez-Caniego}, {Lubin}, {Ma}, {Mac{\'\i}as-P{\'e}rez},
  {Maggio}, {Maino}, {Mandolesi}, {Mangilli}, {Marcos-Caballero}, {Maris},
  {Martin}, {Mart{\'\i}nez-Gonz{\'a}lez}, {Matarrese}, {Mauri}, {McEwen},
  {Melchiorri}, {Mennella}, {Migliaccio}, {Miville-Desch{\^e}nes}, {Molinari},
  {Moneti}, {Montier}, {Morgante}, {Moss}, {Natoli}, {Pagano}, {Paoletti},
  {Patanchon}, {Perrotta}, {Pettorino}, {Piacentini}, {Polastri}, {Polenta},
  {Puget}, {Rachen}, {Reinecke}, {Remazeilles}, {Renzi}, {Ristorcelli},
  {Rocha}, {Rosset}, {Roudier}, {Rubi{\~n}o-Mart{\'\i}n}, {Ruiz-Granados},
  {Salvati}, {Sandri}, {Savelainen}, {Scott}, {Sirignano}, {Sunyaev},
  {Suur-Uski}, {Tauber}, {Tavagnacco}, {Tenti}, {Toffolatti}, {Tomasi},
  {Trombetti}, {Valiviita}, {Van Tent}, {Vielva}, {Villa}, {Vittorio},
  {Wandelt}, {Wehus}, {Zacchei}, \& {Zonca}}]{plan18}
{Planck Collaboration}, {Aghanim}, N., {Akrami}, Y., {et~al.} 2018, ArXiv
  e-prints, arXiv:1807.06212

\bibitem[{{Poglitsch} {et~al.}(2010){Poglitsch}, {Waelkens}, {Geis},
  {Feuchtgruber}, {Vandenbussche}, {Rodriguez}, {Krause}, {Renotte}, {van
  Hoof}, {Saraceno}, {Cepa}, {Kerschbaum}, {Agn{\`e}se}, {Ali}, {Altieri},
  {Andreani}, {Augueres}, {Balog}, {Barl}, {Bauer}, {Belbachir}, {Benedettini},
  {Billot}, {Boulade}, {Bischof}, {Blommaert}, {Callut}, {Cara}, {Cerulli},
  {Cesarsky}, {Contursi}, {Creten}, {De Meester}, {Doublier}, {Doumayrou},
  {Duband}, {Exter}, {Genzel}, {Gillis}, {Gr{\"o}zinger}, {Henning},
  {Herreros}, {Huygen}, {Inguscio}, {Jakob}, {Jamar}, {Jean}, {de Jong},
  {Katterloher}, {Kiss}, {Klaas}, {Lemke}, {Lutz}, {Madden}, {Marquet},
  {Martignac}, {Mazy}, {Merken}, {Montfort}, {Morbidelli}, {M{\"u}ller},
  {Nielbock}, {Okumura}, {Orfei}, {Ottensamer}, {Pezzuto}, {Popesso},
  {Putzeys}, {Regibo}, {Reveret}, {Royer}, {Sauvage}, {Schreiber}, {Stegmaier},
  {Schmitt}, {Schubert}, {Sturm}, {Thiel}, {Tofani}, {Vavrek}, {Wetzstein},
  {Wieprecht}, \& {Wiezorrek}}]{Poglitsch2010}
{Poglitsch}, A., {Waelkens}, C., {Geis}, N., {et~al.} 2010, \aap, 518, L2

\bibitem[{{Poidevin} {et~al.}(2011){Poidevin}, {Bastien}, \& {Jones}}]{poid11}
{Poidevin}, F., {Bastien}, P., \& {Jones}, T.~J. 2011, \apj, 741, 112

\bibitem[{{Price-Whelan} {et~al.}(2018){Price-Whelan}, {Sip{\'{o}}cz},
  {G{\"u}nther}, {Lim}, {Crawford}, {Conseil}, {Shupe}, {Craig}, {Dencheva},
  {Ginsburg}, {VanderPlas}, {Bradley}, {P{'e}rez-Su{'a}rez}, {de Val-Borro},
  {Paper Contributors}, {Aldcroft}, {Cruz}, {Robitaille}, {Tollerud},
  {Coordination Committee}, {Ardelean}, {Babej}, {Bach}, {Bachetti}, {Bakanov},
  {Bamford}, {Barentsen}, {Barmby}, {Baumbach}, {Berry}, {Biscani}, {Boquien},
  {Bostroem}, {Bouma}, {Brammer}, {Bray}, {Breytenbach}, {Buddelmeijer},
  {Burke}, {Calderone}, {Cano Rodr{'i}guez}, {Cara}, {Cardoso}, {Cheedella},
  {Copin}, {Corrales}, {Crichton}, {D{ extquoteright}Avella}, {Deil},
  {Depagne}, {Dietrich}, {Donath}, {Droettboom}, {Earl}, {Erben}, {Fabbro},
  {Ferreira}, {Finethy}, {Fox}, {Garrison}, {Gibbons}, {Goldstein}, {Gommers},
  {Greco}, {Greenfield}, {Groener}, {Grollier}, {Hagen}, {Hirst}, {Homeier},
  {Horton}, {Hosseinzadeh}, {Hu}, {Hunkeler}, {Ivezi{'c}}, {Jain}, {Jenness},
  {Kanarek}, {Kendrew}, {Kern}, {Kerzendorf}, {Khvalko}, {King}, {Kirkby},
  {Kulkarni}, {Kumar}, {Lee}, {Lenz}, {Littlefair}, {Ma}, {Macleod},
  {Mastropietro}, {McCully}, {Montagnac}, {Morris}, {Mueller}, {Mumford},
  {Muna}, {Murphy}, {Nelson}, {Nguyen}, {Ninan}, {N{"o}the}, {Ogaz}, {Oh},
  {Parejko}, {Parley}, {Pascual}, {Patil}, {Patil}, {Plunkett}, {Prochaska},
  {Rastogi}, {Reddy Janga}, {Sabater}, {Sakurikar}, {Seifert}, {Sherbert},
  {Sherwood-Taylor}, {Shih}, {Sick}, {Silbiger}, {Singanamalla}, {Singer},
  {Sladen}, {Sooley}, {Sornarajah}, {Streicher}, {Teuben}, {Thomas},
  {Tremblay}, {Turner}, {Terr{'o}n}, {van Kerkwijk}, {de la Vega}, {Watkins},
  {Weaver}, {Whitmore}, {Woillez}, {Zabalza}, \& {Contributors}}]{astropy:2018}
{Price-Whelan}, A.~M., {Sip{\'{o}}cz}, B.~M., {G{\"u}nther}, H.~M., {et~al.}
  2018, \aj, 156, 123

\bibitem[{{Roussel}(2013)}]{Roussel2013}
{Roussel}, H. 2013, PASP, 125, 1126

\bibitem[{{Sadavoy} {et~al.}(2013){Sadavoy}, {Di Francesco}, {Johnstone},
  {Currie}, {Drabek}, {Hatchell}, {Nutter}, {Andr{\'e}}, {Arzoumanian},
  {Benedettini}, {Bernard}, {Duarte-Cabral}, {Fallscheer}, {Friesen},
  {Greaves}, {Hennemann}, {Hill}, {Jenness}, {K{\"o}nyves}, {Matthews},
  {Mottram}, {Pezzuto}, {Roy}, {Rygl}, {Schneider- Bontemps}, {Spinoglio},
  {Testi}, {Tothill}, {Ward-Thompson}, {White}, {JCMT}, \& {Herschel Gould Belt
  Survey Teams}}]{Sadavoy2013}
{Sadavoy}, S.~I., {Di Francesco}, J., {Johnstone}, D., {et~al.} 2013, \apj,
  767, 126

\bibitem[{{Schleuning}(1998)}]{Schleuning1998}
{Schleuning}, D.~A. 1998, \apj, 493, 811

\bibitem[{Serkowski(1974)}]{Serkowski1974}
Serkowski, K. 1974, in Methods in Experimental Physics, Vol.~12, Astrophysics,
  ed. N.~Carleton (Academic Press), 361 -- 414.
\newblock
  \url{http://www.sciencedirect.com/science/article/pii/S0076695X08605001}

\bibitem[{{Shetty} {et~al.}(2009){Shetty}, {Kauffmann}, {Schnee}, {Goodman}, \&
  {Ercolano}}]{Shetty2009}
{Shetty}, R., {Kauffmann}, J., {Schnee}, S., {Goodman}, A.~A., \& {Ercolano},
  B. 2009, \apj, 696, 2234

\bibitem[{{Snell} {et~al.}(1984){Snell}, {Scoville}, {Sanders}, \&
  {Erickson}}]{Snell1984}
{Snell}, R.~L., {Scoville}, N.~Z., {Sanders}, D.~B., \& {Erickson}, N.~R. 1984,
  \apj, 284, 176

\bibitem[{{Soler} {et~al.}(2013){Soler}, {Hennebelle}, {Martin}, {Miville-
  Desch{\^e}nes}, {Netterfield}, \& {Fissel}}]{Soler2013}
{Soler}, J.~D., {Hennebelle}, P., {Martin}, P.~G., {et~al.} 2013, \apj, 774,
  128

\bibitem[{{Tang} {et~al.}(2010){Tang}, {Ho}, {Koch}, \& {Rao}}]{Tang2010}
{Tang}, Y.-W., {Ho}, P.~T.~P., {Koch}, P.~M., \& {Rao}, R. 2010, \apj, 717,
  1262

\bibitem[{{Vaillancourt}(2002)}]{Vaillancourt2002}
{Vaillancourt}, J.~E. 2002, \apjs, 142, 53

\bibitem[{{Vall{\'e}e} \& {Bastien}(1999)}]{Vallee1999}
{Vall{\'e}e}, J.~P., \& {Bastien}, P. 1999, \apj, 526, 819

\bibitem[{{Valtchanov}(2017)}]{HESV_4}
{Valtchanov}, I. 2017, "Herschel Explanatory Supplement volume IV: The Spectral
  and Photometric Imaging Receiver (SPIRE) Handbook".
\newblock \url{http://herschel.esac.esa.int/Docs/SPIRE/spire_handbook.pdf}

\bibitem[{{van der Walt} {et~al.}(2011){van der Walt}, Colbert, \&
  Varoquaux}]{vanderWalt2011}
{van der Walt}, S., Colbert, S.~C., \& Varoquaux, G. 2011, CSE, 13,
  doi:10.1109/MCSE.2011.37

\bibitem[{{Ward-Thompson} {et~al.}(2017){Ward-Thompson}, {Pattle}, {Bastien},
  {Furuya}, {Kwon}, {Lai}, {Qiu}, {Berry}, {Choi}, {Coud{\'e}}, {Di Francesco},
  {Hoang}, {Franzmann}, {Friberg}, {Graves}, {Greaves}, {Houde}, {Johnstone},
  {Kirk}, {Koch}, {Kwon}, {Lee}, {Li}, {Matthews}, {Mottram}, {Parsons}, {Pon},
  {Rao}, {Rawlings}, {Shinnaga}, {Sadavoy}, {van Loo}, {Aso}, {Byun},
  {Eswaraiah}, {Chen}, {Chen}, {Chen}, {Ching}, {Cho}, {Chrysostomou}, {Chung},
  {Doi}, {Drabek-Maunder}, {Eyres}, {Fiege}, {Friesen}, {Fuller}, {Gledhill},
  {Griffin}, {Gu}, {Hasegawa}, {Hatchell}, {Hayashi}, {Holland}, {Inoue},
  {Inutsuka}, {Iwasaki}, {Jeong}, {Kang}, {Kang}, {Kang}, {Kawabata}, {Kemper},
  {Kim}, {Kim}, {Kim}, {Kim}, {Kim}, {Kim}, {Lacaille}, {Lee}, {Lee}, {Li},
  {Li}, {Liu}, {Liu}, {Liu}, {Liu}, {Lyo}, {Mairs}, {Matsumura},
  {Moriarty-Schieven}, {Nakamura}, {Nakanishi}, {Ohashi}, {Onaka}, {Peretto},
  {Pyo}, {Qian}, {Retter}, {Richer}, {Rigby}, {Robitaille}, {Savini}, {Scaife},
  {Soam}, {Tamura}, {Tang}, {Tomisaka}, {Wang}, {Wang}, {Whitworth}, {Yen},
  {Yoo}, {Yuan}, {Zhang}, {Zhang}, {Zhou}, {Zhu}, {Andr{\'e}}, {Dowell},
  {Falle}, \& {Tsukamoto}}]{Ward-Thompson2017}
{Ward-Thompson}, D., {Pattle}, K., {Bastien}, P., {et~al.} 2017, \apj, 842, 66

\end{thebibliography}

\end{document}